\documentclass[journal,12pt,draftclsnofoot,onecolumn]{IEEEtran}
\usepackage{color}
\usepackage{balance}
\usepackage{graphicx}
\usepackage{enumerate}
\usepackage{subfigure}
\usepackage{algorithm}
\usepackage{algorithmic}
\usepackage{multicol}     
\usepackage{multirow} 
\usepackage{booktabs}
\usepackage{makecell}
\usepackage{hyperref}
\hypersetup{colorlinks=true,
	linkcolor=black,
	citecolor=black}
\usepackage{cite}
\usepackage{amssymb}
\usepackage{amsmath}
\usepackage{bm}
\usepackage{authblk}
\usepackage{stfloats}
\usepackage[utf8]{inputenc}
\usepackage{amsthm}
\usepackage{setspace}
\newtheoremstyle{note}
{3pt}
{3pt}
{\itshape}
{1em}
{\itshape}
{:}
{.5em}
{}

\renewenvironment{proof}{{$\quad $\it Proof:}$\ $}{$\hfill\blacksquare$}

\begin{document}
	
\title{Outage Performance of Uplink Rate Splitting Multiple Access with Randomly Deployed Users}
	
%
		
\author{
 \quad Huabing Lu,~\IEEEmembership{Member,~IEEE,}
Xianzhong Xie,~\IEEEmembership{Member,~IEEE,}\qquad \qquad\qquad
Zhaoyuan Shi,~\IEEEmembership{Member,~IEEE,}
Hongjiang Lei,~\IEEEmembership{Senior Member,~IEEE,} \qquad\qquad\qquad \qquad
Nan Zhao,~\IEEEmembership{Senior Member,~IEEE,}
and Jun Cai,~\IEEEmembership{Senior Member,~IEEE}
		        \vspace{-35pt} 
\thanks{H. Lu is with the Key Laboratory of Intelligent Control and Optimization for Industrial Equipment of Ministry of Education, Dalian University of Technology, Dalian 116024, China, and also with the School of Computer Science and Technology, Chongqing University of Posts and Telecommunications, Chongqing 400065, China (e-mail: luhuabing@dlut.edu.cn).}
\thanks{X. Xie is with the School of Computer Science and Technology, Chongqing University of Posts and Telecommunications, Chongqing 400065, China (e-mail: xiexzh@cqupt.edu.cn).}
\thanks{Z. Shi is with the Key Laboratory of Intelligent Perception and Computing of Anhui Province, Anqing Normal University, Anqing 246011, China (e-mail: shizy@stu.cqupt.edu.cn).}
\thanks{H. Lei is with the School of Communication and Information Engineering, Chongqing University of Posts and Telecommunications, Chongqing 400065, China (e-mail: leihj@cqupt.edu.cn).}
\thanks{N. Zhao is with the Key Laboratory of Intelligent Control and Optimization for Industrial Equipment of Ministry of Education, Dalian University of Technology, Dalian 116024, China (e-mail: zhaonan@dlut.edu.cn).}
\thanks{J. Cai is with the Network Intelligence and Innovation Lab (NI$^2$L), Department of Electrical and Computer Engineering, Concordia University, Montreal, QC H3G 1M8, Canada (e-mail: jun.cai@concordia.ca).}

	}
	
	
	\maketitle
	
\begin{abstract}
With the rapid proliferation of smart devices in wireless networks, more {\color{black}powerful} technologies are expected to fulfill the network requirements of high throughput, massive connectivity, and diversify quality of service. To this end, rate splitting multiple access (RSMA) is proposed as a promising solution to improve spectral efficiency and provide better fairness for the next-generation mobile networks. In this paper, the outage performance of uplink RSMA transmission with randomly deployed users is investigated, taking both user scheduling schemes and power allocation strategies into consideration. Specifically, the greedy user scheduling (GUS) and cumulative distribution function (CDF) based user scheduling (CUS) schemes are considered, which could maximize the rate performance and guarantee scheduling fairness, respectively. Meanwhile, we re-investigate cognitive power allocation (CPA) strategy, and propose a new rate fairness-oriented power allocation (FPA) strategy to enhance the scheduled users' rate fairness. By employing order statistics and stochastic geometry, an analytical expression of the outage probability for each scheduling scheme combining power allocation is derived to characterize the performance. To get more insights, the achieved diversity order of each scheme is also derived. Theoretical results demonstrate that both GUS and CUS schemes applying CPA or FPA strategy can achieve full diversity orders, and the application of CPA strategy in RSMA can effectively eliminate the secondary user's diversity order constraint from the primary user. Simulation results corroborate the accuracy of the analytical expressions, and show that the proposed FPA strategy can achieve excellent rate fairness performance in high signal-to-noise ratio region.

		
	\end{abstract}
	
	\IEEEpeerreviewmaketitle
	\vspace{-15pt}
\section{Introduction}
	\vspace{-5pt}
	\renewcommand{\baselinestretch}{0.75} 
	\setlength{\textfloatsep}{3pt}

\IEEEPARstart{W}{ith} the rapid development of Internet of Things (IoT), the number of devices required to connect is exploding, which poses a great challenge to the limited available spectrum resource. It is urgent to develop new technologies to improve the spectrum efficiency and connectivity for the next-generation mobile networks. To this end, a novel non-orthogonal transmission scheme, called rate splitting multiple access (RSMA), has gained significant attention from both academia and industry in recent years \cite{Clerckx_2016_RS_COMM}. {\color{black}The main concept of RSMA is to employ rate splitting (RS) and supercoding at the transmitters, and then apply successive interference cancellation (SIC) at the receivers \cite{2021_Clerckx_OJCS_Critical}. RS can split a user's message (e.g., information bits) into two or multiple sub-messages, so that each of those sub-messages can be decoded individually at the receivers, and the receivers can reconstruct the original message of that user by retrieving each sub-message. Hence, compared with conventional non-orthogonal multiple access (NOMA) \cite{2017_JSAC_Ding_survey_NOMA,2019_TMC_Cai,2019_TVT_Zaidi_UAV_Ground}, RSMA can manage inter-user interference more flexibly by allowing each user's signal to be partially decoded and partially treated as noise \cite{2019_Mao_RS_multiantenna}.} Taking uplink RSMA as an example, the message of each user is split into two sub-messages before transmitting, then one of the sub-messages can be first decoded and eliminated in the SIC, hence the interference of decoding other users' messages can be reduced. The flexible utilization of RS enables RSMA to achieve the whole capacity region of uplink multiple access channel (MAC) \cite{1996_Remoldi_TIT,2022_Mao_RSMA_SURVEY}. Comparatively, in the decoding process of NOMA, the signal of one user can be eliminated only after all its messages are successfully decoded, and hence only several corner points of the uplink MAC capacity region are achievable for NOMA \cite{2017_Zhu_CD_RS,2005_Tse}.

\vspace{-5pt}
\subsection{Related Work}
\vspace{-5pt}
Extensive research attention has been devoted to RSMA transmission. Particularly, in \cite{2016_Joudeh_TCOM_Partial_CSIT}, Joudeh \textit{et al.} employed RS to maximize the ergodic sum rate (ESR) of downlink multiple input single output (MISO) system based on partial channel state information (CSI). They showed that in addition to ESR gains, the RSMA can relax CSI quality requirements compared with conventional transmission schemes. Clerckx \textit{et al.} \cite{2020_Clarckx_WCL_unifying} considered a two-user MISO broadcast  channel and showed that RSMA can unify orthogonal multiple access (OMA), NOMA, space division multiple access, and multicasting. Considering the power  constraint of SIC, Yang \textit{et al.} \cite{2021_Yang_TCOM_Optimization} investigated the data rate splitting and power allocation schemes with the objective of maximizing the sum rate of the users. Regarding the uplink transmission, Zeng \textit{et al.} \cite{2019_Zeng_up_RS} designed a low-complexity algorithm to maximize the minimum data rate of uplink RSMA and verified its superiority over NOMA. Yang \textit{et al.} \cite{2020_Yang_TMC_Sum-rate_optimization_UL} investigated the sum rate maximization problem by adjusting users' transmit power and decoding order at the base station (BS).


The aforementioned studies \cite{2016_Joudeh_TCOM_Partial_CSIT,2020_Clarckx_WCL_unifying,2021_Yang_TCOM_Optimization,2019_Zeng_up_RS,2020_Yang_TMC_Sum-rate_optimization_UL} mainly focused on the design of practical RSMA systems by optimizing the resource allocation strategies. In order to get some theoretical insights, a number of references dedicated to analyze the performance of RSMA systems. Specifically, Zhu \textit{et al.} \cite{2017_Zhu_RS} first incorporated RS with uplink NOMA and analyzed users' outage performance. Liu \textit{et al.} \cite{2020_Liu_RS} proposed two kinds of RS schemes to enhance the rate fairness\footnote{Note that, both scheduling fairness and rate fairness are considered in this paper, where scheduling fairness refers to the equal probability of scheduling each user, and rate fairness refers to the equal rate of each scheduled user.} and outage performance of delay-limited uplink RSMA transmissions. Singh \textit{et al.} \cite{2021_Singh_WCL_UAV-RSMA} studied the outage probability of unmanned aerial vehicle assisted downlink RSMA system, and highlighted the importance of judicious selection of the power allocation coefficients. Kong \textit{et al.} \cite{2022_Kong_WCL_throughput_analysis} combined RSMA with beamforming to improve the spectral efficiency of satellite communication system, and analyzed the average throughput performance. In \cite{2022_Abbasi_arxiv_C-RSMA}, Abbasi \textit{et al.} proposed cooperative NOMA and cooperative RSMA schemes, and maximized the minimum rates of the two schemes by considering proportional fairness. Moreover, they also theoretically proved that both schemes can achieve the diversity orders of two. Tegos \textit{et al.} \cite{2022_Tegos_CL_Performance_of_UL_RSMA} presented closed-form expressions for the outage probability of uplink RSMA transmission considering all possible decoding orders. Following the concept of cognitive radio, Liu \textit{et al.} \cite{2022_Liu_TVT_RS-CR-NOMA} proposed a dynamic power allocation strategy and analyzed the outage performance for uplink RSMA system. Note that, fixed user locations were assumed in \cite{2017_Zhu_RS,2020_Liu_RS,2022_Abbasi_arxiv_C-RSMA,2021_Singh_WCL_UAV-RSMA,2022_Kong_WCL_throughput_analysis,2022_Tegos_CL_Performance_of_UL_RSMA,2022_Liu_TVT_RS-CR-NOMA}, which helps to reduce the performance analysis complexity for RSMA transmission. However, in practical systems, user locations are often randomly distributed.

{\color{black}Recently, some studies investigated the performance of RSMA transmissions by considering the effect of spatial user locations. Specifically, Demarchou \textit{et al.} derived closed-form expressions for the receiver's coverage probability and average sum rate for two-user single input single output (SISO) \cite{2020_Demarchou_ICC_Spatial_randomness} and MISO \cite{2021_Demarchou_ISWCS_MISO} downlink RSMA systems. Recently, Demarchou \textit{et al.} \cite{2022_Demarchou_TCOM_RS_Caching} and Chen \textit{et al.} \cite{2023_Chen_JSAC_RSMA-MEC} respectively integrated RSMA with wireless edge caching and mobile edge computing by considering spatial user locations. All these work showed that RSMA can bring significant performance gains compared to NOMA, but the users are divided into cell center and edge parts which did not fully applied the potential of opportunistic scheduling. The comparison of our work with the  pertinent literature on uplink RSMA is shown in Table \ref{literature_comparison}.}

\begin{table}
\caption{~~Comparison of this work with related literature on uplink RSMA (UAV: unmanned aerial vehicle,~WEC: wireless edge caching, MEC: mobile edge computing)}
\small	
\label{literature_comparison}
\centering
\renewcommand\arraystretch{1.3} 
\vspace{-10pt}
\begin{tabular}{|c|c|c|c|c|c|c|c|}
	\hline 
 &\multicolumn{1}{m{1.6cm}<{\centering}|}{\centering Random user locations} & \multicolumn{1}{m{1.6cm}<{\centering}|}{Unified scheduling} & \multicolumn{1}{m{1.6cm}<{\centering}|}{User pairing/ scheduling} &\multicolumn{1}{m{1.6cm}<{\centering}|}{Dynamic power allocation} &\multicolumn{1}{m{1.6cm}<{\centering}|}{Outage probability} &\multicolumn{1}{m{1.6cm}<{\centering}|}{ High SNR approximation} & \multicolumn{1}{m{1.6cm}<{\centering}|}{Network}  \\
	\hline 	
	\hline 		
	\cite{2017_Zhu_RS}  &  &  &  & $\checkmark$ & $\checkmark$ &&  Cellular\\
	\hline
	\cite{2020_Liu_RS}&  &  &  & $\checkmark$ & $\checkmark$ & $\checkmark$ & Cellular \\
	\hline  
	\cite{2021_Singh_WCL_UAV-RSMA}  &  &  & $\checkmark$ & $\checkmark$ & $\checkmark$ &&  UAV\\
	\hline
	\cite{2022_Kong_WCL_throughput_analysis}&  &  &  $\checkmark$ &  & $\checkmark$ & & Satellite  \\
	\hline
	\cite{2022_Abbasi_arxiv_C-RSMA}&  &  &  & $\checkmark$ & $\checkmark$ & $\checkmark$ &  Cellular  \\
	\hline
	\cite{2022_Tegos_CL_Performance_of_UL_RSMA}&  &  &  &  & $\checkmark$ &  &    Cellular \\
	\hline
	\cite{2022_Liu_TVT_RS-CR-NOMA}&  &  &  & $\checkmark$ & $\checkmark$ &  &   Cellular  \\
	\hline
	\cite{2020_Demarchou_ICC_Spatial_randomness}&  $\checkmark$&  &  &  & $\checkmark$ &  &   Cellular \\
	\hline
    \cite{2021_Demarchou_ISWCS_MISO}&  &  &  &  & $\checkmark$ & $\checkmark$ &   Cellular \\
	\hline	
	\cite{2022_Demarchou_TCOM_RS_Caching}& $\checkmark$  &  & $\checkmark$ &  & $\checkmark$ &$\checkmark$ &   WEC\\
	\hline
	\cite{2023_Chen_JSAC_RSMA-MEC}& $\checkmark$ &  & $\checkmark$ & $\checkmark$ &  &  &  MEC \\
	\hline
	Ours& $\checkmark$ & $\checkmark$ & $\checkmark$ & $\checkmark$ & $\checkmark$ & $\checkmark$ &   Cellular\\
	\hline
\end{tabular}
\end{table}

\vspace{-5pt}
\subsection{Motivation and Contributions}
\vspace{-5pt}
{\color{black}Although it has been shown that uplink RSMA transmission can theoretically achieve the capacity region \cite{1996_Remoldi_TIT}, there is a paucity of literature quantifying the impact of random user locations on the performance of uplink RSMA. Moreover, opportunistic user scheduling schemes were not studied in the aforementioned literature  \cite{2020_Liu_RS,2017_Zhu_RS,2022_Abbasi_arxiv_C-RSMA,2021_Singh_WCL_UAV-RSMA,2022_Kong_WCL_throughput_analysis,2022_Tegos_CL_Performance_of_UL_RSMA,2021_Demarchou_ISWCS_MISO,2020_Demarchou_ICC_Spatial_randomness,2022_Liu_TVT_RS-CR-NOMA}, but the performance of RSMA transmission can be greatly improved by leveraging effective scheduling schemes. Furthermore, as fixed power allocation strategies were considered in most of the existing work such as \cite{2021_Singh_WCL_UAV-RSMA,2022_Kong_WCL_throughput_analysis,2022_Tegos_CL_Performance_of_UL_RSMA,2021_Demarchou_ISWCS_MISO,2020_Demarchou_ICC_Spatial_randomness,2022_Demarchou_TCOM_RS_Caching}, the flexibility of RSMA transmission has not been fully explored.}

{\color{black}To fill the above research gap, this paper aims at analyzing the outage performance of uplink RSMA transmission with multiple randomly deployed users. We take into consideration the joint effects of opportunistic user scheduling schemes and dynamic power allocation strategies, as the performance of RSMA is related to the two aspects. Different from most existing work, such as \cite{2021_Demarchou_ISWCS_MISO,2020_Demarchou_ICC_Spatial_randomness,2022_Demarchou_TCOM_RS_Caching,2020_Lu_TVT,2016_Liu_cooperative_SWIPT_NOMA,2023_Chen_JSAC_RSMA-MEC}, in which users are deployed in distinct distributions, all users in this paper are assumed to be randomly deployed in the uniform area to enhance the performance. It should be noted that the performance analysis in this scenario is much more complicated, as multi-user scheduling is considered and the scheduled users' channel gains become dependent.} The main contributions of this paper are outlined as follows.


$\bullet$ We investigate the performance of uplink RSMA transmission taking into account the effect of random user locations. To effectively utilize the multi-user diversity and achieve the optimal performance, greedy user scheduling (GUS) scheme is employed to select the best users for transmitting. Moreover, in order to address the scheduling fairness issue of the GUS scheme, we invoke cumulative distribution function (CDF) based user scheduling (CUS) to provide fair access opportunities for all users. In addition, based on the stochastic geometry and probability theory, we derive the joint CDF expression of the scheduled users' channel gains for the CUS scheme.

$\bullet$ To fully utilize the flexibility of RSMA, we propose a fairness-oriented power allocation (FPA) strategy to maximize the rate fairness of the scheduled users. We also evaluate the cognitive power allocation (CPA) strategy by taking both user scheduling and spatial randomness into consideration. We theoretically analyze the scheduled users' outage performance for both GUS and CUS schemes combined with both CPA and FPA strategies. Specifically, by applying stochastic geometry and order statistics, we derive analytical expressions for the outage probability of the scheduled users. To get more insights, we also analyze the users' achievable diversity orders under these different scenarios.

$\bullet$ Through sufficient theoretical analyses and simulation results, we demonstrate that both scheduling schemes can achieve full diversity orders with each power allocation strategy. Moreover, we illustrate that the secondary user's achievable diversity order for the CPA strategy is not restricted by the primary user's channel quality in the RSMA transmission. In addition, we show that the application of FPA strategy to RSMA can achieve rate fairness in high signal-to-noise ratio (SNR) region, and hence the combination of CUS scheme with FPA strategy can achieve both scheduling fairness and rate fairness in the RSMA transmission.


 \vspace{-5pt}
\subsection{Organization}
\vspace{-5pt}

The rest of this paper is organized as follows. Section \ref{system model} introduces the system model. The power allocation strategies, scheduling schemes and the corresponding scheduled users' channel gains' joint distribution functions are derived in Section \ref{sec_joint_CDF}. Sections \ref{cog_PA} and \ref{fair_PA} analyze the outage performance of CPA and FPA strategies, respectively. In Section \ref{Simulation and discussion}, simulation results are presented to corroborate the theoretical analyses, and Section \ref{conclusion} concludes this paper.

\vspace{-3pt}
\section{System Model}\label{system model}
\vspace{-5pt}


We consider a single-cell uplink cellular network, where a single-antenna BS is located at the center of the coverage disc area with radius $R$. {\color{black}To be more practical, we assume that $K$ single-antenna users,}{\footnote{As \cite{2022_Abbasi_arxiv_C-RSMA,2022_Tegos_CL_Performance_of_UL_RSMA,2020_Demarchou_ICC_Spatial_randomness,2022_Liu_TVT_RS-CR-NOMA} for the sake of analytical tractability, we consider the scenario that each node is equipped with a single antenna, and two users are scheduled at each time slot. Note that, equipping more antennas on each node will further enhance the performance of RSMA transmissions, but it is beyond the scope of this paper.}} {\color{black}denoted as U$_k$, $k=1, 2, \dots, K$, are randomly distributed in the coverage area of the BS following a homogeneous Binomial point process (HBPP) $\Upsilon_\text{B}$ \cite{2012_Haenggin_stochastic_geometry,2017Afshang_TWC_BPP,2018Yue_TCOM_unified}.}\footnote{{\color{black}It is shown in \cite{2020_Ali_Partial} that, for opportunistic scheduling, the optimal scheduling strategy is to select the users with lower channel disparity but overall better channel conditions. Hence, in this paper, we consider a different system model from \cite{2022_Demarchou_TCOM_RS_Caching,2020_Demarchou_ICC_Spatial_randomness,2023_Chen_JSAC_RSMA-MEC}, in which the scheduled two users are selected from different regions.}} The time is slotted into intervals of equal length. At the beginning of each time slot, every user can estimate its CSI based on the pilots sent by the BS. The main notations used in this paper are summarized in Table \ref{notations}.

\begin{table}[t!]
	\small
	\centering
	\caption{List of the main notations}
	\renewcommand\arraystretch{1.3} 
	\vspace{-11pt}
	\begin{tabular}{|l|l||l|l|}		
		\hline
		 \textbf{Notation} & \textbf{Description}& \textbf{Notation} &\textbf{Description}\\
		\hline
		$h_i$&Channel of $\text{U}_i$ with the BS&$P_i$ $(\rho_i)$  &Transmit power (SNR) of $\text{U}_i$\\		
		$\eta_i$  &Effective received SNR of $\text{U}_i$&
		$r_i$&Distance between $\text{U}_i$ and the BS\\
		$\beta$  &Power allocation coefficient&
		$f_X(\cdot)$ & Probability density function of $X$\\
		$K$ & Total number of users&
		$F_X(\cdot)$ & Cumulative distribution function of $X$  \\				
		$\alpha$&Path loss exponent&
		$\mathcal{G}[a, b;f(x)]$& Applying Gaussian-Chebyshev quadrature to\\
		$R$&Coverage area of the BS&& the integral of $f(x)$ with integral interval $(a,b)$\\		
		$R_i~(\gamma_i)$ & Target rate (SINR) of $\text{U}_i$&$\mathcal{CN}(\mu,\sigma^2)$&Complex Gaussian random variable\\ 
		$\mathbb{P}\{\cdot\}$ & Probability of an event&&with mean $\mu$ and variance $\sigma^2$ \\ 
		
		\hline
	\end{tabular}%
	\label{notations}%
\end{table}%

{\color{black}To avoid singularity at small distances from the BS to users, a bounded path-loss model is employed, namely, the channel between the $k$-th user U$_k$ and the BS is modeled as $h_k=\frac{\iota_k}{\sqrt{1+r_k^{\alpha}}}$ \cite{2022_Demarchou_TCOM_RS_Caching}}, where $r_k$ represents the distance between U$_k$ and the BS, $\alpha$ denotes the path-loss exponent, and $\iota_k$ represents Rayleigh fading coefficient with $\iota_k\sim\mathcal{CN}(0,1)$. The channel coefficients are assumed to be invariant in each time slot but can change independently among slots. Without loss of generality, we assume that the users' channel gains are ordered as\footnote{Note that, this assumption is used to facilitate performance analysis, and all nodes in the system do not know this order.}
\begin{equation}\label{channel order}
	\setlength{\abovedisplayskip}{3pt}
	\setlength{\belowdisplayskip}{3pt}
	\begin{aligned}
	|h_1|^2\geq\dots\geq|h_K|^2.
	\end{aligned}
\end{equation}

 In each time slot, two of the $K$ users, denoted as U$_i$ and U$_j$, are scheduled to communicate with the BS. {\color{black}It has been shown in  \cite{1996_Remoldi_TIT,2005_Tse,2022_Mao_RSMA_SURVEY} that, a single user's message splitting can guarantee the two-user uplink RSMA transmission to achieve the whole capacity region of uplink MAC.} Hence, without loss of generality, we assume that the message of U$_i$, M$_i$, is split into two sub-messages M$_{i1}$ and M$_{i2}$. The two sub-messages are independently encoded into steams $s_{i1}$ and $s_{i2}$. Then, the two streams are allocated with certain powers and superposed at U$_i$. Thus, the received signal at the BS can be expressed as
\begin{equation}\label{received_signal}
\setlength{\abovedisplayskip}{3pt}
\setlength{\belowdisplayskip}{3pt}
\begin{aligned}	
y=\sqrt{\beta P_{i}}h_is_{i1}+\sqrt{\left(1-\beta\right)P_{i}}h_is_{i2}+\sqrt{P_{j}}h_js_{j}+n_0,
\end{aligned}
\end{equation}
where $s_{v}$, $v\in\{i1,i2,j\}$, satisfies $\mathbb{E}[|s_{v}^2|]=1$, $P_{i}\ (P_{j})$ represents the transmit power of U$_i$ (U$_j$), $\beta\ (0\leq\beta\leq 1)$ denotes the power allocation coefficient of U$_i$, and $n_0$ represents the additive white Gaussian noise (AWGN) at the BS with zero mean and variance $\sigma^2$. For ease of representation, in the rest of the paper, we denote the transmit SNR of U$_i$ (U$_j$) as $\rho_i=\frac{P_{i}}{\sigma^2}$ ($\rho_j=\frac{P_{j}}{\sigma^2}$), and use $\eta_i=\rho_i|h_i|^2$ ($\eta_j=\rho_j|h_j|^2$) to represent the effective received SNRs of U$_i$ (U$_j$). Assume that all users have the same maximal transmit SNR $\rho_m$.

\begin{figure}[t]
	\setlength{\abovecaptionskip}{-7pt}
	\setlength{\belowcaptionskip}{-12pt} 
	\centering
	\includegraphics[width=0.45\textwidth]{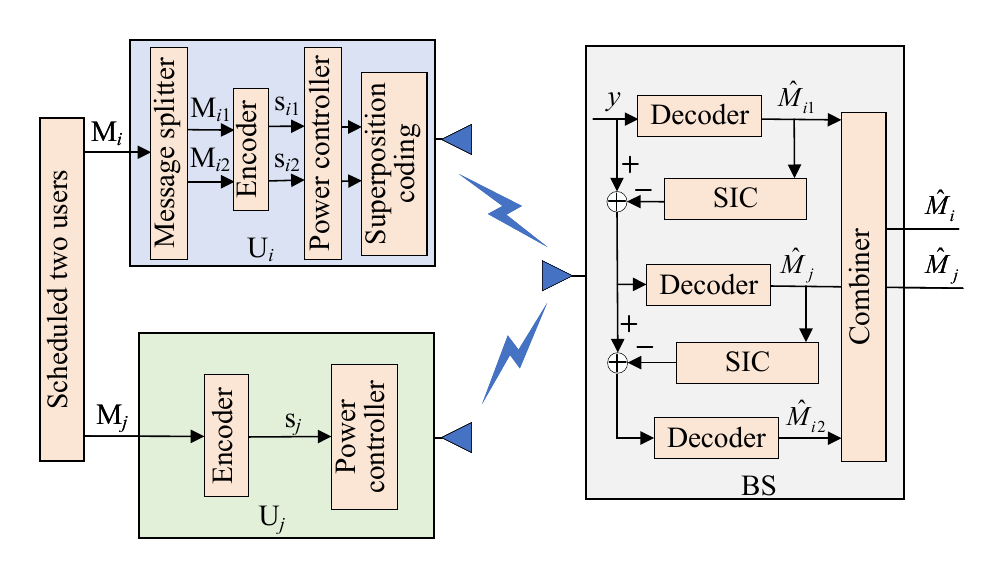}
	\caption{The transceiver architecture of 2-user uplink RSMA.}
	\label{fig1_system model}
\end{figure}

The transceiver architecture of a 2-user uplink RSMA is illustrated in Fig. \ref{fig1_system model}, and SIC is executed at the BS to decode the transmitted messages. In this paper, we assume that the decoding order is $s_{i1}\rightarrow s_{j}\rightarrow s_{i2}$ as in \cite{2020_Liu_RS}. Particularly, if $\beta=0$ ($\beta=1$), the RSMA transmission will degrade to NOMA transmission, and the decoding order will become $s_{j}\rightarrow s_{i}$ ($s_{i}\rightarrow s_{j}$). Based on (\ref{received_signal}), the instantaneous signal to interference plus noise ratio (SINR) for decoding $s_{i1}$ can be denoted as
\begin{equation}\label{SINR_Ui1}
	\setlength{\abovedisplayskip}{3pt}
	\setlength{\belowdisplayskip}{3pt}
	\begin{aligned}	
		\text{SINR}_{i1}=\frac{\beta \rho_i|h_i|^2}{\left(1-\beta\right)\rho_i|h_i|^2+\rho_j|h_j|^2+1}.
	\end{aligned}
\end{equation}
After decoding $s_{i1}$, the BS will reconstruct and eliminate the signal of $s_{i1}$ from the received compound signal. Then the remaining signal can be represented as
\begin{equation}\label{remaining_signal}
	\setlength{\abovedisplayskip}{3pt}
	\setlength{\belowdisplayskip}{3pt}
	\begin{aligned}	
		\tilde{y}=\sqrt{\left(1-\beta\right)P_{i}}h_is_{i2}+\sqrt{P_{j}}h_js_{j}+n_0.
	\end{aligned}
\end{equation}
From (\ref{remaining_signal}), the SINR for decoding $s_j$ can be expressed as
\begin{equation}\label{SINR_Uj}
	\setlength{\abovedisplayskip}{3pt}
	\setlength{\belowdisplayskip}{3pt}
	\begin{aligned}	
		\text{SINR}_{j}=\frac{\rho_j|h_j|^2}{\left(1-\beta\right)\rho_i|h_i|^2+1}.
	\end{aligned}
\end{equation}
Similarly, after recovering $s_{j}$ and subtracting its reconstructed signal from (\ref{SINR_Uj}), the SNR of decoding $s_{i2}$ can be denoted as
\begin{equation}\label{SINR_Ui2}
	\setlength{\abovedisplayskip}{3pt}
	\setlength{\belowdisplayskip}{3pt}
	\begin{aligned}	
		\text{SINR}_{i2}=\left(1-\beta\right)\rho_i|h_i|^2.
	\end{aligned}
\end{equation}

\section{Scheduling Schemes and Power Allocation Strategies}\label{sec_joint_CDF}

In this section, the GUS and CUS schemes are first presented, and the joint CDFs of the scheduled users' channel gains are given which will be applied to analyze the users' performance. Then, the CPA and FPA strategies are introduced.

\vspace{-5pt}
\subsection{Scheduling Schemes}
\vspace{-5pt}

Two scheduling schemes are investigated in this paper. In particular, the GUS scheme is first considered in order to obtain the maximal system performance. Then, CUS scheme is introduced to tackle the scheduling fairness problem of the GUS scheme, which can also effectively utilize the multi-user diversity gain.


 \subsubsection{GUS Scheme}
 In the GUS scheme, two users with the largest channel gains, namely, U$_{1}$ and U$_2$, are scheduled, and the joint probability density function (PDF) of their channel gains is \cite{2003_order_statistics}
 \begin{equation}\label{PDF_largest2}
 	\setlength{\abovedisplayskip}{3pt}
 	\setlength{\belowdisplayskip}{3pt}
 	\begin{aligned}
 		f_{|h_2|^2,|h_1|^2}^{\text{GUS}}(x,y)
 		&=K(K-1)f_{|h|^2}(x)f_{|h|^2}(y)F_{|h|^2}(x)^{K-2},\\
 	\end{aligned}
 \end{equation}
 where $x\leq y$. $F_{|h|^2}(x)$ and $f_{|h|^2}(x)$ respectively represent the CDF and PDF of the randomly selected user's channel gain. According to \cite{2014_Ding_random_deployed}, we have
 \begin{equation}\label{CDF_unorder user}
 	\setlength{\abovedisplayskip}{3pt}
 	\setlength{\belowdisplayskip}{3pt}
 	\begin{aligned}
 		F_{|h|^2}(x)\approx\sum_{l=1}^{L}\Psi_l(1-e^{-\mu_lx}),
 	\end{aligned}
 \end{equation}
 and
 \begin{equation}\label{PDF_unorder user}
 	\setlength{\abovedisplayskip}{3pt}
 	\setlength{\belowdisplayskip}{3pt}
 	\begin{aligned}
 		f_{|h|^2}(x)\approx\sum_{l=1}^{L}\Psi_l\mu_le^{-\mu_lx},
 	\end{aligned}
 \end{equation}
where Gaussian-Chebyshev quadrature is applied\cite{Gaussian_Chebyshev}, $\mu_l=1+\left(\frac{R}{2}+\frac{R}{2}\psi_l\right)^\alpha$, $\Psi_l=\frac{\pi}{L}\sqrt{1-\psi_l^2}(1+\psi_l)$, $\psi_l=\mathrm{cos}\left(\frac{2l-1}{2L}\pi \right)$, and $L$ is the parameter for making complexity-accuracy tradeoff.
 
{\color{black}The GUS scheme can achieve the best system performance, since it schedules the users with the largest channel gains \cite{2015_Tabassum_TWC_user_scheduling}.} However, as the users closer to the BS have smaller path-losses and then are scheduled more often than others, the scheduling fairness issue arises. However, in practical applications, apart from rate performance, admission fairness is also a pivotal system performance indicator for wireless networks \cite{2013_Sediq_TWC,2014_Shi_CST}, especially in opportunistic scheduling scenarios\cite{2005_Tse}. Hence, in the following, CUS is invoked to provide fair scheduling opportunities for the users.

 \subsubsection{CUS Scheme}
 In CUS scheme, the two users with the largest CDF values\footnote{In this paper, we use ``CDF" to denote the cumulative distribution function, for example, $F_k(x)$, and use ``CDF value" to represent the corresponding output value of a CDF with a specific input $x$.} with respect to their channel gains, namely, the users whose channels are good enough relative to their own statistics \cite{2005_Park_packet_CDF,2020_Lu_TVT}, will be selected to transmit. {\color{black}Assume that the CDF of $\text{U}_k$'s channel gain $|h_k|^2$ is denoted as $F_k(x)$, which can be obtained by the user's observation \cite{2015-Jin-Fundamental_limits_CDF}, and the two users with the largest CDF values can be identified by distributed contention \cite{2006-Bletsas-a-simple,2005-Zhao-Opprotunistic}.} For Rayleigh small-scale fading, the CDF of $\text{U}_k$'s channel gain with a given distance $r_k$ can be expressed as
	\begin{equation}\label{CDF_Uk}
	\setlength{\abovedisplayskip}{3pt}
	\setlength{\belowdisplayskip}{3pt}
	F_k(x|r_k)=1-e^{-(1+r_k^\alpha)x}.
	\end{equation}
Let $\mathcal{U}_k=F_k(x|r_k)$. Since the user's channel gain is a random variable, $\mathcal{U}_k$ is a random variable as well, and it is uniformly distributed in $[0, 1]$ \cite{2015-Jin-Fundamental_limits_CDF}. Moreover, since all users' channels are assumed to be independent of each other, every user has the same probability to have the largest CDF value and hence the scheduling fairness can be guaranteed\cite{2005_Park_packet_CDF,2015-Jin-Fundamental_limits_CDF}.


In practical applications, assuming the BS sends pilot signals at the beginning of each time slot. In time division duplexing mode, each  user can estimate its channel gains by measuring these pilot signals \cite{2017_Choi_ALOHA}. By applying non-parametric CDF based scheduling (NPCS) or parametric CDF based scheduling (PCS) algorithm \cite{2014_Nguyen_TSP_Leaning_methods}, each user can estimate its CDF based on the estimation of its channel gains \cite{2015-Jin-Fundamental_limits_CDF}. If one user wants to transmit data at a specific time slot, it will estimate the instantaneous channel gain at the beginning of that time slot, and then obtain the corresponding CDF value by substituting the estimated result into its CDF. But employing distributed contention control strategy \cite{2006_Bletsas_network_path_selection}, the users with the largest CDF values can be admitted.  The joint CDF of the admitted two users' channel gains in CUS scheme is given in the following lemma.

\textit{Lemma 1:}
	Let random variables $X$ and $Y$ denote the channel gains of users with the largest and the second-largest CDF values, respectively. Then, the joint CDF of $X$ and $Y$ in CUS scheme can be approximated as the following 4 cases.
	
	Case 1: If $x\geq (R^\alpha+1)y$,
	\begin{equation}\label{CDF_joint_xy0}
		\setlength{\abovedisplayskip}{3pt}
		\setlength{\belowdisplayskip}{3pt}
		\begin{aligned}
			F_{X,Y}^{\text{CUS}}(x,y)
			\approx&K\sum_{l_1=1}^{L}\Psi_{l_1}(1-e^{-\mu_{l_1}y})^{K-1}\sum_{l_2=1}^{L}\Psi_{l_2}(1-e^{-\mu_{l_2}x})\\
			&-(K-1)\sum_{l_1=1}^{L}\Psi_{l_1}(1-e^{-\mu_{l_1}y})^K.
		\end{aligned}
	\end{equation}

	Case 2: If $y\leq x<(R^\alpha+1)y$,
	\begin{equation}\label{CDF_joint_xy}
		\setlength{\abovedisplayskip}{3pt}
		\setlength{\belowdisplayskip}{3pt}
		\begin{aligned}
			F_{X,Y}^{\text{CUS}}(x,y)
			\approx&Ka_1w_M\sum_{m=1}^{M}t_m[1-e^{-(1+t_m^\alpha)y}]^{K-1}\sum_{l=1}^{L}\Psi_l(1-e^{-\mu_lx})\\
			&+K(R-a_1)w_Q\sum_{q=1}^{Q}\Theta_q[1-e^{-(1+\Theta_q^\alpha)y}]^{K-1}(R-I_1(\Theta_q))\\
			&\quad \times w_N\sum_{n=1}^{N}\Phi_n(\Theta_q)[1-e^{-(1+\Phi_n(\Theta_q)^\alpha)x}]\\
			&-(K-1)a_1w_M\sum_{m=1}^{M}t_m[1-e^{-(1+t_m^\alpha)y}]^{K}\\
			&-(K-1)(R-a_1)w_Q\sum_{q=1}^{Q}\Theta_q[1-e^{-(1+\Theta_q^\alpha)y}]^{K}[1-\frac{1}{R^2}I_1(\Theta_q)^2]\\
			&+a_2w_B\sum_{b=1}^{B}\Xi_b[1-e^{-(1+\Xi_b^\alpha)x}]^K[1-\frac{1}{R^2}I_2(\Xi_b)^2].
		\end{aligned}
	\end{equation}

Case 3: If $x<y<(R^\alpha+1)x$,
\begin{equation}\label{CDF_joint_yx}
	\setlength{\abovedisplayskip}{3pt}
	\setlength{\belowdisplayskip}{3pt}
	\begin{aligned}
		F_{X,Y}^{\text{CUS}}(x,y)
		\approx
		&a_1w_M\sum_{m=1}^{M}t_m[1-e^{-(1+t_m^\alpha)x}]^{K}+Ka_2w_B\sum_{b=1}^{B} \Xi_b[1-e^{-(1+\Xi_b^\alpha)y}]^{K-1}\\
		&\quad\times(R-I_1(\Xi_b))w_N\sum_{n=1}^{N} \Phi_n(\Xi_b)[1-e^{-(1+\Phi_n(\Xi_b)^\alpha)x}]\\
		&-(K-1)a_2w_B\sum_{b=1}^{B} \Xi_b[1-e^{-(1+\Xi_b^\alpha)y}]^K[1-\frac{1}{R^2}I_1(\Xi_b)^2]\\
		&+(R-a_1)w_Q\sum_{q=1}^{Q} \Theta_q[1-e^{-(1+\Theta_q^\alpha)x}]^{K}[1-\frac{1}{R^2}I_2(\Theta_q)^2].\\
	\end{aligned}
\end{equation}

Case 4: If $y\geq (R^\alpha+1)x$,
\begin{equation}\label{CDF_joint_yx0}
	\setlength{\abovedisplayskip}{3pt}
	\setlength{\belowdisplayskip}{3pt}
	\begin{aligned}
		F_{X,Y}^{\text{CUS}}(x,y)
		\approx&\sum_{l=1}^{L}\Psi_l(1-e^{-\mu_lx})^{K}.
	\end{aligned}
\end{equation}
Where, $a_1=(\frac{x}{y}-1)^{\frac{1}{\alpha}}$,  $a_2=[\frac{y}{x}(R^\alpha+1)-1]^{\frac{1}{\alpha}}$, if $x\geq y$; and $a_1=(\frac{y}{x}-1)^{\frac{1}{\alpha}}$,  $a_2=[\frac{x}{y}(R^\alpha+1)-1]^{\frac{1}{\alpha}}$, if $y>x$. $I_1(z)=[\frac{y}{x}(1+z^\alpha)-1]^{\frac{1}{\alpha}}$, $I_2(z)=[\frac{x}{y}(1+z^\alpha)-1]^{\frac{1}{\alpha}}$, $w_M=\frac{\pi}{MR^2}\sqrt{1-\varphi_m^2}$, $\varphi_m=\mathrm{cos}(\frac{2m-1}{2M}\pi)$, $t_m=\frac{a_1}{2}(1+\varphi_m)$, $w_Q=\frac{\pi}{QR^2}\sqrt{1-\theta_q^2}$, $\theta_q=\mathrm{cos}(\frac{2q-1}{2Q}\pi)$, $\Theta_q=\frac{R+a_1}{2}-\frac{R-a_1}{2}\theta_q$, $w_N=\frac{\pi}{NR^2}\sqrt{1-\phi_n^2}$, $\phi_n=\mathrm{cos}(\frac{2n-1}{2N}\pi)$, $\Phi_n(z)=\frac{R+I_1(z)}{2}+\frac{R-I_1(z)}{2}\phi_n$, $w_B=\frac{\pi}{BR^2}\sqrt{1-\xi_b^2}$, $\xi_b=\mathrm{cos}(\frac{2b-1}{2B}\pi)$, $\Xi_b=\frac{a_2}{2}(1+\xi_b)$, $M$, $Q$, $N$, and $B$ are parameters for complexity-accuracy trade-off.

\begin{proof}
	Please refer to Appendix \ref{Proof of lemma1}.
\end{proof}


\textit{Remark 1:}
{\color{black}It can be seen from Lemma 1 that the parameters $R$ and $\alpha$ as well as the values of $X$ and $Y$ have an essential impact on the expression of the joint CDF. Because of this, the joint CDF is a piecewise function, and the derivation of the outage probability for the CUS scheme is quite involved.}

\textit{Remark 2:}
When $y\geq (R^\alpha+1)x$, we can see from (\ref{CDF_joint_yx0}) that the joint CDF is only related to the channel gain of the user with the largest CDF value. That is because $y\geq (R^\alpha+1)x$ leads to $F_i(x|r_i)<F_j(y|r_j)$ according to (\ref{CDF_Uk}). And since the CUS scheme always schedules the two users with the largest CDF values, according to the joint distribution of two order statistics \cite{2003_order_statistics}, the constraint of $y$ can be replaced by that of $x$. Actually, (\ref{CDF_joint_yx0}) also denotes the CDF of channel gain for the user with the largest CDF value. Moreover, by applying the joint CDF in (\ref{CDF_joint_xy0}) and setting $x\to \infty$, we can obtain the CDF of channel gain for the user with the second-largest CDF value as
	\begin{equation}\label{CDF_y}
		\setlength{\abovedisplayskip}{3pt}
		\setlength{\belowdisplayskip}{3pt}
		\begin{aligned}
			F_{Y}^{\text{CUS}}(y)
			\approx&K\sum_{l_1=1}^{L}\Psi_{l_1}(1-e^{-\mu_{l_1}y})^{K-1}-(K-1)\sum_{l_1=1}^{L}\Psi_{l_1}(1-e^{-\mu_{l_1}y})^K.
		\end{aligned}
	\end{equation}

\textit{Remark 3:}
Note that, both the CUS scheme and random user scheduling (RUS) scheme\cite{2020_Demarchou_ICC_Spatial_randomness,2016_Liu_cooperative_SWIPT_NOMA} can provide equal admission probability to all users, but the scheduled users' joint CDFs of the two scheduling schemes have different characteristics. Specifically, the scheduled {\color{black}users'} channel gains are independent for the RUS scheme. However, by applying Lemma 1 and Remark 2, we can see that the scheduled two users' channel gains are dependent for the CUS scheme, from which the performance of the CUS scheme is enhanced as shown in the simulation results.

\vspace{-5pt}
\subsection{Power Allocation Strategies}
\vspace{-5pt}

{\color{black}As uplink RSMA can achieve the maximal capacity with any power allocation strategies \cite{2022_Mao_RSMA_SURVEY,2005_Tse,1996_Remoldi_TIT}, two power allocation schemes are considered in this paper which can achieve different targets. Specifically, the CPA strategy intends to prioritize the target rate of one user, while the FPA strategy aims to balance the scheduled users' achievable rates.}

\subsubsection{CPA Strategy}


In CPA strategy \cite{2016_Ding_impact_of_pairing,2022_Liu_TVT_RS-CR-NOMA}, one of the scheduled two users is deemed as primary user (denoted as U$_\text{p}$), and the other user is regarded as secondary user (denoted as U$_\text{s}$). Particularly, the rate of U$_\text{p}$ should be satisfied with priority, while U$_\text{s}$ can be served on the premise that its access can not affect the performance of U$_\text{p}$. Hence, compared with the conventional OMA transmission in which the spectrum is solely occupied by one user, applying CPA strategy in RSMA can greatly enhance the connectivity and improve the spectrum efficiency. In order not to affect the transmission of U$_\text{p}$, U$_\text{s}$ is deemed as the splitting-user (namely, its message may be split into two parts). Based on the instantaneous channel gains of the scheduled users, the CPA strategy is designed with the following three cases \cite{2022_Liu_TVT_RS-CR-NOMA}:\footnote{Note that, the CPA strategy for RSMA was first proposed in \cite{2022_Liu_TVT_RS-CR-NOMA}, in which only two users with fixed user locations are considered. In this paper, we further investigate the CPA strategy by taking into account both random user locations and user scheduling schemes with multiple users, and new insights are revealed.}
\begin{enumerate}
	\item The signal of  U$_\text{p}$ cannot be successfully decoded even if it solely occupies the channel, i.e., $\text{log}_2\left(1+\eta_{\text{p}}\right)<\hat{R}_\text{p}$, where $\eta_{\text{p}}=\rho_{\text{p}}|h_{\text{p}}|^2$ and $\hat{R}_\text{p}$ denotes the target data rate of U$_\text{p}$: In this case, U$_\text{p}$ will be in an outage definitely. In order to maximize U$_\text{s}$'s data rate, the power allocation coefficient $\beta$ should be set as 1 to make U$_\text{s}$'s signal be decoded at the first stage of SIC. Hence, U$_\text{s}$ should not be split, and its achievable rate is $\text{log}_2\left(1+\frac{\eta_\text{s}}{\eta_{\text{p}}+1}\right)$, where $\eta_{\text{s}}=\rho_{\text{s}}|h_{\text{s}}|^2$.
	\item The signal of U$_\text{p}$ can be successfully decoded even if all the received power of U$_\text{s}$ is regarded as noise, namely, $\text{log}_2\left(1+\frac{\eta_{\text{p}}}{\eta_{\text{s}}+1}\right)\geq\hat{R}_\text{p}$: In this case, to maximize the performance of U$_\text{s}$, $\beta$ should be set as 0, and the achievable rate of U$_\text{s}$ is $\text{log}_2\left(1+\eta_{\text{s}}\right)$.
	\item The signal of  U$_\text{p}$ can be successfully decoded if it solely occupies the channel, but cannot be decoded if all the received power of U$_\text{s}$ is regarded as noise, i.e., $\text{log}_2\left(1+\eta_{\text{p}}\right)\geq \hat{R}_\text{p}$ and $\text{log}_2\left(1+\frac{\eta_{\text{p}}}{\eta_{\text{s}}+1}\right)<\hat{R}_\text{p}$: In this case, RS should be conducted at U$_\text{s}$. Specifically, in order to guarantee the target rate of U$_\text{p}$ and to maximize the achievable data rate of U$_\text{s}$, $\beta$ should be set to meet $\text{log}_2\left(1+\frac{\eta_{\text{p}}}{\left(1-\beta\right)\eta_{\text{s}}+1}\right)=\hat{R}_\text{p}$, namely, $\beta=1-\frac{\eta_{\text{p}}-\hat{\gamma_{\text{p}}}}{\eta_{\text{s}}\hat{\gamma_{\text{p}}}}$, where $\hat{\gamma_{\text{p}}}=2^{\hat{R}_\text{p}}-1$. According to (\ref{SINR_Ui1}) and (\ref{SINR_Ui2}), the achievable data rate of U$_\text{s}$ is $\text{log}_2\left(\frac{\hat{\gamma_{\text{p}}}\left(1+\eta_{\text{p}}+\eta_{\text{s}}\right)}{\eta_{\text{p}}\left(1+\hat{\gamma_{\text{p}}}\right)}\right)+\text{log}_2\left(\frac{\eta_{\text{p}}}{\hat{\gamma_{\text{p}}}}\right)=\text{log}_2\left(1+\eta_{\text{p}}+\eta_{\text{s}}\right)-\hat{R}_\text{p}$.
\end{enumerate}



\subsubsection{FPA Strategy}\label{FPA_strategy_subsec}
We propose FPA strategy to improve the scheduled users rate fairness by  fully utilizing the flexibility of RSMA. It has been shown that two corner points on the capacity region of uplink MAC can be achieved by NOMA transmission \cite{2005_Tse}, i.e., $\left(R_i^\text{F},R_j^\text{S}\right)$ and $\left(R_i^\text{S},R_j^\text{F}\right)$, where $R_i^\text{F}=\text{log}(1+\frac{\eta_i}{1+\eta_j})$ and $R_i^\text{S}=\text{log}(1+\eta_i)$ denote the achievable rates of decoding $\text{U}_i$'s signal at the first and second stages of SIC, respectively. In addition to these two corner points, RSMA transmission can achieve all the points on the line of the two corner points by adjusting the power allocation factor $\beta$,\footnote{It should be noted that the whole MAC capacity region can also be achieved via time sharing \cite{1996_Remoldi_TIT,2005_Tse,2022_Mao_RSMA_SURVEY}. However, time sharing requires stringent synchronization among users, which is difficult to achieve in practice \cite{1996_Remoldi_TIT,2022_Mao_RSMA_SURVEY}.} where all these points can achieve the same sum rate performance \cite{2020_Liu_RS,2017_Zhu_RS,1996_Remoldi_TIT}. In other words, by using RSMA, the sum rate $R_i^\text{F}+R_j^\text{S}=\text{log}_2(1+\eta_i+\eta_j)$ can be flexibly allocated to the scheduled users by adjusting $\beta$. Hence, we utilize this feature to enhance the scheduled users' rate fairness performance on the premise of achieving the optimal sum rate.

Assuming $\eta_i>\eta_j$, we know rate-pair $\left(R_i^\text{F},R_j^\text{S}\right)$ can achieve much better rate fairness than $\left(R_i^\text{S},R_j^\text{F}\right)$, since the first decoded user's achievable data rate will be interfered by the second decoded user's signal. Furthermore, if $R_i^\text{F}\geq R_j^\text{S}$, both NOMA and RSMA schemes can achieve the same fairness performance by setting $\beta=0$ and decoding $\text{U}_i$'s signal at the first stage of SIC. In other words, when $\text{log}_2\left(1+\frac{\eta_i}{\eta_j+1}\right)\geq \text{log}_2\left(1+\eta_j\right)$ (i.e., $\eta_i\geq \eta_j+\eta_j^2$), the data rates of U$_i$ and U$_j$ are $\text{log}_2\left(1+\frac{\eta_i}{\eta_j+1}\right)$ and $\text{log}_2\left(1+\eta_j\right)$, respectively. 

On the other hand, if $R_i^\text{F}< R_j^\text{S}$ (i.e., $\eta_i<\eta_j+\eta_j^2$), RSMA can achieve better fairness than NOMA by applying proper power allocation strategy to U$_j$. In this case, the scheduled users can achieve the same data rate for RSMA scheme, namely, $R_i=R_j=\frac{1}{2}\text{log}_2(1+\eta_i+\eta_j)$. By setting $\text{log}_{2}\left( 1+\frac{{{\eta }_{i}}}{\left( 1-{{\beta }} \right){{\eta }_{j}}+1} \right)=\frac{1}{2}\text{log}_{2}\left( 1+\eta _{i}+\eta_{j} \right)$, the power allocation coefficient can be derived as ${{\beta }}=1+\frac{1}{{{\eta }_{j}}}-\frac{{{\eta }_{i}}}{{{\eta }_{j}}\left( \sqrt{1+{{\eta }_{i}}+{{\eta }_{j}}}-1 \right)}$. As such, the transmission rates of U$_{j1}$ and U$_{j2}$ are $\text{log}_2\left(1+\frac{\beta\eta_{j}}{\left(1-\beta\right)\eta_j+\eta_{i}+1}\right)$ and $\text{log}_2(1+(1-\beta)\eta_j)$, respectively. Hence, in the case $\eta_i>\eta_j$, the power allocation coefficient of the splitting-user U$_j$ is
\begin{equation}\label{FPA_power allocation}
	\setlength{\abovedisplayskip}{3pt}
	\setlength{\belowdisplayskip}{3pt}
	\begin{aligned}
\beta=
\begin{cases}
	0, &\text{if}\ \eta_i\geq \eta_j+\eta_j^2\\
	1+\frac{1}{{{\eta }_{j}}}-\frac{{{\eta }_{i}}}{{{\eta }_{j}}\left( \sqrt{1+{{\eta }_{i}}+{{\eta }_{j}}}-1 \right)}, &\text{otherwise}	
\end{cases}.
	\end{aligned}
\end{equation}

For the case $\eta_i\leq\eta_j$, U$_i$ is regarded as the splitting-user, and the power allocation coefficient can be derived similar to the case $\eta_i>\eta_j$. 

\vspace{-5pt}
\subsection{Performance Metric}
\vspace{-5pt}

In this paper, outage probability is used as the performance metric, which is defined as the probability that the instantaneous achievable rate of a user is less than a target one. To gain more insights, diversity order will also be derived. The diversity order highlights the scaling law of the outage probability with respect to the transmit SNR, which can be defined as \cite{diversity_order}
\begin{equation}\label{define_DO}
	\setlength{\abovedisplayskip}{3pt}
	\setlength{\belowdisplayskip}{3pt}
	\begin{aligned}
		d=-\underset{\rho\rightarrow\infty}{\text{lim}}\frac{\text{log}\mathcal{P}(\rho)}{\text{log}\rho},
	\end{aligned}
\end{equation}
where $\mathcal{P}(\rho)$ and $\rho$ denote the outage probability and transmit SNR, respectively. In the following two sections, the outage performance of the two power allocation strategies will be analyzed for each user scheduling scheme.

\vspace{-5pt}
\section{Outage Performance Analysis of The CPA Strategy}\label{cog_PA}
\vspace{-5pt}

In this section, the outage probability framework of CPA strategy is derived first. Then, the outage performance of the GUS and CUS scheme with CPA strategy is analyzed.

\vspace{-5pt}
\subsection{Outage Probability of the CPA Strategy}
\vspace{-5pt}

In CPA strategy, as U$_\text{p}$ can achieve the same performance as that achieved in OMA transmission, we only evaluate the outage performance of U$_\text{s}$ \cite{2016_Ding_impact_of_pairing}, which can be expressed as
\begin{equation}\label{pout_cog1}
	\setlength{\abovedisplayskip}{3pt}
	\setlength{\belowdisplayskip}{3pt}
	\begin{aligned}
	\mathcal{P}^{\text{CPA}}=&\mathbb{P}\left[\text{log}_2\left(1+\eta_{\text{p}}\right)\leq\hat{R}_{\text{p}},\text{log}_2\left(1+\frac{\eta_{\text{s}}}{\eta_{\text{p}}+1}\right)<\hat{R}_{\text{s}}\right]\\
	&+\mathbb{P}\left[\text{log}_2\left(1+\frac{\eta_{\text{p}}}{\eta_{\text{s}}+1}\right)>\hat{R}_{\text{p}},\text{log}_2\left(1+\eta_{\text{s}}\right)<\hat{R}_{\text{s}}\right]\\
	&+\mathbb{P}\left[\text{log}_2\left(1+\eta_{\text{p}}\right)>\hat{R}_{\text{p}},\text{log}_2\left(1+\frac{\eta_{\text{p}}}{\eta_{\text{s}}+1}\right)<\hat{R}_{\text{p}},\right.\\
	&\left.\qquad\ \text{log}_2\left(1+\eta_{\text{p}}+\eta_{\text{s}}\right)-\hat{R}_{\text{p}}<\hat{R}_{\text{s}}\right],\\
	\end{aligned}
\end{equation}
where $\hat{R}_\text{s}$ denotes the target data rate of U$_\text{s}$. The first term denotes both U$_\text{p}$ and U$_\text{s}$ are in outage (in this case $\beta=1$, namely, U$_\text{s}$'s signal is decoded at the first stage of SIC). The second term represents that U$_\text{p}$'s signal is successfully decoded at the first stage of SIC, and U$_\text{s}$'s signal is in an outage (in this case $\beta=0$). The third term shows that U$_\text{p}$'s signal can be successfully decoded at the second stage of SIC, but U$_\text{s}$ is still in an outage (in this case $0<\beta<1$).

After some manipulations, (\ref{pout_cog1}) can be expressed as
\begin{equation}\label{pout_cog2}
	\setlength{\abovedisplayskip}{3pt}
	\setlength{\belowdisplayskip}{3pt}
\begin{aligned}
\mathcal{P}^{\text{CPA}}=&\mathbb{P}\left[\frac{\eta_{\text{s}}}{\hat{\gamma}_{\text{s}}}-1<\eta_{\text{p}}\leq\hat{\gamma}_{\text{p}}\right]+\mathbb{P}\left[\eta_{\text{p}}>\hat{\gamma}_{\text{p}}\left(\eta_{\text{s}}+1\right),\eta_{\text{s}}<\hat{\gamma}_{\text{s}}\right]\\
&+\mathbb{P}\left[\hat{\gamma}_{\text{p}}<\eta_{\text{p}}<\hat{\gamma}_{\text{p}}\left(\eta_{\text{s}}+1\right),\eta_{\text{p}}<\hat{\gamma}_{\text{p}}+\hat{\gamma}_{\text{s}}+\hat{\gamma}_{\text{p}}\hat{\gamma}_{\text{s}}-\eta_{\text{s}}\right],\\
\end{aligned}
\end{equation}
where $\hat{\gamma_{\text{s}}}=2^{\hat{R}_\text{s}}-1$ and $\hat{\gamma}_{\text{p}}+\hat{\gamma}_{\text{s}}+\hat{\gamma}_{\text{p}}\hat{\gamma}_{\text{s}}=2^{\hat{R}_{\text{p}}+\hat{R}_{\text{s}}}-1$ is applied. With some manipulations, we find that the first term in (\ref{pout_cog2}) holds if $\eta_{\text{s}}<\hat{\gamma}_{\text{s}}(1+\hat{\gamma}_{\text{p}})$. For the third term in (\ref{pout_cog2}), we know that $\hat{\gamma}_{\text{p}}\left(\eta_{\text{s}}+1\right)<\hat{\gamma}_{\text{p}}+\hat{\gamma}_{\text{s}}+\hat{\gamma}_{\text{p}}\hat{\gamma}_{\text{s}}-\eta_{\text{s}}$ holds if $\eta_\text{s}<\hat{\gamma}_{\text{s}}$, and vice versa. Hence, (\ref{pout_cog2}) can be converted to
\begin{equation}\label{pout_cog3}
	\setlength{\abovedisplayskip}{3pt}
	\setlength{\belowdisplayskip}{3pt}
\begin{aligned}
\mathcal{P}^{\text{CPA}}
=&\mathbb{P}\left[\frac{\eta_{\text{s}}}{\hat{\gamma}_{\text{s}}}-1<\eta_{\text{p}}\leq\hat{\gamma}_{\text{p}},\eta_{\text{s}}<\hat{\gamma}_{\text{s}}(1+\hat{\gamma}_{\text{p}})\right]+\mathbb{P}\left[\eta_{\text{p}}>\hat{\gamma}_{\text{p}}\left(\eta_{\text{s}}+1\right),\eta_{\text{s}}<\hat{\gamma}_{\text{s}}\right]\\
&+\mathbb{P}\left[\hat{\gamma}_{\text{p}}<\eta_{\text{p}}<\hat{\gamma}_{\text{p}}\left(\eta_{\text{s}}+1\right),\eta_\text{s}<\hat{\gamma}_{\text{s}}\right]\\
&+\mathbb{P}\left[\hat{\gamma}_{\text{p}}<\eta_{\text{p}}<\hat{\gamma}_{\text{p}}+\hat{\gamma}_{\text{s}}+\hat{\gamma}_{\text{p}}\hat{\gamma}_{\text{s}}-\eta_{\text{s}},\eta_\text{s}>\hat{\gamma}_{\text{s}}\right].\\
\end{aligned}
\end{equation}

To guarantee $\hat{\gamma}_{\text{p}}<\hat{\gamma}_{\text{p}}+\hat{\gamma}_{\text{s}}+\hat{\gamma}_{\text{p}}\hat{\gamma}_{\text{s}}-\eta_{\text{s}}$ in the fourth term of (\ref{pout_cog3}), $\eta_{\text{s}}$ should be less than $\hat{\gamma}_{\text{s}}(1+\hat{\gamma}_{\text{p}})$. Based on that, and combining the second and third terms of (\ref{pout_cog3}), the outage probability for the CPA strategy can be further denoted as
\begin{equation}\label{pout_cog31}
	\setlength{\abovedisplayskip}{3pt}
	\setlength{\belowdisplayskip}{3pt}
\begin{aligned}
\mathcal{P}^{\text{CPA}}
=&\mathbb{P}\left[\frac{\eta_{\text{s}}}{\hat{\gamma}_{\text{s}}}-1<\eta_{\text{p}}\leq\hat{\gamma}_{\text{p}},\eta_{\text{s}}<\hat{\gamma}_{\text{s}}(1+\hat{\gamma}_{\text{p}})\right]+\mathbb{P}\left[\eta_{\text{p}}>\hat{\gamma}_{\text{p}},\eta_{\text{s}}<\hat{\gamma}_{\text{s}}\right]\\
&+\mathbb{P}\left[\hat{\gamma}_{\text{p}}<\eta_{\text{p}}<\hat{\gamma}_{\text{p}}+\hat{\gamma}_{\text{s}}+\hat{\gamma}_{\text{p}}\hat{\gamma}_{\text{s}}-\eta_{\text{s}},\hat{\gamma}_{\text{s}}<\eta_\text{s}<\hat{\gamma}_{\text{s}}(1+\hat{\gamma}_{\text{p}})\right]\\
\overset{(a)}{=}&\mathbb{P}\left[\eta_{\text{p}}>\frac{\eta_{\text{s}}}{\hat{\gamma}_{\text{s}}}-1,\eta_{\text{s}}<\hat{\gamma}_{\text{s}}\right]\\
&+\mathbb{P}\left[\frac{\eta_{\text{s}}}{\hat{\gamma}_{\text{s}}}-1<\eta_{\text{p}}<\hat{\gamma}_{\text{p}}+\hat{\gamma}_{\text{s}}+\hat{\gamma}_{\text{p}}\hat{\gamma}_{\text{s}}-\eta_{\text{s}},\hat{\gamma}_{\text{s}}<\eta_\text{s}<\hat{\gamma}_{\text{s}}(1+\hat{\gamma}_{\text{p}})\right],\\
\end{aligned}
\end{equation}
where $(a)$ is obtained by first applying
\begin{equation}
	\setlength{\abovedisplayskip}{3pt}
	\setlength{\belowdisplayskip}{3pt}
    \begin{aligned}
    	&\mathbb{P}\left[\frac{\eta_{\text{s}}}{\hat{\gamma}_{\text{s}}}-1<\eta_{\text{p}}\leq\hat{\gamma}_{\text{p}},\eta_{\text{s}}<\hat{\gamma}_{\text{s}}(1+\hat{\gamma}_{\text{p}})\right]\\
    	=&\mathbb{P}\left[\frac{\eta_{\text{s}}}{\hat{\gamma}_{\text{s}}}-1<\eta_{\text{p}}\leq\hat{\gamma}_{\text{p}},\eta_{\text{s}}<\hat{\gamma}_{\text{s}}\right]+\mathbb{P}\left[\frac{\eta_{\text{s}}}{\hat{\gamma}_{\text{s}}}-1<\eta_{\text{p}}\leq\hat{\gamma}_{\text{p}},\hat{\gamma}_{\text{s}}<\eta_{\text{s}}<\hat{\gamma}_{\text{s}}(1+\hat{\gamma}_{\text{p}})\right],
    \end{aligned}
\end{equation}
 and then combining some of the terms. Since $\frac{\eta_{\text{s}}}{\hat{\gamma}_{\text{s}}}-1<0$ if $\eta_{\text{s}}<\hat{\gamma}_{\text{s}}$, it means the constraint $\eta_{\text{p}}>\frac{\eta_{\text{s}}}{\hat{\gamma}_{\text{s}}}-1$ in the first term of (\ref{pout_cog31}) can always be satisfied. Finally, $\mathcal{P}^{\text{CPA}}$ can be denoted as
\begin{equation}\label{pout_cog_cdf}
	\setlength{\abovedisplayskip}{3pt}
	\setlength{\belowdisplayskip}{3pt}
	\begin{aligned}
	\mathcal{P}^{\text{CPA}}
	=&\mathbb{P}\left[\eta_{\text{s}}<\hat{\gamma}_{\text{s}}\right]\\
	&+\mathbb{P}\left[\frac{\eta_{\text{s}}}{\hat{\gamma}_{\text{s}}}-1<\eta_{\text{p}}<\hat{\gamma}_{\text{p}}+\hat{\gamma}_{\text{s}}+\hat{\gamma}_{\text{p}}\hat{\gamma}_{\text{s}}-\eta_{\text{s}},\hat{\gamma}_{\text{s}}<\eta_\text{s}<\hat{\gamma}_{\text{s}}(1+\hat{\gamma}_{\text{p}})\right]\\
	=&\mathbb{P}\left[|h_{\text{s}}|^2<\hat{\tau}_{\text{s}}\right]\\
	&+\mathbb{P}\left[\frac{|h_{\text{s}}|^2}{\hat{\gamma}_{\text{s}}}-\frac{1}{\rho_m}<|h_{\text{p}}|^2<\hat{\tau}_{\text{p}}+\hat{\tau}_{\text{s}}+\hat{\tau}_{\text{p}}\hat{\gamma}_{\text{s}}-|h_{\text{s}}|^2,\hat{\tau}_{\text{s}}<|h_{\text{s}}|^2<\hat{\tau}_{\text{s}}(1+\hat{\gamma}_{\text{p}})\right],\\
	\end{aligned}
\end{equation}
where $\hat{\tau}_{\text{s}}=\frac{\hat{\gamma}_{\text{s}}}{\rho_{\text{s}}}$ and $\hat{\tau}_{\text{p}}=\frac{\hat{\gamma}_{\text{p}}}{\rho_{\text{p}}}$. In the following, the outage performance of the two scheduling schemes with CPA strategy will be analyzed based on (\ref{pout_cog_cdf}).

\vspace{-5pt}
\subsection{Outage Performance Analysis of the GUS Scheme with CPA Strategy}
\vspace{-5pt}

According to (\ref{channel order}), U$_1$ and U$_2$ will be scheduled in GUS scheme. For space limitations, we only consider the case regarding U$_1$ as the primary user in this paper, and the case regarding U$_2$ as the primary user can be derived similarly. Hence, in the following, we focus on investigating the outage performance of U$_2$, whose outage probability is given in the following theorem.

\textit{Theorem 1:} For GUS scheme with CPA strategy, the outage probability of the secondary user U$_2$ can be approximated as the following 3 cases.

Case 1: if $\hat{\gamma_{\text{s}}}\geq 1$,
\begin{equation}\label{pout_cog4_bus5}
	\setlength{\abovedisplayskip}{3pt}
	\setlength{\belowdisplayskip}{3pt}
	\begin{aligned}
		\mathcal{P}^{\text{CPA}}_\text{GUS}
		\approx&K(-\sum_{l_1=0}^{L}\Psi_{l_1}e^{-\mu_{l_1}\hat{\tau}_\text{s}})^{K-1}-(K-1)(-\sum_{l_1=0}^{L}\Psi_{l_1}e^{-\mu_{l_1}\hat{\tau}_\text{s}})^{K}\\
		&+K(K-1)\Xi_{K-2}e^{-\mu_{l_1}(\hat{\tau}_\text{s}+\hat{\tau}_\text{p}+\hat{\tau}_\text{p}\hat{\gamma}_\text{s})}\nu(\Delta_1;\hat{\tau}_{\text{s}},m_{i1})\\
		&+K(K-1)\Xi_{K-2}(e^{-\Delta_2m_{i1}}-e^{-\Delta_2\hat{\tau}_{\text{s}}})/\Delta_2.\\
	\end{aligned}
\end{equation}

Case 2: if $\hat{\gamma}_{\text{s}}< 1$ and $\hat{\gamma}_{\text{p}}>\frac{\hat{\gamma}_{\text{s}}}{1-\hat{\gamma}_{\text{s}}}$,
\begin{equation}\label{pout_cog4_bus51}
	\setlength{\abovedisplayskip}{3pt}
	\setlength{\belowdisplayskip}{3pt}
	\begin{aligned}
		\mathcal{P}^{\text{CPA}}_\text{GUS}
		\approx&K(-\sum_{l_1=0}^{L}\Psi_{l_1}e^{-\mu_{l_1}\hat{\tau}_\text{s}})^{K-1}-(K-1)(-\sum_{l_1=0}^{L}\Psi_{l_1}e^{-\mu_{l_1}\hat{\tau}_\text{s}})^{K}\\
		&+K(K-1)\Xi_{K-2}e^{-\mu_{l_1}(\hat{\tau}_\text{s}+\hat{\tau}_\text{p}+\hat{\tau}_\text{p}\hat{\gamma}_\text{s})}(\nu(\Delta_1;\hat{\tau}_{\text{s}},m_{i2})+\nu(\Delta_1;\frac{\hat{\tau}_{\text{s}}}{1-\hat{\gamma}_{\text{s}}},\hat{\tau}_{\text{s}}(1+\hat{\gamma}_{\text{p}})))\\
		&+K(K-1)\Xi_{K-2}(\frac{e^{-\Delta_2m_{i2}}-e^{-\Delta_2\hat{\tau}_{\text{s}}}}{\Delta_2}+e^{\frac{\mu_{l_1}}{\rho_m}}\frac{e^{-\Delta_3\hat{\tau}_{\text{s}}(1+\hat{\gamma}_{\text{p}})}-e^{-\Delta_3\frac{\hat{\tau}_{\text{s}}}{1-\hat{\gamma}_{\text{s}}}}}{\Delta_{3}}).\\
	\end{aligned}
\end{equation}

Case 3: if $\hat{\gamma}_{\text{s}}< 1$ and $\hat{\gamma}_{\text{p}}\leq\frac{\hat{\gamma}_{\text{s}}}{1-\hat{\gamma}_{\text{s}}}$,
\begin{equation}\label{pout_cog4_bus52}
	\setlength{\abovedisplayskip}{3pt}
	\setlength{\belowdisplayskip}{3pt}
	\begin{aligned}
		\mathcal{P}^{\text{CPA}}_\text{GUS}
		\approx&K(-\sum_{l_1=0}^{L}\Psi_{l_1}e^{-\mu_{l_1}\hat{\tau}_\text{s}})^{K-1}-(K-1)(-\sum_{l_1=0}^{L}\Psi_{l_1}e^{-\mu_{l_1}\hat{\tau}_\text{s}})^{K}\\
		&+K(K-1)\Xi_{K-2}e^{-\mu_{l_1}(\hat{\tau}_\text{s}+\hat{\tau}_\text{p}+\hat{\tau}_\text{p}\hat{\gamma}_\text{s})}\nu(\Delta_1;\hat{\tau}_{\text{s}},m_{i2})\\
		&+K(K-1)\Xi_{K-2}(e^{-\Delta_2m_{i2}}-e^{-\Delta_2\hat{\tau}_{\text{s}}})/\Delta_2.\\
	\end{aligned}
\end{equation}
Where, $m_{i1}=\text{min}\{\hat{\tau}_{\text{s}}+\hat{\tau}_{\text{s}}\hat{\gamma}_{\text{p}},\frac{1}{2}(\hat{\tau}_{\text{s}}+\hat{\tau}_{\text{p}}+\hat{\tau}_{\text{p}}\hat{\gamma}_{\text{s}})\}$, $m_{i2}=\text{min}\{m_{i1},\frac{\hat{\tau}_{\text{s}}}{1-\hat{\gamma}_{\text{s}}}\}$, $\Delta_1=\mu_{l_1}-\mu_{l_2}-\sum_{l=0}^{L}p_l\mu_l$, $\Xi_{K-2}=\sum_{l_1=0}^{L}\Psi_{l_1}\sum_{l_2=1}^{L}\Psi_{l_2}\mu_{l_2}\left(-1\right)^{K-1}\sum_{\sum_{l=0}^{L}p_l=K-2}\binom{K-2}{p_0,\dots,p_L}\left(\prod_{l=0}^{L}\Psi_l^{p_l}\right)$, $\Delta_2=\mu_{l_1}+\mu_{l_2}+\sum_{l=0}^{L}p_l\mu_l$, $\Delta_3=\frac{1}{\hat{\gamma}_{\text{s}}}\mu_{l_1}+\mu_{l_2}+\sum_{l=0}^{L}p_l\mu_l$, and
\begin{equation}
	\setlength{\abovedisplayskip}{3pt}
	\setlength{\belowdisplayskip}{3pt}
	\nu(\Delta;a,b)=
	\begin{cases}
		\frac{1}{\Delta}(e^{\Delta a}-e^{\Delta b}), &\text{if}\ \Delta\neq0\\
		b-a, &\text{otherwise}
	\end{cases}.
\end{equation}

\begin{proof}
	Please refer to Appendix \ref{proof_theorem1}.
\end{proof}


The expressions in Theorem 1 are quite complicated. To get some insights, we further derive the outage approximation expressions at the high SNR region.

\textit{Corollary 1:} When $\rho_m\to\infty$, the high SNR approximation of $\mathcal{P}^{\text{CPA}}_\text{GUS}$ is $\vec{\mathcal{P}}^{\text{CPA}}_\text{GUS}=K(S_L\hat{\tau}_\text{s})^{K-1}$ for all the three cases in Theorem 1, where $S_L=\sum_{l=1}^{L}\Psi_l\mu_l$. The secondary user U$_2$ can achieve a diversity order of $K-1$ in all three cases.

\begin{proof}
	Please refer to Appendix \ref{proof_corollary1}.
\end{proof}


{\color{black}\textit{Remark 4:} Corollary 1 shows that, in high SNR region, the outage performance of U$_2$ is only related to $\hat{\tau_{\text{s}}}$ and the total number of users $K$, but irrelevant to the target rate of the primary user. Specifically, the outage probability of U$_2$ is increasing with the uplift of $\hat{\tau_{\text{s}}}$, and decreasing with the increase of $K$.}

\vspace{-5pt}
\subsection{Outage Performance Analysis of the CUS Scheme with CPA Strategy}
\vspace{-5pt}

Due to space limitations, we only consider the case regarding the user with the largest CDF value as the primary user in this paper, and the case regarding the second-largest CDF user as the primary user can be analyzed similarly.

\textit{Theorem 2:}	For CUS scheme with CPA strategy, the outage probability of the user with the second-largest CDF value can be approximated as
	\begin{equation}
		\setlength{\abovedisplayskip}{3pt}
		\setlength{\belowdisplayskip}{3pt}
		\begin{aligned}
		\mathcal{P}_{\text{CUS}}^{\text{CPA}}\approx	K\sum_{l_1=1}^{L}\Psi_{l_1}(1-e^{-\mu_{l_1}\hat{\tau}_\text{s}})^{K-1}-(K-1)\sum_{l_1=1}^{L}\Psi_{l_1}(1-e^{-\mu_{l_1}\hat{\tau}_\text{s}})^K +T_1-T_2,
		\end{aligned}
	\end{equation}	
where $T_1$ and $T_2$ are calculated with Tables \ref{tableT1} and \ref{tableT2}, respectively. $\epsilon_1=\frac{\hat{\tau}_\text{s}+\hat{\tau}_\text{p}+\hat{\tau}_\text{p}\hat{\gamma}_\text{s}}{R^\alpha+2}$, $\epsilon_2=\frac{1}{2}(\hat{\tau}_\text{s}+\hat{\tau}_\text{p}+\hat{\tau}_\text{p}\hat{\gamma}_\text{s})$, $\epsilon_3=\frac{R^\alpha+1}{R^\alpha+2}(\hat{\tau}_\text{s}+\hat{\tau}_\text{p}+\hat{\tau}_\text{p}\hat{\gamma}_\text{s})$, $\epsilon_4=\frac{\hat{\tau}_\text{s}}{1-\hat{\gamma}_\text{s}-R^\alpha\hat{\gamma}_\text{s}}$, $\epsilon_5=\frac{\hat{\tau}_\text{s}}{1-\hat{\gamma}_\text{s}}$, $\epsilon_6=\frac{(R^\alpha+1)\hat{\tau}_\text{s}}{R^\alpha+1-\hat{\gamma}_\text{s}}$, and $x$ represents $\hat{\tau}_\text{s}+\hat{\tau}_\text{p}(1+\hat{\gamma}_\text{s})-y$ and  $\frac{1}{\hat{\gamma}_\text{s}}y-\frac{1}{\rho_m}$ in Table \ref{tableT1} and Table \ref{tableT2}, respectively. $F_{X,Y'}^{\text{CUS},I}(x,y)$, $F_{X,Y'}^{\text{CUS},II}(x,y)$, and  $F_{X,Y'}^{\text{CUS},III}(x,y)$ denote the partial derivatives of (\ref{CDF_joint_xy0}), (\ref{CDF_joint_xy}) and (\ref{CDF_joint_yx}) with respect to $y$, respectively, and we omit the expressions due to space limitations. $\mathcal{G}[a, b;F_{X,Y'}^{\text{CUS}}(x,y)]$ denotes applying Gaussian-Chebyshev quadrature \cite{Gaussian_Chebyshev} to the integral of $F_{X,Y'}^{\text{CUS}}(x,y)$ with integral interval $(a,b)$, namely, $\mathcal{G}[a, b;F_{X,Y'}^{\text{CUS}}(x,y)]\approx \int_{a}^{b}F_{X,Y'}^{\text{CUS}}(x,y)dy$.



\begin{proof}
	Please refer to Appendix \ref{proof_theorem_cog_cdf}.
\end{proof}


\begin{table*}
	\small
	\caption{Calculation of $T_1$}
	\label{tableT1}
	\centering
	\renewcommand\arraystretch{1.8} 
	\begin{tabular}{|c|m{4cm}<{\centering}|m{4.5cm}<{\centering}|}
		\hline
		Assumption 1&Assumption 2&$T_1$\\
		\hline
		\hline
		\multirow{4}{*}{$ \epsilon_2>\hat{\tau}_{\text{s}}(1+\hat{\gamma}_{\text{p}})$} &$ \epsilon_1>\hat{\tau}_{\text{s}}(1+\hat{\gamma}_{\text{p}})$&$\mathcal{G}[\hat{\tau}_{\text{s}},\hat{\tau}_{\text{s}}(1+\hat{\gamma}_{\text{p}});F_{X,Y'}^{\text{CUS,I}}(x,y)]$\\
		\cline{2-3}
		&$\hat{\tau}_\text{s}< \epsilon_1<\hat{\tau}_{\text{s}}(1+\hat{\gamma}_{\text{p}})$&$\mathcal{G}[\hat{\tau}_\text{s}, \epsilon_1;F_{X,Y'}^{\text{CUS,I}}(x,y)]$ + $\mathcal{G} [\epsilon_1,\hat{\tau}_\text{s}(1+\hat{\gamma}_\text{p});F_{X,Y'}^{\text{CUS,II}}(x,y)]$\\
		\cline{2-3}
		&$ \epsilon_1<\hat{\tau}_{\text{s}}$&$\mathcal{G}[\hat{\tau}_\text{s},\hat{\tau}_\text{s}(1+\hat{\gamma}_\text{p});F_{X,Y'}^{\text{CUS,II}}(x,y)]$\\
		\hline
		\multirow{7}{*}{$\hat{\tau}_{\text{s}}< \epsilon_2<\hat{\tau}_{\text{s}}(1+\hat{\gamma}_{\text{p}})$}& $\epsilon_1<\hat{\tau}_{\text{s}}$ and $ \epsilon_3<\hat{\tau}_{\text{s}}(1+\hat{\gamma}_{\text{p}})$& $\mathcal{G}[\hat{\tau}_{\text{s}}, \epsilon_2;F_{X,Y'}^{\text{CUS,II}}(x,y)]$  + $\mathcal{G}[\epsilon_2, \epsilon_3;F_{X,Y'}^{\text{CUS,III}}(x,y)]$\\
		\cline{2-3}
		&$ \epsilon_1<\hat{\tau}_{\text{s}}$ and $ \epsilon_3>\hat{\tau}_{\text{s}}(1+\hat{\gamma}_{\text{p}})$&$\mathcal{G}[\hat{\tau}_{\text{s}}, \epsilon_2;F_{X,Y'}^{\text{CUS,II}}(x,y)]$ + $\mathcal{G}[\epsilon_2,\hat{\tau}_{\text{s}}(1+\hat{\gamma}_{\text{p}});F_{X,Y'}^{\text{CUS,III}}(x,y)]$\\
		\cline{2-3}
		&$\hat{\tau}_{\text{s}}< \epsilon_1< \epsilon_2$ and $ \epsilon_3<\hat{\tau}_{\text{s}}(1+\hat{\gamma}_{\text{p}})$&$\mathcal{G}[\hat{\tau}_\text{s}, \epsilon_1;F_{X,Y'}^{\text{CUS,I}}(x,y)]$ + $\mathcal{G}[\epsilon_1, \epsilon_2;F_{X,Y'}^{\text{CUS,II}}(x,y)]$ + $\mathcal{G}[\epsilon_2, \epsilon_3;F_{X,Y'}^{\text{CUS,III}}(x,y)]$\\
		\cline{2-3}
		&$\hat{\tau}_{\text{s}}< \epsilon_1< \epsilon_2$ and $ \epsilon_3>\hat{\tau}_{\text{s}}(1+\hat{\gamma}_{\text{p}})$& $\mathcal{G}[\hat{\tau}_\text{s}, \epsilon_1;F_{X,Y'}^{\text{CUS,I}}(x,y)]$ + $\mathcal{G}[\epsilon_1, \epsilon_2;F_{X,Y'}^{\text{CUS,II}}(x,y)]$ + $\mathcal{G}[\epsilon_2,\hat{\tau}_{\text{s}}(1+\hat{\gamma}_{\text{p}});F_{X,Y'}^{\text{CUS,III}}(x,y)]$\\
		\hline
		\multirow{4}{*}{$ \epsilon_2<\hat{\tau}_{\text{s}}$}&$ \epsilon_3<\hat{\tau}_{\text{s}}$&$0$\\
		\cline{2-3}
		&$\hat{\tau}_{\text{s}}< \epsilon_3<\hat{\tau}_{\text{s}}(1+\hat{\gamma}_{\text{p}})$&$\mathcal{G}[\hat{\tau}_{\text{s}}, \epsilon_3;F_{X,Y'}^{\text{CUS,III}}(x,y)]$\\
		\cline{2-3}
		&$ \epsilon_3>\hat{\tau}_{\text{s}}(1+\hat{\gamma}_{\text{p}})$&$\mathcal{G}[\hat{\tau}_{\text{s}},\hat{\tau}_{\text{s}}(1+\hat{\gamma}_{\text{p}});F_{X,Y'}^{\text{CUS,III}}(x,y)]$\\
		\hline
	\end{tabular}
\end{table*}

\begin{table*}
	\small
	\caption{Calculation of $T_2$}
	\label{tableT2}
	\centering
	\renewcommand\arraystretch{1.8} 
	\begin{tabular}{|c|c| m{3.5cm}<{\centering}|m{4.5cm}<{\centering}|}
		\hline
		Assumption 1&Assumption 2&Assumption 3&$T_2$\\
		\hline
		\hline
		\multirow{4}{*}{$\hat{\gamma}_\text{s}>1$}  &$R^\alpha+1<\hat{\gamma}_\text{s}$&---&$0$\\
		\cline{2-4}
		&\multirow{3}{*}{$R^\alpha+1>\hat{\gamma}_\text{s}$}&$\epsilon_6>\hat{\tau}_s(1+\hat{\gamma}_p)$&$0$\\
		\cline{3-4}
		&&$\epsilon_6<\hat{\tau}_s(1+\hat{\gamma}_p)$&$\mathcal{G}[\epsilon_6,\hat{\tau}_{\text{s}}(1+\hat{\gamma}_{\text{p}});F_{X,Y'}^{\text{CUS,III}}(x,y)]$\\
		\hline
		\multirow{7}{*}{$\hat{\gamma}_\text{s}<1$}& \multirow{4}{*}{$\hat{\gamma}_\text{s}<\frac{\hat{\gamma}_\text{p}}{1+\hat{\gamma}_\text{p}}$}&$\hat{\gamma}_\text{s}>\frac{1}{(R^\alpha+1)(\hat{\gamma}_\text{p}+1)}$ &$\mathcal{G}[\epsilon_6,\epsilon_5;F_{X,Y'}^{\text{CUS,III}}(x,y)]$ +$\mathcal{G}[\epsilon_5,\hat{\tau}_\text{s}(1+\hat{\gamma}_\text{p});F_{X,Y'}^{\text{CUS,II}}(x,y)]$\\
		\cline{3-4}
		 &&$\hat{\gamma}_\text{s}<\frac{1}{(R^\alpha+1)(\hat{\gamma}_\text{p}+1)}$&$\mathcal{G}[\epsilon_6,\epsilon_5;F_{X,Y'}^{\text{CUS,III}}(x,y)]$ + $\mathcal{G}[\epsilon_5,\epsilon_4;F_{X,Y'}^{\text{CUS,II}}(x,y)]$ + $\mathcal{G}[\epsilon_4,\hat{\tau}_{\text{s}}(1+\hat{\gamma}_{\text{p}});F_{X,Y'}^{\text{CUS,I}}(x,y)]$\\
		 \cline{2-4}
		 &\multirow{2}{*}{$\hat{\gamma}_\text{s}>\frac{\hat{\gamma}_\text{p}}{1+\hat{\gamma}_\text{p}}$}&$\hat{\gamma}_\text{s}>\frac{(R^\alpha+1)\hat{\gamma}_\text{p}}{1+\hat{\gamma}_\text{p}}$&$\mathcal{G}[\epsilon_6,\hat{\tau}_{\text{s}}(1+\hat{\gamma}_{\text{p}});F_{X,Y'}^{\text{CUS,III}}(x,y)]$\\
		 \cline{3-4}
		 &&$\hat{\gamma}_\text{s}<\frac{(R^\alpha+1)\hat{\gamma}_\text{p}}{1+\hat{\gamma}_\text{p}}$&$0$\\
		 \hline
	\end{tabular}
\end{table*}


	\textit{Corollary 2:} In CUS scheme with CPA strategy, the user with the second-largest CDF value can achieve a diversity order $K-1$.

\begin{proof}
	Please refer to Appendix \ref{proof_corollary2}.
\end{proof}


In CPA strategy, the primary user, U$_\text{p}$, can achieve the same performance as it solely occupies the channel. By applying the results in \cite{2020_Lu_TVT,2016_Ding_impact_of_pairing}, it can be easily demonstrated that a diversity order of $K$ is achievable for U$_\text{p}$ in both GUS and CUS schemes. Meanwhile, Corollaries 1 and 2 illustrate that the secondary users in both GUS and CUS schemes can achieve the diversity orders of $K-1$. Hence, it can be concluded that full diversity orders can be achieved for CPA strategy in both GUS and CUS schemes. 

The CUS scheme not only can provide a fair access opportunity for all the users, but Corollary 2 illustrates that it can also achieve full diversity order as GUS scheme for CPA strategy. It is known that random user scheduling can also provide a fair access opportunity for the users, but it can not effectively employ multi-user diversity. Hence the CUS scheme will achieve much better outage performance than the random user selection scheme, especially for the system with many users, which will be shown in the simulation results.

\textit{Remark 5:} Compared with conventional NOMA transmission in \cite{2020_Lu_TVT,2016_Ding_impact_of_pairing}, an interesting phenomenon can be observed from RSMA transmission in the application of CPA strategy. Specifically, in \cite{2020_Lu_TVT,2016_Ding_impact_of_pairing}, the achieved diversity order of the secondary user is restricted by the channel quality of the primary user, however, such restriction vanishes in RSMA transmission. For example, in GUS scheme, assuming U$_2$ is the primary user, it can be demonstrated that the achieved diversity order of the secondary user, U$_\text{1}$, is also $K$ by applying (\ref{pout_cog_upbound}). However, in \cite{2020_Lu_TVT,2016_Ding_impact_of_pairing}, U$_1$ in GUS scheme can only achieve a diversity order of $K-1$. In other words, for RSMA transmission with CPA strategy, the secondary user's achieved diversity order is only determined by its own channel quality.



\vspace{-5pt}
\section{Outage Performance Analysis of The FPA Strategy}\label{fair_PA}
\vspace{-5pt}


In this section, we first derive the outage probability expression of FPA strategy. Then, we analyze the outage performance of each scheduling scheme with FPA strategy.

\vspace{-5pt}
\subsection{Outage Probability of the FPA Strategy}
\vspace{-5pt}

According to Section \ref{FPA_strategy_subsec}, the outage probability of $\text{U}_i$ with FPA strategy can be denoted as
\begin{equation}\label{Pout_fair0}
	\setlength{\abovedisplayskip}{3pt}
	\setlength{\belowdisplayskip}{3pt}
\begin{aligned}
\mathcal{P}_i^{\text{FPA}}=&\mathbb{P}\left[\eta_j<\eta_i<\eta_j+\eta_j^2,\frac{1}{2}\text{log}_2(1+\eta_i+\eta_j)<\hat{R}_i\right]\\
&+\mathbb{P}\left[\eta_j+\eta_j^2<\eta_i,\text{log}_2(1+\frac{\eta_i}{\eta_j+1})<\hat{R}_i\right]\\
&+\mathbb{P}\left[\eta_i<\eta_j<\eta_i+\eta_i^2,\frac{1}{2}\text{log}_2(1+\eta_i+\eta_j)<\hat{R}_i\right]\\
&+\mathbb{P}\left[\eta_i+\eta_i^2<\eta_j,\text{log}_2(1+\eta_i)<\hat{R}_i\right],\\
\end{aligned}
\end{equation}
where $\hat{R}_i$ denotes the target data rate of U$_i$. It is noted that $2^{2\hat{R}_i}-1=\hat{\gamma}_i^2+2\hat{\gamma}_i$, and $\hat{\gamma_i}=2^{\hat{R}_i}-1$ represents the target SINR of U$_i$. After some manipulations, (\ref{Pout_fair0}) can be converted to
\begin{equation}\label{Pout_fair1}
	\setlength{\abovedisplayskip}{3pt}
	\setlength{\belowdisplayskip}{3pt}
\begin{aligned}
\mathcal{P}_i^{\text{FPA}}=&\mathbb{P}\left[\eta_j<\eta_i<\eta_j+\eta_j^2,\eta_j<\hat{\gamma}_i\right]+\mathbb{P}\left[\eta_j<\eta_i<\hat{\gamma}_i^2+2\hat{\gamma}_i-\eta_j,\hat{\gamma}_i<\eta_j<\hat{\gamma}_i+\frac{1}{2}\hat{\gamma}_i^2\right]\\
&+\mathbb{P}\left[\eta_j+\eta_j^2<\eta_i<\hat{\gamma}_i(\eta_j+1),\eta_j<\hat{\gamma}_i\right]\\
&+\mathbb{P}\left[\eta_i<\eta_j<\eta_i+\eta_i^2,\eta_i<\hat{\gamma}_i\right]+\mathbb{P}\left[\eta_i<\eta_j<\hat{\gamma}_i^2+2\hat{\gamma}_i-\eta_i,\hat{\gamma}_i<\eta_i<\hat{\gamma}_i+\frac{1}{2}\hat{\gamma}_i^2\right]\\
&+\mathbb{P}\left[\eta_i+\eta_i^2<\eta_j,\eta_i<\hat{\gamma}_i\right].\\
\end{aligned}
\end{equation}
By combining some of the terms in (\ref{Pout_fair1}), we have
\begin{equation}\label{Pout_fair2}
	\setlength{\abovedisplayskip}{3pt}
	\setlength{\belowdisplayskip}{3pt}
\begin{aligned}
\mathcal{P}^{\text{FPA}}_i=&\mathbb{P}\left[\eta_j<\eta_i<\hat{\gamma}_i(\eta_j+1),\eta_j<\hat{\gamma}_i\right]+\mathbb{P}\left[\eta_j<\eta_i<\hat{\gamma}_i^2+2\hat{\gamma}_i-\eta_j,\hat{\gamma}_i<\eta_j<\hat{\gamma}_i+\frac{1}{2}\hat{\gamma}_i^2\right]\\
&+\mathbb{P}\left[\eta_i<\eta_j,\eta_i<\hat{\gamma}_i\right]+\mathbb{P}\left[\eta_i<\eta_j<\hat{\gamma}_i^2+2\hat{\gamma}_i-\eta_i,\hat{\gamma}_i<\eta_i<\hat{\gamma}_i+\frac{1}{2}\hat{\gamma}_i^2\right].\\
\end{aligned}
\end{equation}

\begin{figure}[t]
		\setlength{\abovecaptionskip}{-7pt}
		\setlength{\belowcaptionskip}{-12pt} 
	\centering
	\includegraphics[width=0.4\textwidth]{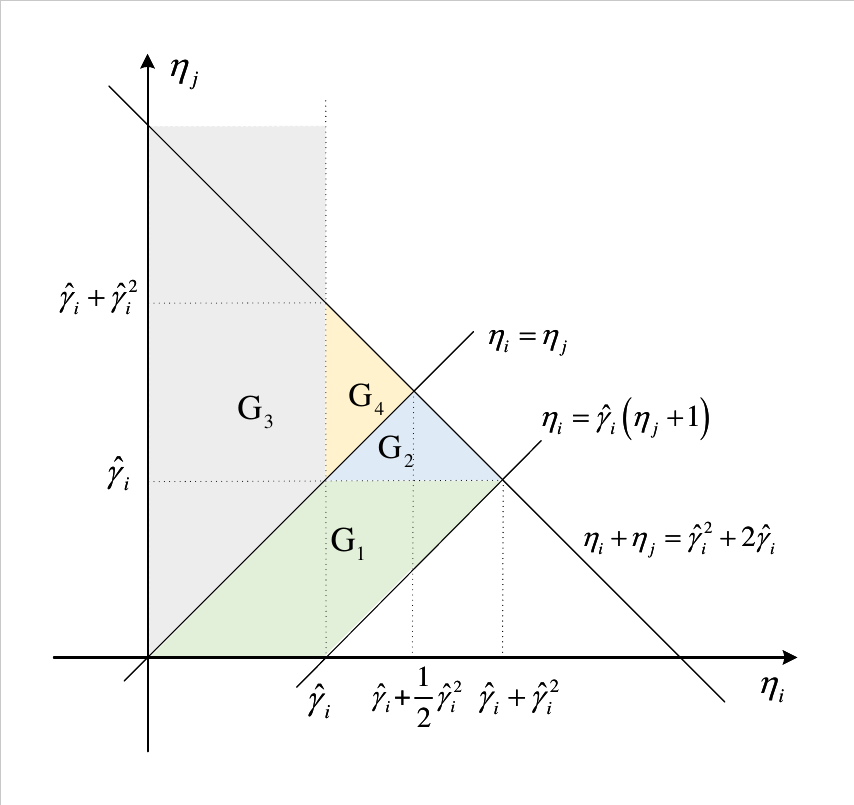}
	\caption{The integral regions of (\ref{Pout_fair2}), where G$_1$, G$_2$, G$_3$, and G$_4$ correspond to the integral regions of the first, second, third, and fourth terms of (\ref{Pout_fair2}), respectively.}
	\label{fig2_int_region}
\end{figure}

To further simplify the calculation complexity, we draw the integral regions of (\ref{Pout_fair2}) in Fig. \ref{fig2_int_region}. It can be observed from Fig. \ref{fig2_int_region} that $\mathcal{P}^{\text{FPA}}_i$ can be finally expressed as
\begin{equation}\label{Pout_fair3}
	\setlength{\abovedisplayskip}{3pt}
	\setlength{\belowdisplayskip}{3pt}
\begin{aligned}
\mathcal{P}^{\text{FPA}}_i
=&\mathbb{P}\left[\hat{\gamma}_i<\eta_i<\hat{\gamma}_i^2+\hat{\gamma}_i,\frac{1}{\hat{\gamma}_i}\eta_i-1<\eta_j<\hat{\gamma}_i^2+2\hat{\gamma}_i-\eta_i\right]+\mathbb{P}\left[\eta_i<\hat{\gamma}_i\right]\\
=&\mathbb{P}\left[\hat{\tau}_i<|h_i|^2<\hat{\gamma}_i\hat{\tau}_i+\hat{\tau}_i,\frac{1}{\hat{\gamma}_i}|h_i|^2-\frac{1}{\rho_m}<|h_j|^2<\hat{\tau}_i\hat{\gamma}_i+2\hat{\tau}_i-|h_i|^2\right]+\mathbb{P}\left[|h_i|^2<\hat{\tau}_i\right].\\
\end{aligned}
\end{equation}
Similarly, the outage probability of $\text{U}_j$ can be expressed as
\begin{equation}\label{Pout_fair2_i}
	\setlength{\abovedisplayskip}{3pt}
	\setlength{\belowdisplayskip}{3pt}
\begin{aligned}
\mathcal{P}_j^{\text{FPA}}=&\mathbb{P}\left[\eta_j<\eta_i<\eta_j+\eta_j^2,\frac{1}{2}\text{log}_2(1+\eta_i+\eta_j)<\hat{R}_j\right]\\
&+\mathbb{P}\left[\eta_j+\eta_j^2<\eta_i,\text{log}_2(1+\eta_j)<\hat{R}_j\right]\\
&+\mathbb{P}\left[\eta_i<\eta_j<\eta_i+\eta_i^2,\frac{1}{2}\text{log}_2(1+\eta_i+\eta_j)<\hat{R}_j\right]\\
&+\mathbb{P}\left[\eta_i+\eta_i^2<\eta_j,\text{log}_2(1+\frac{\eta_j}{1+\eta_i})<\hat{R}_j\right].\\
\end{aligned}
\end{equation}
By applying the same steps of deriving (\ref{Pout_fair3}), $\mathcal{P}^{\text{FPA}}_j$ can be finally denoted as
\begin{equation}\label{Pout_fair23}
	\setlength{\abovedisplayskip}{3pt}
	\setlength{\belowdisplayskip}{3pt}
\begin{aligned}
\mathcal{P}^{\text{FPA}}_j
=&\mathbb{P}\left[\hat{\tau}_j<|h_j|^2<\hat{\gamma}_j\hat{\tau}_j+\hat{\tau}_j,\frac{1}{\hat{\gamma}_j}|h_j|^2-\frac{1}{\rho_m}<|h_i|^2<\hat{\tau}_j\hat{\gamma}_j+2\hat{\tau}_j-|h_j|^2\right]+\mathbb{P}\left[|h_j|^2<\hat{\tau}_j\right],\\
\end{aligned}
\end{equation}
where $\hat{R}_j$ and $\hat{\gamma_j}=2^{\hat{R}_j}-1$ denote the target data rate and target SINR of U$_j$, respectively.

\vspace{-5pt}
\subsection{Outage Performance Analysis of the GUS Scheme with FPA Strategy}
\vspace{-5pt}

In the GUS scheme, U$_1$ and U$_2$ are scheduled and their channel gains are ordered as $|h_1|^2\geq|h_2|^2$. The following theorem shows the two users' outage probabilities.


\textit{Theorem 3:}
For GUS scheme with FPA strategy, the scheduled two users' outage probabilities can be expressed as
	\begin{equation}\label{thm_fair_GS1}
		\setlength{\abovedisplayskip}{3pt}
		\setlength{\belowdisplayskip}{3pt}
		\begin{aligned}
			\mathcal{P}^{\text{FPA}}_{\text{GUS},1}
			\approx&K\Xi_{K-1,l}\sum_{n=1}^{L}\Psi_n\mu_n(e^{-(\hat{\gamma}_1\hat{\tau}_1+2\hat{\tau}_1)\sum_{l=0}^{L}p_l\mu_l}\nu(\chi_3;\hat{\tau}_1+\frac{1}{2}\hat{\tau}_1\hat{\gamma}_1,\hat{\tau}_1+\hat{\tau}_1\hat{\gamma}_1)\\
			&+\frac{1}{\chi_4}e^{\frac{1}{\rho_m}\sum_{l=0}^{L}p_l\mu_l}(e^{-\chi_4(\hat{\tau}_1+\hat{\tau}_1\hat{\gamma}_1)}-e^{-\chi_4\hat{\tau}_1})\\
			&+\frac{1}{\chi_5}(e^{-\chi_5\hat{\tau}_1}-e^{-\chi_5(\hat{\tau}_1+\frac{1}{2}\hat{\tau}_1\hat{\gamma}_1)}))+(-\sum_{l=0}^{L}\Psi_le^{-\mu_l\hat{\tau}_1})^K\\
		\end{aligned}
	\end{equation}
and
\begin{equation}\label{thm_fair_GS2}
	\setlength{\abovedisplayskip}{3pt}
	\setlength{\belowdisplayskip}{3pt}
	\begin{aligned}
		\mathcal{P}^{\text{FPA}}_{\text{GUS},2}
		\approx&K(K-1)\Xi_{K-2}(e^{-\mu_{l_1}(\hat{\gamma}_2\hat{\tau}_2+2\hat{\tau}_2)}\nu(\Delta_1;\hat{\tau}_2,\hat{\tau}_2+\frac{1}{2}\hat{\tau}_2\hat{\gamma}_2)+\frac{e^{-\Delta_2(\hat{\tau}_2+\frac{1}{2}\hat{\tau}_2\hat{\gamma}_2)}-e^{-\Delta_2\hat{\tau}_2}}{\Delta_2})\\
		&+K(-\sum_{l=0}^{L}\Psi_le^{-\mu_l\hat{\tau}_2})^{K-1}-(K-1)(-\sum_{l=0}^{L}\Psi_le^{-\mu_l\hat{\tau}_2})^K,\\
	\end{aligned}
\end{equation}
 respectively, where $\chi_3=\sum_{l=0}^{L}p_l\mu_l-\mu_n$, $\chi_4=\frac{1}{\hat{\gamma}_{1}}\sum_{l=0}^{L}p_l\mu_l+\mu_n$, $\chi_5=\sum_{l=0}^{L}p_l\mu_l+\mu_n$.

\begin{proof}
	Please refer to Appendix \ref{proof_theorem3}.
\end{proof}


When the transmit SNR approaches infinity, $\hat{\tau}_1$, $\hat{\tau}_2$, $\hat{\tau}_1+\hat{\tau}_1\hat{\gamma}_1$, and $\hat{\tau}_2+\frac{1}{2}\hat{\tau}_2\hat{\gamma}_2$ approach zero. Then, substituting (\ref{CDF_unorder user_app}) and (\ref{PDF_unorder user_app}) into (\ref{Pout_fair4_greedy}) and (\ref{Pout_fair7_greedy}), we can obtain the high SNR approximation of the two users' outage probabilities. Similar with the derivation of Corollary 1, we can derive the following corollary.

\textit{Corollary 3:} When the transmit SNR $\rho_m\to \infty$, the high SNR approximation of $\mathcal{P}^{\text{FPA}}_{\text{GUS},1}$ and $\mathcal{P}^{\text{FPA}}_{\text{GUS},2}$ can be expressed as

\begin{equation}\label{Pout_fair4_gs_app}
	\setlength{\abovedisplayskip}{3pt}
	\setlength{\belowdisplayskip}{3pt}
		\begin{aligned}
			\vec{\mathcal{P}}^{\text{FPA}}_{\text{GUS},1}
			=&K(S_L\hat{\tau}_1)^{K}\sum_{k_1=0}^{K-1}\binom{K-1}{k_1}\frac{(\hat{\gamma}_1+2)^{K-1-k_1}(-1)^{k_1}}{k_1+1}[(1+\hat{\gamma}_1)^{k_1+1}-(1+\frac{1}{2}\hat{\gamma}_1)^{k_1+1}]\\
			&-K(\frac{S_L}{\rho_m})^{K}\sum_{k_2=0}^{K-1}\binom{K-1}{k_2}\frac{\hat{\gamma}_1}{k_2+1}(-1)^{K-1-k_2}[(1+\hat{\gamma}_1)^{k_2+1}-1]\\
			&+(S_L\hat{\tau}_1)^{K}(1+\frac{1}{2}\hat{\gamma}_1)^{K},
		\end{aligned}
	\end{equation}
 and 
\begin{equation}\label{Pout_fair7_app}
	\setlength{\abovedisplayskip}{3pt}
	\setlength{\belowdisplayskip}{3pt}
	\begin{aligned}
		\vec{\mathcal{P}}^{\text{FPA}}_{\text{GUS},2}
		=&K(S_L\hat{\tau}_2)^{K-1},\\
	\end{aligned}
\end{equation}
respectively. U$_1$ and U$_2$ can achieve diversity orders of $K$ and $K-1$, respectively.

{\color{black}From Corollary 3, we can conclude that for GUS scheme with FPA strategy, the high SNR outage probability of each user is related to it's own target rate, the number of users for scheduling, and the transmit SNR, and irrelevant with another user's target rate.}

\vspace{-5pt}
\subsection{Outage Performance Analysis of the CUS Scheme with FPA Strategy}
\vspace{-5pt}

Assuming the users with the largest and the second-largest CDF values are denoted as U$_i$ and U$_j$, respectively. The following theorem shows their outage probability expressions.

\textit{Theorem 4:} In the CUS scheme with FPA strategy, the outage probability of the user with the largest CDF value can be approximated as  
\begin{equation}\label{Pout_fair_CDF}
	\setlength{\abovedisplayskip}{3pt}
	\setlength{\belowdisplayskip}{3pt}
	\begin{aligned}
		\mathcal{P}^{\text{FPA}}_{\text{CUS},i}
		=&\mathcal{G}[\hat{\tau}_i,\epsilon_8;F_{X',Y}^{\text{CUS},III}(x,y)]+\mathcal{G}[\epsilon_8,\hat{\tau}_i+\hat{\tau}_i\hat{\gamma}_i;F_{X',Y}^{\text{CUS},II}(x,y)]\\
		&-\mathcal{G}[\hat{\tau}_i,\epsilon_{11};F_{X',Y}^{\text{CUS},I}(x,y)]-\mathcal{G}[\epsilon_{11},\hat{\tau}_i+\hat{\tau}_i\hat{\gamma}_i;F_{X',Y}^{\text{CUS},II}(x,y)],\\
	\end{aligned}
\end{equation}
if $R^{\alpha}>\hat{\gamma}_i$, otherwise
\begin{equation}\label{Pout_fair_CDF2}
	\setlength{\abovedisplayskip}{3pt}
	\setlength{\belowdisplayskip}{3pt}
	\begin{aligned}
		\mathcal{P}^{\text{FPA}}_{\text{CUS},i}
		=&\mathcal{G}[\hat{\tau}_i,\epsilon_7;F_{X',Y}^{\text{CUS},IV}(x,y)]+\mathcal{G}[\epsilon_7,\epsilon_8;F_{X',Y}^{\text{CUS},III}(x,y)]\\
		&+\mathcal{G}[\epsilon_8,\epsilon_{9};F_{X',Y}^{\text{CUS},II}(x,y)]+\mathcal{G}[\epsilon_{9},\hat{\tau}_i+\hat{\tau}_i\hat{\gamma}_i;F_{X',Y}^{\text{CUS},I}(x,y)]\\
		&-\mathcal{G}[\hat{\tau}_i,\hat{\tau}_i+\hat{\tau}_i\hat{\gamma}_i;F_{X',Y}^{\text{CUS},I}(x,y)],\\
	\end{aligned}
\end{equation}
where $\epsilon_7=\frac{2\hat{\tau}_i+\hat{\tau}_i\hat{\gamma}_i}{R^{\alpha}+2}$, $\epsilon_8=\hat{\tau}_i+\frac{1}{2}\hat{\tau}_i\hat{\gamma}_i$, $\epsilon_9=\frac{(2\hat{\tau}_i+\hat{\tau}_i\hat{\gamma}_i)(R^{\alpha}+1)}{R^{\alpha}+2}$, $\epsilon_{10}=\frac{\hat{\tau}_i}{1-\hat{\gamma}_i}$, and $\epsilon_{11}=\frac{\left(R^{\alpha}+1\right)\hat{\tau}_i}{R^{\alpha}+1-\hat{\gamma}_i}$. $F_{X',Y}^{\text{CUS},I}(x,y)$, $F_{X',Y}^{\text{CUS},II}(x,y)$, $F_{X',Y}^{\text{CUS},III}(x,y)$, and $F_{X',Y}^{\text{CUS},IV}(x,y)$ represent the partial derivatives of (\ref{CDF_joint_xy0}), (\ref{CDF_joint_xy}), (\ref{CDF_joint_yx}), and (\ref{CDF_joint_yx0}) with respect to $x$, respectively. Note that, in (\ref{Pout_fair_CDF}) and (\ref{Pout_fair_CDF2}), $y$ should be replaced with $2\hat{\tau}_i+\hat{\tau}_i\hat{\gamma}_i-x$ for the positive terms, while $y$ should be substituted with $\frac{1}{\hat{\gamma}_i}x-\frac{1}{\rho_m}$ for the negative terms. 



The outage probability of the user with the second-largest CDF value can be calculated with the results of Theorem 2 by setting $\hat{\tau}_\text{s}=\hat{\tau}_\text{p}=\hat{\tau}_j$ and $\hat{\gamma}_\text{s}=\hat{\gamma}_\text{p}=\hat{\gamma}_j$.

\begin{proof}
	Please refer to Appendix \ref{proof_theorem_fair_CDF}.
\end{proof}


Similar to the derivation of Corollary 2, we have the following corollary.

\textit{Corollary 4:} In the CUS scheme with FPA strategy, the scheduled two users can achieve diversity orders of $K$ and $K-1$, respectively.

\textit{Remark 6:} Corollaries 1 - 4 show that the scheduled users in both GUS and CUS schemes can achieve full diversity orders for both CPA and FPA strategies, and the achieved diversity orders are closely related to the total number of users $K$. In other words, the uplink RSMA transmission can achieve the same diversity orders as downlink NOMA transmission \cite{2014_Ding_random_deployed,2020_Lim_OP_CS_NOMA}. Moreover, compared with the schemes that divide all users in the system into two parts \cite{2020_Lu_TVT,2021_Yang_opportunistic_TWC,2018_Janghel_adaptive_COML,2020_Ye_TCOM_MEC_NNNF}, the scheduling schemes considered in this paper can achieve higher diversity orders.

\section{Simulation Results and Discussions}\label{Simulation and discussion}
\vspace{-5pt}

	

In this section, the accuracy of the theoretical analyses and the performance of different transmission schemes are examined through computer simulations. All the simulation results are averaged over $10^6$ independent trials, and the users are randomly distributed in the coverage area of the BS in each trial. Hereinafter, unless other specified, the simulation parameters are set similar with \cite{2020_Wei_Gain_NOMA_OMA} as $\alpha=3.76$, $K=4$, $R=500$ m, $\hat{R}_i=\hat{R}_j=\hat{R}_\text{p}=\hat{R}_\text{s}=1$ bps/Hz. The noise power is set as -100 dBm. Since the multinomial theorem is applied in this paper, to simplify the calculation complexity, all the complexity-accuracy tradeoff parameters are set as $10$ \cite{2014_Ding_random_deployed}, which is sufficient to maintain the calculation accuracy. It should be noted that, for each specific scheduling scheme, the users' achievable sum rates in both CPA and FPA strategies are the same, hence we do not compare the performance of the two power allocation strategies.

\vspace{-5pt}
\subsection{Simulation Results of the CPA Strategy}
\vspace{-5pt}

\begin{figure}[t]
	\setlength{\abovecaptionskip}{-3pt}
	\centering
	\subfigure[Different target rate pairs ]{\includegraphics[width=0.4\textwidth]{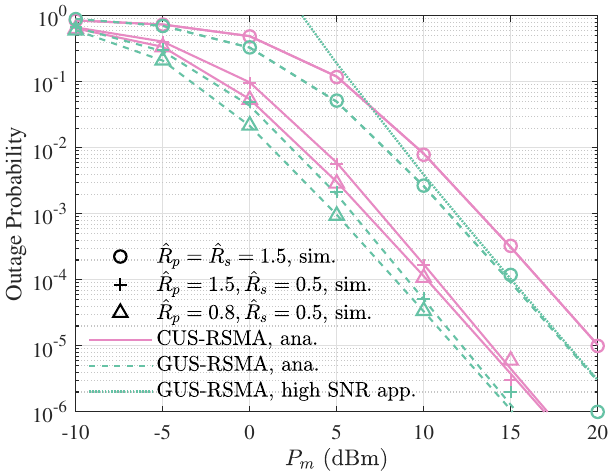}}\qquad
	\subfigure[Different coverage radii and path-loss exponents]{\includegraphics[width=0.4\textwidth]{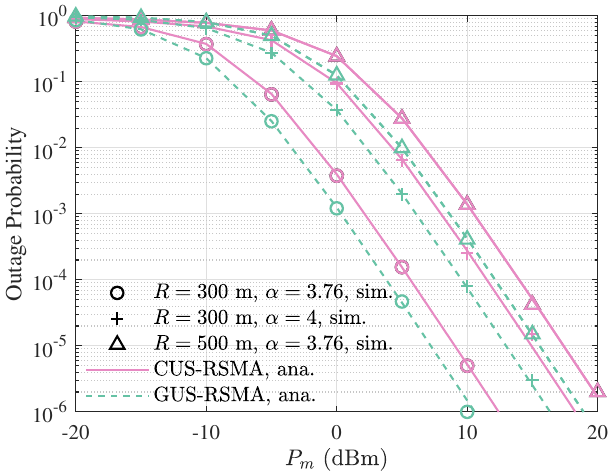}}
	\caption{Accuracy of the developed analytical results for CPA strategy.}
	\label{fig2_verify}
\end{figure}

The accuracy of the developed analytical results for the CPA strategy is examined in Fig. \ref{fig2_verify}, in which the high SNR approximation curve of the GUS-RSMA scheme is obtained based on Corollary 1. The analytical results of GUS-RSMA and CUS-RSMA schemes are plotted based on the expressions in Theorems 1 and 2, respectively. It should be noted that, the three target rate pairs in Fig. \ref{fig2_verify}(a) correspond to the three cases in Theorem 1. It can be observed that the developed analytical expressions match well with the simulation results in all three cases, which verifies the correctness of the theoretical analyses. Moreover, in the high SNR region, the approximation results of Corollary 1 for the GUS-RSMA scheme also tightly match the simulation results. Note that, for both scheduling schemes, the curves of rate pair ``$\hat{R_{\text{p}}}=1.5, \hat{R_{\text{s}}}=0.5$'' and ``$\hat{R_{\text{p}}}=0.8, \hat{R_{\text{s}}}=0.5$'' coincide in the high SNR region. This phenomenon can be explained by Corollary 1 that the outage probability in high SNR is only related to $K$, $\hat{R}_\text{s}$, and the transmit power, but irrelevant with $\hat{R}_\text{p}$. In Fig. \ref{fig2_verify}(b), the outage probability is compared with different coverage radii and path-loss exponents for the CPA strategy. It can be seen that the analytical results match well with the simulation results in all cases. Moreover, the outage probabilities of the two scheduling schemes are increasing with the increase of the coverage radius and path-loss exponent. Both result from the severe of the path-loss, which is proportional to the path-loss exponent and scheduled users' distances to the BS.

\begin{figure}[t]
	\setlength{\abovecaptionskip}{-3pt}
	\centering
	\subfigure[Users' admission probabilities comparison]{\includegraphics[width=0.4\textwidth]{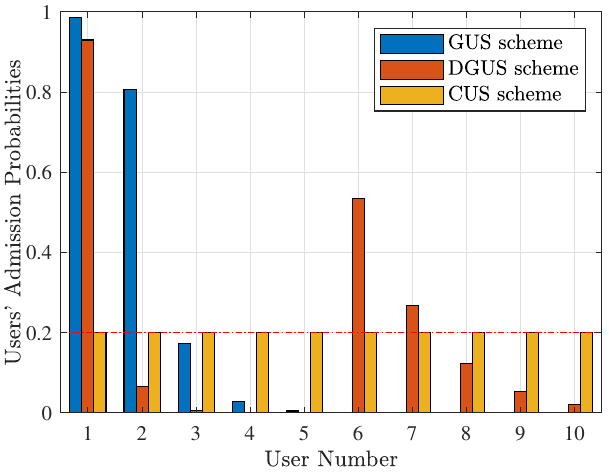}}
	\subfigure[Outage Probability Comparison]{\includegraphics[width=0.4\textwidth]{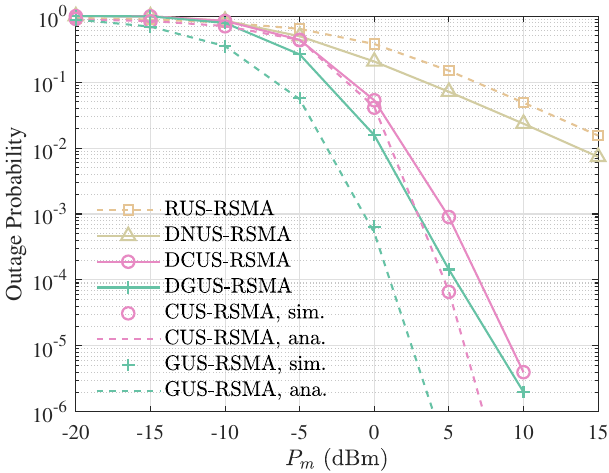}}\qquad
	\caption{Performance comparison of different scheduling schemes for CPA strategy.}
	\label{fig1_COG_OP}
\end{figure}

{\color{black}Fig. \ref{fig1_COG_OP} compares the users' outage probabilities and admission probabilities of different scheduling schemes with the CPA strategy for a 10-user case. The widely considered RUS scheme \cite{2020_Demarchou_ICC_Spatial_randomness,2013_Tabassum_ICI} is used as benchmark. In addition, we simulate the widely considered scheduling schemes, which first divide the BS's coverage area into two parts, and then scheduling one user from each part. Specifically, 5 users are randomly distributed in the disc region with radius 250 m, and the other 5 users are randomly distributed within the ring region with inner radius 250 m and outer radius 500 m. Based on which user is scheduled, we specifically compare three scheduling schemes, namely, 1) divided nearest user scheduling (DNUS) scheme \cite{2023_Chen_JSAC_RSMA-MEC}: the user closest to the BS in each region is scheduled; 2) divided GUS (DGUS) scheme: the user with the best channel gain in each region is scheduled; 3) divided CUS (DCUS) scheme \cite{2020_Lu_TVT}: the user with the largest CDF value in each region is scheduled.} 

{\color{black}Fig. \ref{fig1_COG_OP}(a) compares the admission fairness performance of the GUS, DGUS, and CUS schemes, where the 10 users' distances to the BS vary from 50 m to 500 m with an interval of 50 m. For ease of concision, the RUS scheme which can achieve fair admission probability and the DNUS scheme which always schedule user 1 and user 6 are not plotted here. We can see that the CUS scheme can achieve equal admission probability. However, in GUS and DGUS schemes, the users closer to the BS are scheduled with much higher probability, which means scheduling unfairness. Fig. \ref{fig1_COG_OP}(b) compares the outage performance of all scheduling schemes. It can be observed that the GUS and RUS schemes achieve the best and the worst outage performance, respectively. That is because the GUS scheme always schedules the users with the best channel gains and can effectively employ multi-user diversity, while the users are randomly scheduled in RUS scheme without utilizing multi-user diversity. In order to provide fair access opportunities to the users, the CUS scheme achieves inferior outage probability to the GUS scheme, whereas it can also effectively exploit multi-user diversity. Moreover, we can observe that the GUS (CUS) scheme can achieve better performance than the DGUS (DCUS) scheme, which demonstrate the advantage of unify scheduling considered in this paper.}


%
%

	\begin{figure}[t]
			\setlength{\abovecaptionskip}{-3pt}
		\centering
		\subfigure[Outage Probability Comparison]{\includegraphics[width=0.4\textwidth]{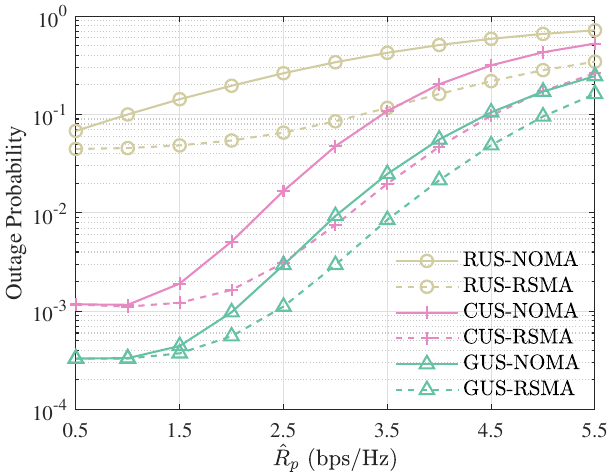}}\qquad
		\subfigure[Ergodic Data Rate Comparison]{\includegraphics[width=0.4\textwidth]{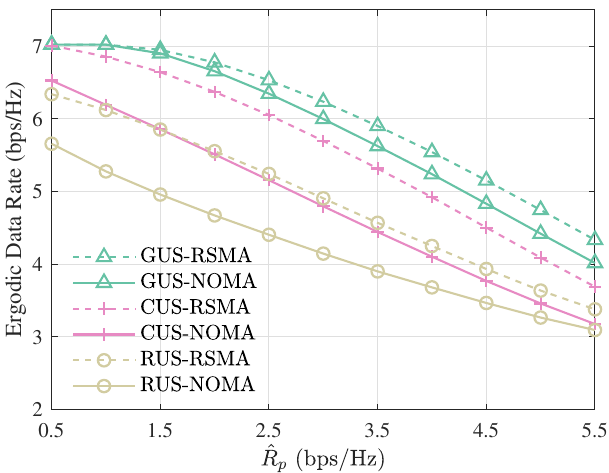}}
		\caption{Performance comparison versus the target rate of U$_\text{p}$ for CPA strategy ($P_m=15$ dBm).}
		\label{performance_ver_up}
	\end{figure}
	
	In Fig. \ref{performance_ver_up}, the outage probability and ergodic data rate performance are compared with a different target rate of U$_\text{p}$, where $P_m=15$ dBm. The newest NOMA transmission with HSIC decoding strategy [49] is used as a benchmark scheme, in which the secondary user’s transmit power is tuned for guaranteeing the primary user’s target rate. It can be observed that for all three scheduling strategies, RSMA can achieve better performance than NOMA in most cases. That is because RSMA can utilize the transmit power of U$_\text{s}$ more efficiently by introducing an additional SIC procedure. Moreover, both sub-figures illustrate that the performance of all schemes deteriorates with the uplift of $\hat{R_{\text{p}}}$. That is because the sum rates of the two users are fixed for RSMA. Thus, the higher $\hat{R_{\text{p}}}$ is, the lower rate is reserved for U$_\text{s}$. For NOMA, the interference threshold of U$_\text{p}$ determines the achievable rate of U$_\text{s}$, and the interference threshold is decreasing with the increase of $\hat{R_{\text{p}}}$. Conclusively, RSMA can achieve much better outage probability and ergodic data rate performance than NOMA for CPA strategy.

	

	
	\subsection{Simulation Results of the FPA Strategy}
	\vspace{-5pt}
	
		\begin{figure}[t]
		\setlength{\abovecaptionskip}{-3pt}
		\centering
		\includegraphics[width=0.4\textwidth]{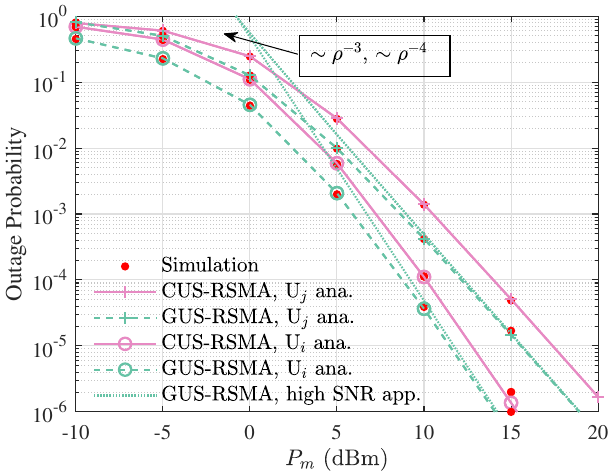}
		\caption{Accuracy of the developed analytical results for FPA strategy.}
		\label{accuracy_FPA}
	\end{figure}

\begin{figure}[t]
			\setlength{\abovecaptionskip}{-3pt}
		\centering
		\includegraphics[width=0.4\textwidth]{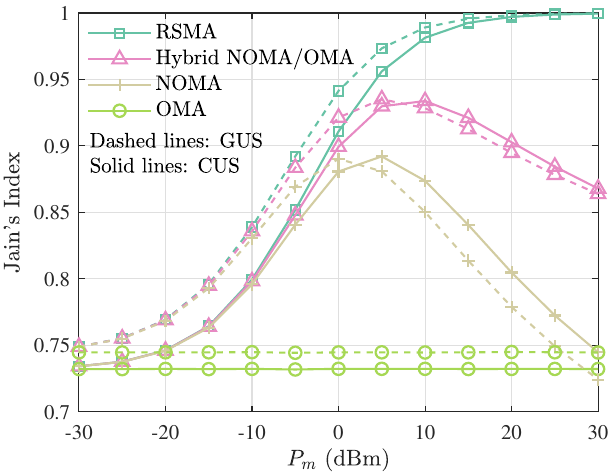}
		\caption{Rate fairness comparison of different transmission schemes.}
		\label{fig5_jain_index}
	\end{figure}

	Fig. \ref{accuracy_FPA} depicts the outage probability versus SNR for GUS-RSMA and CUS-RSMA schemes with FPA policy. It can be observed that the analytical results match well with the simulation results, which verifies the theoretical expressions in Theorems 3 and 4. It can also be observed from Fig. \ref{accuracy_FPA} that in the high SNR region, the outage probability curves of U$_i$ and U$_j$ are respectively parallel with $\rho^{-4}$ and $\rho^{-3}$ for both scheduling schemes, which demonstrates the conclusions of Corollaries 3 and 4 that both schemes can achieve full diversity orders. An interesting observation is that, in the high SNR region, the rate fairness performance of RSMA scheme is approximate 1 as shown in Fig. \ref{fig5_jain_index}. However, there is still some probability that U$_j$ owns extremely low channel gain, leading the outage performance of U$_j$ inferior to U$_i$.
	
The rate fairness of different schemes are compared in Fig. \ref{fig5_jain_index}, where the OMA, NOMA, and hybrid NOMA/OMA schemes \cite{2017_Wei_fairness_comparision} are used as benchmarks. Specifically, for the NOMA scheme, the user with strong channel gain is decoded first to maximize the rate fairness. For the OMA scheme, in order to achieve the same sum rate as the NOMA scheme, the optimal DOF allocation is applied \cite{2017_Wei_fairness_comparision,2005_Tse}, namely, the time-sharing factor for U$_k$ is $t_k=\frac{|h_k|^2}{|h_i|^2+|h_j|^2}, k=i,j$. The hybrid NOMA/OMA scheme \cite{2017_Wei_fairness_comparision} always chooses the transmission scheme between NOMA and OMA with higher fairness performance. For the RSMA scheme, the proposed FPA strategy is applied. It should be noted that all the four schemes are compared under the condition of achieving the same sum rate, denoted as $R_{\text{sum}}=\text{log}_2(1+\eta_i+\eta_j)$. 
In the simulations, Jain's index \cite{1984_jain} is employed as the metric for rate fairness evaluation,  which is widely used as a quantitative fairness measurement and is defined as $J=\frac{(\sum_{k=1}^{K}R_k)^2}{K\sum_{k=1}^{K}(R_k)^2}$. Note that $\frac{1}{K}\leq J\leq 1$, and a larger $J$ represents a higher fairness level. Particularly, we have $J=1$ when all users achieve the same data rate. 

It can be observed from Fig. \ref{fig5_jain_index} that the fairness performance of OMA scheme is constant, which is irrelevant to the transmit power. That is because the achievable rate of U$_k$ for OMA scheme is $R_k=t_kR_{\text{sum}}$ \cite{2017_Wei_fairness_comparision,2005_Tse}, and the Jain's index for OMA scheme is $J_\text{OMA}=\frac{R_{\text{sum}}^2}{2(t_i^2R_{\text{sum}}^2+t_j^2R_{\text{sum}}^2)}=\frac{1}{2}+\frac{|h_i|^2|h_j|^2}{|h_i|^4+|h_j|^4}$, which only depends on the users' channel gains. While the fairness performance of both NOMA and hybrid NOMA/OMA schemes increases with the increasing of the transmit power at the beginning, and then deteriorates as the transmit power increasing further. That is because in NOMA scheme, the strong user (with higher received power) is interfered by the weak user (with lower received power), so that the higher the transmit SNR is, the larger interference the strong user will encounter. Therefore, the weak user will achieve a much higher data rate than the strong user in the high SNR region, which greatly degrades the fairness performance of NOMA and hybrid NOMA/OMA schemes. However, the RSMA scheme can always achieve the best fairness performance by adaptively allocating the weak user's total transmit power to two sub-messages, by which the interference to the strong user can be flexibly adjusted. 

It should be noted that in Fig. \ref{accuracy_FPA} the GUS scheme shows better outage performance, and also in Fig. \ref{fig5_jain_index} it exhibits better rate fairness than the CUS scheme in low-to-moderate SNR region. That is because the GUS scheme always schedules the users with the best channel gains, hence it can achieve better outage performance than the CUS scheme. Moreover, as the variance of the scheduled two users' channel gains for the CUS scheme is larger than that of the GUS scheme, the condition of $\eta_i\geq \eta_j+\eta_j^2$ in (\ref{FPA_power allocation}) will be satisfied with much higher probability for the CUS scheme in low-to-moderate SNR region, which leads the fairness of the CUS scheme is worse than that of the GUS scheme. However, with the uplift of the transmit SNR, the impact of users' channel gains is decline, which leads to the probability of $\eta_i\geq \eta_j+\eta_j^2$ for the CUS scheme is decreasing faster than that of the GUS scheme, and the two scheduling schemes achieve the same fairness performance in high SNR region.

	\begin{figure}[t]
		\setlength{\abovecaptionskip}{-3pt}
	\centering
	\subfigure[CUS scheme ]{\includegraphics[width=0.4\textwidth]{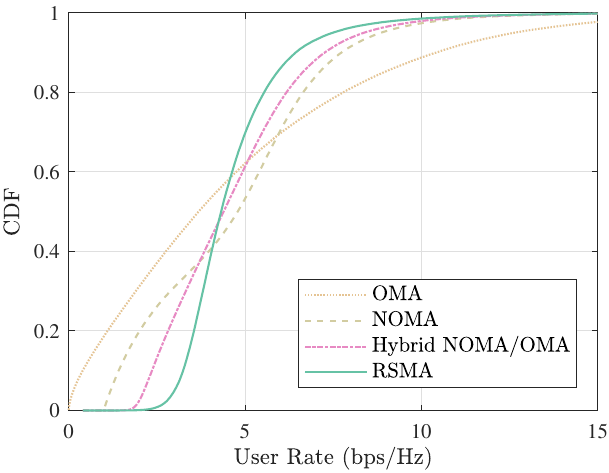}}\qquad
	\subfigure[GUS scheme]{\includegraphics[width=0.4\textwidth]{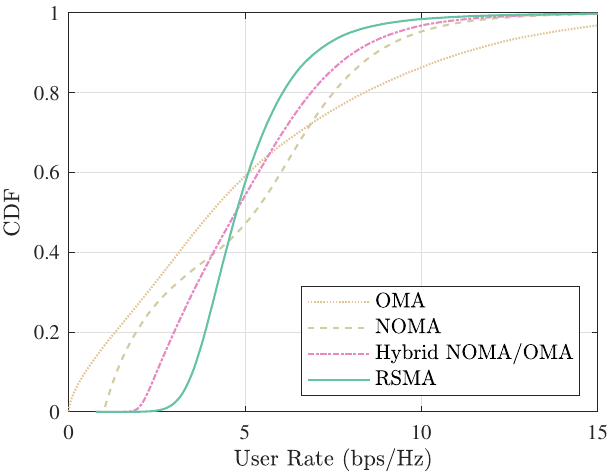}}
	\caption{User rate's CDF comparison of different transmission schemes ($P_m=15$ dBm).}
	\label{fig7_CDF}
\end{figure}

Finally, the CDF of user rate for each transmission scheme is simulated and compared in Fig. \ref{fig7_CDF}, where $P_m$ is set as 15 dBm.  It can be observed that the user rate distribution of RSMA scheme is more concentrated than those of the other three schemes, which indicates that the RSMA scheme can achieve the best fairness performance. Since the 10th-percentile of the user rate is highly related to fairness and user experience \cite{2017_Wei_fairness_comparision}, we can observe from Fig. \ref{fig7_CDF} that, compared with the hybrid NOMA/OMA scheme, the 10th-percentile of user rate increases by 0.9 bps/Hz (1 bps/Hz) for CUS scheme (GUS scheme). In summary, by combining the proposed FPA strategy with RSMA scheme, the low-rate user's performance and quality of experience can be greatly improved.

	
	\section{Conclusions}\label{conclusion}
In this paper, we studied the outage performance of the uplink RSMA system with randomly deployed users. Specifically, two scheduling schemes were studied, which can maximize the sum rate of opportunistic scheduling with different scheduling fairness requirements. Moreover, two kinds of dynamic power allocation strategies were studied, i.e., CPA strategy to guarantee one user's target rate, and FPA strategy to enhance the users' rate fairness, to further explore the flexibility of RSMA. By utilizing stochastic geometry and probability theory, analytical expressions of the users' outage probabilities were derived for each combination of scheduling scheme and power allocation strategy. In addition, we theoretically demonstrated that all combinations can achieve full diversity orders. Simulation results were provided to verify the accuracy of the analytical results and the superior performance of RSMA to NOMA. Hence, RSMA is a promising multiple access technique for the next-generation mobile networks, and the CUS-RSMA and GUS-RSMA schemes can be applied if the operators concern more about the admission fairness and the rate performance, respectively. In the future, it is interesting to investigate the performance of RSMA with multi-antenna transceivers or with other scheduling schemes. {\color{black}It is also interesting to investigate more general case, such as scheduling more users, considering imperfect SIC.}


	 \appendices

\setcounter{equation}{0}
\renewcommand{\theequation}{A.\arabic{equation}}

\section{Proof of Lemma 1}\label{Proof of lemma1}

For given the distances of all users to the BS $\mathbf{r}=\{r_1,\dots,r_K\}$, the joint CDF of the scheduled two users' channel gains can be denoted as \cite{2020_Lim_OP_CS_NOMA}
\begin{equation}\label{CDF_joint_CDF1}
	\setlength{\abovedisplayskip}{3pt}
	\setlength{\belowdisplayskip}{3pt}
	\begin{aligned}
		F_{X,Y}^{\text{CUS}}(x,y|\mathbf{r})
		=&\mathbb{P}[X\leq x,Y\leq y|\mathbf{r}]\\
		=&\sum_{i=1}^{K}\sum_{j=1,j\neq i}^{K}\mathbb{P}[X\leq x,Y\leq y, \text{U$_i$ and U$_j$ have the largest} \\ 
		&\qquad\qquad\quad\ \text{and the second-largest CDF values, respectively}|\mathbf{r}]\\
		\overset{(a)}{=}&\sum_{i=1}^{K}\sum_{j=1,j\neq i}^{K}\mathbb{P}[X\leq x,Y\leq y,F_i(X|r_i)\geq F_j(Y|r_j),F_j(Y|r_j)\geq U_k (k\neq i,j)]\\
		=&\sum_{i=1}^{K}\sum_{j=1,j\neq i}^{K}\int_{X}\int_{Y}F_j(Y|r_j)^{K-2}dF_j(Y|r_j)dF_i(X|r_i),\\
	\end{aligned}
\end{equation}
where $U_k$ in $(a)$ represents the random variable of U$_k$'s channel gain's CDF value, which is uniformly distributed in $[0, 1]$ \cite{2015-Jin-Fundamental_limits_CDF}. In the following, the integral interval will be analyzed first, then the joint CDF can be obtained by averaging over $r_i$ and $r_j$.

\subsection{Integral Interval Analysis}
 Firstly, we consider the case $F_i(x|r_i)>F_j(y|r_j)$. In this case, for these $X$ satisfying $X\geq F_i^{-1}(F_j(y|r_j))$, all the values of $Y\leq y$ satisfy the requirement of $F_i(X|r_i)\geq F_j(Y|r_j)$, thus the integral interval of $X$ and $Y$ are $[F_i^{-1}(F_j(y|r_j)),x]$ and $[0,y]$, respectively; on the other hand, for these $X<F_i^{-1}(F_j(y|r_j))$, to meet the requirement of $F_i(X|r_i)\geq F_j(Y|r_j)$, the integral interval of $Y$ should be $[0,F_j^{-1}(F_i(X|r_i))]$. Therefore, if $F_i(x|r_i)>F_j(y|r_j)$, the joint CDF can be expressed as
\begin{equation}\label{CDF_joint_CDF2}
	\setlength{\abovedisplayskip}{3pt}
	\setlength{\belowdisplayskip}{3pt}
	\begin{aligned}
		F_{X,Y}^{\text{CUS}}(x,y|\mathbf{r})
		=&\sum_{i=1}^{K}\sum_{j=1,j\neq i}^{K}\int_{F_i^{-1}(F_j(y|r_j))}^x\int_{0}^{y}F_j(Y|r_j)^{K-2}dF_j(Y|r_j)dF_i(X|r_i)\\
		&+\sum_{i=1}^{K}\sum_{j=1,j\neq i}^{K}\int_{0}^{F_i^{-1}(F_j(y|r_j))}\int_{0}^{F_j^{-1}(F_i(X|r_i))}F_j(Y|r_j)^{K-2}dF_j(Y|r_j)dF_i(X|r_i)\\
		=&\sum_{i=1}^{K}\sum_{j=1,j\neq i}^{K}\frac{1}{K-1}F_j(y|r_j)^{K-1}F_i(x|r_i)-\sum_{i=1}^{K}\sum_{j=1,j\neq i}^{K}\frac{1}{K}F_j(y|r_j)^K.
	\end{aligned}
\end{equation}

For the case $F_i(x|r_i)\leq F_j(y|r_j)$, to meet the requirement of $F_i(X|r_i)\geq F_j(Y|r_j)$, the integral interval of $Y$ should be $[0,F_j^{-1}(F_i(X|r_i))]$ for each $X\leq x$. In this case, the joint CDF can be expressed as
\begin{equation}\label{CDF_joint_CDF3}
	\setlength{\abovedisplayskip}{3pt}
	\setlength{\belowdisplayskip}{3pt}
	\begin{aligned}
		F_{X,Y}^{\text{CUS}}(x,y|\mathbf{r})
		=&\sum_{i=1}^{K}\sum_{j=1,j\neq i}^{K}\int_{0}^{x}\int_{0}^{F_j^{-1}(F_i(X|r_i))}F_j(Y|r_j)^{K-2}dF_j(Y|r_j)dF_i(X|r_i)\\
		=&\sum_{i=1}^{K}\sum_{j=1,j\neq i}^{K}\frac{1}{K(K-1)}F_i(x|r_i)^K.
	\end{aligned}
\end{equation}

\subsection{Averaging over the Distances}

Since all users are randomly distributed in the disc area with radius $R$, every user's distance to the BS is independent of each other and the PDF of each user's distance is $f_{r_k}(z)=\frac{2z}{R^2}$ \cite{1993_Kingman_PP}. Based on that, we can derive the joint CDF by averaging (\ref{CDF_joint_CDF2}) and (\ref{CDF_joint_CDF3}) over the distances. Note that, the CDF of U$_k\  (k\in\{1,\dots,K\})$ is $F_k(x|r_k)=1-e^{-(1+r_k^\alpha)x}$. Thus, $F_i(x|r_i)>F_j(y|r_j)$ means $(1+r_i^\alpha)x>(1+r_j^\alpha)y$, which can be converted to $r_i>[(1+r_j^\alpha)\frac{y}{x}-1]^{\frac{1}{\alpha}}$. Note that, when $r_j\in(0,R)$, if $x>y$, $[(1+r_j^\alpha)\frac{y}{x}-1]^{\frac{1}{\alpha}}$ will be less than $R$ and it may be even less than 0 (Specially, when $r_j=(\frac{x}{y}-1)^{\frac{1}{\alpha}}$, $[(1+r_j^\alpha)\frac{y}{x}-1]^{\frac{1}{\alpha}}=0$ holds); on the other hand, if $x<y$, $[(1+r_j^\alpha)\frac{y}{x}-1]^{\frac{1}{\alpha}}$ should be larger than $0$ and may even larger than $R$ (Specially, when $r_j=[(R^{\alpha}+1)\frac{x}{y}-1]^{\frac{1}{\alpha}}$, $[(1+r_j^\alpha)\frac{y}{x}-1]^{\frac{1}{\alpha}}=R$ holds). For $F_i(x|r_i)\leq F_j(y|r_j)$, similar conclusions can be obtained for $r_i$. Base on that, the joint CDF can be derived with two cases.

First, we consider the case $x>y$. Substituting $F_k(x|r_k)=1-e^{-(1+r_k^\alpha)x}$ into (\ref{CDF_joint_CDF2}) and (\ref{CDF_joint_CDF3}), the joint CDF can be expressed as
\begin{equation}\label{CDF_joint_CDF4}
	\setlength{\abovedisplayskip}{3pt}
	\setlength{\belowdisplayskip}{3pt}
	\begin{aligned}
		F_{X,Y}^{\text{CUS}}(x,y)
		=&K\int_{0}^{a_1}\frac{2r_j}{R^2}[1-e^{-(1+r_j^\alpha)y}]^{K-1}\int_{0}^{R}\frac{2r_i}{R^2}[1-e^{-(1+r_i^\alpha)x}]dr_idr_j\\
		&+K\int_{a_1}^{R}\frac{2r_j}{R^2}[1-e^{-(1+r_j^\alpha)y}]^{K-1}\int_{I_1(r_j)}^{R}\frac{2r_i}{R^2}[1-e^{-(1+r_i^\alpha)x}]dr_idr_j\\
		&-(K-1)\int_{0}^{a_1}\frac{2r_j}{R^2}[1-e^{-(1+r_j^\alpha)y}]^{K}\int_{0}^{R}\frac{2r_i}{R^2}dr_idr_j\\
		&-(K-1)\int_{a_1}^{R}\frac{2r_j}{R^2}[1-e^{-(1+r_j^\alpha)y}]^{K}\int_{I_1(r_j)}^{R}\frac{2r_i}{R^2}dr_idr_j\\
		&+\int_{0}^{a_2}\frac{2r_i}{R^2}[1-e^{-(1+r_i^\alpha)x}]^K\int_{I_2(r_i)}^{R}\frac{2r_j}{R^2}dr_jdr_i.
	\end{aligned}
\end{equation}
 It is quite challenging to obtain an insightful expression with (\ref{CDF_joint_CDF4}). So, we use Gaussian-Chebyshev quadrature \cite{Gaussian_Chebyshev} to develop an approximation expression. By applying Gaussian-Chebyshev quadrature and with some manipulations, we can obtain (\ref{CDF_joint_xy}).

Note that, (\ref{CDF_joint_CDF4}) and (\ref{CDF_joint_xy}) require $a_1<R$ and $a_2>0$ (namely, $\frac{x}{y}<R^{\alpha}+1$). If $\frac{x}{y}\geq R^{\alpha}+1$, the joint CDF can be simplified to
\begin{equation}\label{CDF_joint_xy01}
	\setlength{\abovedisplayskip}{3pt}
	\setlength{\belowdisplayskip}{3pt}
	\begin{aligned}
		F_{X,Y}^{\text{CUS}}(x,y)
		=&K\int_{0}^{R}\frac{2r_j}{R^2}[1-e^{-(1+r_j^\alpha)y}]^{K-1}\int_{0}^{R}\frac{2r_i}{R^2}[1-e^{-(1+r_i^\alpha)x}]dr_idr_j\\
		&-(K-1)\int_{0}^{R}\frac{2r_j}{R^2}[1-e^{-(1+r_j^\alpha)y}]^{K}\int_{0}^{R}\frac{2r_i}{R^2}dr_idr_j.
	\end{aligned}
\end{equation}
Similarly, by applying Gaussian-Chebyshev quadrature, we can get (\ref{CDF_joint_xy0}).

When $x<y$, the joint CDF can be denoted as
\begin{equation}\label{CDF_joint_CDF6}
	\setlength{\abovedisplayskip}{3pt}
	\setlength{\belowdisplayskip}{3pt}
	\begin{aligned}
		F_{X,Y}^{\text{CUS}}(x,y)
		=&K\int_{0}^{a_2}\frac{2r_j}{R^2}[1-e^{-(1+r_j^\alpha)y}]^{K-1}\int_{I_1(r_j)}^{R}\frac{2r_i}{R^2}[1-e^{-(1+r_i^\alpha)x}]dr_idr_j\\
		&-(K-1)\int_{0}^{a_2}\frac{2r_j}{R^2}[1-e^{-(1+r_j^\alpha)y}]^{K}\int_{I_1(r_j)}^{R}\frac{2r_i}{R^2}dr_idr_j\\
		&+\int_{0}^{a_1}\frac{2r_i}{R^2}[1-e^{-(1+r_i^\alpha)x}]^K\int_{0}^{R}\frac{2r_j}{R^2}dr_jdr_i\\
		&+\int_{a_1}^{R}\frac{2r_i}{R^2}[1-e^{-(1+r_i^\alpha)x}]^K\int_{I_2(r_i)}^{R}\frac{2r_j}{R^2}dr_jdr_i.
	\end{aligned}
\end{equation}
Applying Gaussian-Chebyshev quadrature to (\ref{CDF_joint_CDF6}), we have (\ref{CDF_joint_yx}). Similar with the case $x>y$, (\ref{CDF_joint_CDF6}) and (\ref{CDF_joint_yx}) require $\frac{y}{x}<R^{\alpha}+1$. If $\frac{y}{x}\geq R^{\alpha}+1$, the joint CDF can be simplified as
\begin{equation}\label{CDF_joint_yx01}
	\setlength{\abovedisplayskip}{3pt}
	\setlength{\belowdisplayskip}{3pt}
	\begin{aligned}
		F_{X,Y}^{\text{CUS}}(x,y)
		=
		&\int_{0}^{R}\frac{2r_i}{R^2}[1-e^{-(1+r_i^\alpha)x}]^K\int_{0}^{R}\frac{2r_j}{R^2}dr_jdr_i.
	\end{aligned}
\end{equation}
And we can get (\ref{CDF_joint_yx0}) by utilizing Gaussian-Chebyshev quadrature as well. The proof of Lemma 1 is completed.

\section{Proof of Theorem 1}\label{proof_theorem1}
	Since the relationship of the scheduled users' channel gains is determined, substituting $|h_{\text{s}}|^2\leq|h_{\text{p}}|^2$ into (\ref{pout_cog_cdf}), the outage probability of U$_2$ can be expressed as
	\begin{equation}
		\setlength{\abovedisplayskip}{3pt}
		\setlength{\belowdisplayskip}{3pt}
		\begin{aligned}\label{BS_cog_1}
			\mathcal{P}^{\text{CPA}}_\text{GUS}
			=&\mathbb{P}\left[|h_{\text{s}}|^2<\hat{\tau}_{\text{s}}\right]+\mathbb{P}[\text{max}\{|h_{\text{s}}|^2,\frac{|h_{\text{s}}|^2}{\hat{\gamma}_{\text{s}}}-\frac{1}{\rho_m}\}<|h_{\text{p}}|^2<\hat{\tau}_{\text{p}}+\hat{\tau}_{\text{s}}+\hat{\tau}_{\text{p}}\hat{\gamma}_{\text{s}}-|h_{\text{s}}|^2,\\
			&\hat{\tau}_{\text{s}}<|h_{\text{s}}|^2<\hat{\tau}_{\text{s}}(1+\hat{\gamma}_{\text{p}})].\\
		\end{aligned}
	\end{equation}
	After some manipulations, we can find that in the case $\hat{\gamma}_{\text{s}}\geq 1$, $|h_{\text{s}}|^2$ is always larger than $\frac{|h_{\text{s}}|^2}{\hat{\gamma}_{\text{s}}}-\frac{1}{\rho_m}$. And in the case $\hat{\gamma}_{\text{s}}<1$, $|h_{\text{s}}|^2>\frac{|h_{\text{s}}|^2}{\hat{\gamma}_{\text{s}}}-\frac{1}{\rho_m}$ holds if $|h_{\text{s}}|^2<\frac{\hat{\tau}_{\text{s}}}{1-\hat{\gamma}_{\text{s}}}$; and $|h_{\text{s}}|^2<\frac{|h_{\text{s}}|^2}{\hat{\gamma}_{\text{s}}}-\frac{1}{\rho_m}$ holds if $|h_{\text{s}}|^2>\frac{\hat{\tau}_{\text{s}}}{1-\hat{\gamma}_{\text{s}}}$. 
	
	Therefore, $\mathcal{P}^{\text{CPA}}_\text{GUS}$ can be converted to the following two cases. Case 1, if $\hat{\gamma}_{\text{s}}\geq 1$,
	\begin{equation}
		\setlength{\abovedisplayskip}{3pt}
		\setlength{\belowdisplayskip}{3pt}
		\begin{aligned}\label{BS_cog_2}
			\mathcal{P}^{\text{CPA}}_\text{GUS}
			=&\mathbb{P}\left[|h_{\text{s}}|^2<\hat{\tau}_{\text{s}}\right]+\mathbb{P}[|h_{\text{s}}|^2<|h_{\text{p}}|^2<\hat{\tau}_{\text{p}}+\hat{\tau}_{\text{s}}+\hat{\tau}_{\text{p}}\hat{\gamma}_{\text{s}}-|h_{\text{s}}|^2,\hat{\tau}_{\text{s}}<|h_{\text{s}}|^2<m_{i1}],\\
		\end{aligned}
	\end{equation}
	where $m_{i1}=\text{min}\{\hat{\tau}_{\text{s}}+\hat{\tau}_{\text{s}}\hat{\gamma}_{\text{p}},\frac{1}{2}(\hat{\tau}_{\text{s}}+\hat{\tau}_{\text{p}}+\hat{\tau}_{\text{p}}\hat{\gamma}_{\text{s}})\}$ and the constraint $|h_{\text{s}}|^2<\frac{1}{2}(\hat{\tau}_{\text{s}}+\hat{\tau}_{\text{p}}+\hat{\tau}_{\text{p}}\hat{\gamma}_{\text{s}})$ is obtained from $|h_{\text{s}}|^2<\hat{\tau}_{\text{p}}+\hat{\tau}_{\text{s}}+\hat{\tau}_{\text{p}}\hat{\gamma}_{\text{s}}-|h_{\text{s}}|^2$. Substituting the joint PDF in (\ref{PDF_largest2}) into (\ref{BS_cog_2}), we have
	\begin{equation}\label{pout_cog4_bus}
		\setlength{\abovedisplayskip}{3pt}
		\setlength{\belowdisplayskip}{3pt}
		\begin{aligned}
			\mathcal{P}^{\text{CPA}}_\text{GUS}
			=&F_{|h_{K-1}|^2}(\hat{\tau}_\text{s})\\
			&+K(K-1)\int_{\hat{\tau}_{\text{s}}}^{m_{i1}}[F_{|h|^2}(\hat{\tau}_\text{s}+\hat{\tau}_\text{p}+\hat{\tau}_\text{p}\hat{\gamma}_\text{s}-x)-F_{|h|^2}(x)]f_{|h|^2}(x)F_{|h|^2}(x)^{K-2}dx.\\
		\end{aligned}
	\end{equation}

	
	In the following, the expression of $F_{|h|^2}(x)^{K-2}$ is derived first. For ease of calculation, we rewrite (\ref{CDF_unorder user}) as
	\begin{equation}\label{CDF_GF_re}
		\setlength{\abovedisplayskip}{3pt}
		\setlength{\belowdisplayskip}{3pt}
		\begin{aligned}
			F_{|h|^2}(x)\approx-\sum_{l=0}^{L}\Psi_le^{-\mu_lx},
		\end{aligned}
	\end{equation}
	where $\Psi_0=-\sum_{l=1}^{L}\Psi_l$ and $\mu_0=0$. Based on (\ref{CDF_GF_re}), we can derive the following expression
	\begin{equation}\label{F_F_M}
		\setlength{\abovedisplayskip}{3pt}
		\setlength{\belowdisplayskip}{3pt}
		\begin{aligned}
			\left[F_{|h|^2}(x)\right]^M
			&\approx\left(-\sum_{l=0}^{L}\Psi_le^{-\mu_lx}\right)^M\\
			&\approx\left(-1\right)^M\sum_{\sum_{l=0}^{L}p_l=M}\binom{M}{p_0,\dots,p_L}\left(\prod_{l=0}^{L}\Psi_l^{p_l}\right)e^{-\sum_{l=0}^{L}p_l\mu_lx}.
		\end{aligned}
	\end{equation}
	Furthermore, the CDF of the second-largest order statistics is \cite{2003_order_statistics}
	\begin{equation}\label{CDF_2largest}
		\setlength{\abovedisplayskip}{3pt}
		\setlength{\belowdisplayskip}{3pt}
		\begin{aligned}
			F_{|h_{2}|^2}(x)=KF_{|h|^2}(x)^{K-1}-(K-1)F_{|h|^2}(x)^K.
		\end{aligned}
	\end{equation}
	Substituting (\ref{CDF_2largest}) and (\ref{F_F_M}) into (\ref{pout_cog4_bus}) and with some manipulations, we can obtain (\ref{pout_cog4_bus5}).

	Case 2, if $\hat{\gamma}_{\text{s}}<1$, we have
	\begin{equation}
		\setlength{\abovedisplayskip}{3pt}
		\setlength{\belowdisplayskip}{3pt}
		\begin{aligned}\label{BS_cog_3}
			\mathcal{P}^{\text{CPA}}_\text{GUS}
			=&\mathbb{P}\left[|h_{\text{s}}|^2<\hat{\tau}_{\text{s}}\right]+\mathbb{P}[|h_{\text{s}}|^2<|h_{\text{p}}|^2<\hat{\tau}_{\text{p}}+\hat{\tau}_{\text{s}}+\hat{\tau}_{\text{p}}\hat{\gamma}_{\text{s}}-|h_{\text{s}}|^2,\hat{\tau}_{\text{s}}<|h_{\text{s}}|^2<m_{i2}],\\
			&+\mathbb{P}\left[\frac{|h_{\text{s}}|^2}{\hat{\gamma}_{\text{s}}}-\frac{1}{\rho_m}<|h_{\text{p}}|^2<\hat{\tau}_{\text{p}}+\hat{\tau}_{\text{s}}+\hat{\tau}_{\text{p}}\hat{\gamma}_{\text{s}}-|h_{\text{s}}|^2,\frac{\hat{\tau}_{\text{s}}}{1-\hat{\gamma}_{\text{s}}}<|h_{\text{s}}|^2<\hat{\tau}_{\text{s}}(1+\hat{\gamma}_{\text{p}})\right],\\
		\end{aligned}
	\end{equation}
	where $m_{i2}=\text{min}\{m_{i1},\frac{\hat{\tau}_{\text{s}}}{1-\hat{\gamma}_{\text{s}}}\}$.
	
	With some manipulations, we have $\frac{\hat{\tau}_{\text{s}}}{1-\hat{\gamma}_{\text{s}}}<\hat{\tau}_{\text{s}}(1+\hat{\gamma}_{\text{p}})$, if $\hat{\gamma}_{\text{p}}>\frac{\hat{\gamma}_{\text{s}}}{1-\hat{\gamma}_{\text{s}}}$.
	Hence, for the case $\hat{\gamma}_{\text{s}}< 1$ and $\hat{\gamma}_{\text{p}}>\frac{\hat{\gamma}_{\text{s}}}{1-\hat{\gamma}_{\text{s}}}$, substituting (\ref{PDF_largest2}) into (\ref{BS_cog_3}), we have
	\begin{equation}\label{pout_cog4_bus_xiaoyu1}
		\setlength{\abovedisplayskip}{3pt}
		\setlength{\belowdisplayskip}{3pt}
		\begin{aligned}
			\mathcal{P}^{\text{CPA}}_\text{GUS}
			=&F_{|h_2|^2}(\hat{\tau}_\text{s})+K(K-1)\\
			&\times\int_{\frac{\hat{\tau}_{\text{s}}}{1-\hat{\gamma}_{\text{s}}}}^{\hat{\tau}_{\text{s}}(1+\hat{\gamma}_{\text{p}})}[F_{|h|^2}(\hat{\tau}_\text{s}+\hat{\tau}_\text{p}+\hat{\tau}_\text{p}\hat{\gamma}_\text{s}-x)-F_{|h|^2}(\frac{x}{\hat{\gamma}_{\text{s}}}-\frac{1}{\rho_m})]f_{|h|^2}(x)F_{|h|^2}(x)^{K-2}dx\\
			&+K(K-1)\int_{\hat{\tau}_{\text{s}}}^{m_{i2}}[F_{|h|^2}(\hat{\tau}_\text{s}+\hat{\tau}_\text{p}+\hat{\tau}_\text{p}\hat{\gamma}_\text{s}-x)-F_{|h|^2}(x)]f_{|h|^2}(x)F_{|h|^2}(x)^{K-2}dx.\\
		\end{aligned}
	\end{equation}
	Then, following the same lines of deriving (\ref{pout_cog4_bus5}), $\mathcal{P}^{\text{CPA}}_\text{GUS}$ can be calculated as (\ref{pout_cog4_bus51}). Similarly, for the case $\hat{\gamma}_{\text{s}}< 1$ and $\hat{\gamma}_{\text{p}}\leq\frac{\hat{\gamma}_{\text{s}}}{1-\hat{\gamma}_{\text{s}}}$,
	$\mathcal{P}^{\text{CPA}}_\text{GUS}$ can be calculated as (\ref{pout_cog4_bus52}). The proof of Theorem 1 is completed.

\section{Proof of Corollary 1}\label{proof_corollary1}
	It is quite involved to derive the high SNR approximation expressions with (\ref{pout_cog4_bus5}), (\ref{pout_cog4_bus51}), and (\ref{pout_cog4_bus52}). In the following, (\ref{pout_cog4_bus}) and (\ref{pout_cog4_bus_xiaoyu1}) are used to derive the approximation expressions. When $\rho_m\to \infty$, we have $m_{i1}\to 0$, $m_{i2}\to 0$, and $\hat{\tau}_{\text{s}}(1+\hat{\gamma}_{\text{p}})\to 0$. 
	
	If $x\to 0$, (\ref{CDF_unorder user}) and (\ref{PDF_unorder user}) can be respectively approximated as
	\begin{equation}\label{CDF_unorder user_app}
		\setlength{\abovedisplayskip}{3pt}
		\setlength{\belowdisplayskip}{3pt}
		\begin{aligned}
			F_{|h|^2}(x)\approx\sum_{l=1}^{L}\Psi_l\mu_lx,
		\end{aligned}
	\end{equation}
	and
	\begin{equation}\label{PDF_unorder user_app}
		\setlength{\abovedisplayskip}{3pt}
		\setlength{\belowdisplayskip}{3pt}
		\begin{aligned}
			f_{|h|^2}(x)\approx\sum_{l=1}^{L}\Psi_l\mu_l(1-\mu_lx)=\sum_{l=1}^{L}\Psi_l\mu_l.
		\end{aligned}
	\end{equation}
	
	Hence, substituting  (\ref{CDF_unorder user_app}) and (\ref{PDF_unorder user_app}) into (\ref{pout_cog4_bus_xiaoyu1}), we have
	\begin{equation}\label{pout_cog4_bus_dayu1}
		\setlength{\abovedisplayskip}{3pt}
		\setlength{\belowdisplayskip}{3pt}
		\begin{aligned}
			\vec{\mathcal{P}}^{\text{CPA}}_\text{GUS}
			\approx&K(S_L\hat{\tau}_\text{s})^{K-1}-(K-1)(S_L\hat{\tau}_\text{s})^K\\
			&+K(K-1)\int_{\frac{\hat{\tau}_{\text{s}}}{1-\hat{\gamma}_{\text{s}}}}^{\hat{\tau}_{\text{s}}(1+\hat{\gamma}_{\text{p}})}S_L[(\hat{\tau}_\text{s}+\hat{\tau}_\text{p}+\hat{\tau}_\text{p}\hat{\gamma}_\text{s}-x)-(\frac{x}{\hat{\gamma}_{\text{s}}}-\frac{1}{\rho_m})]S_L(S_Lx)^{K-2}dx\\
			&+K(K-1)\int_{\hat{\tau}_{\text{s}}}^{m_{i2}}S_L[(\hat{\tau}_\text{s}+\hat{\tau}_\text{p}+\hat{\tau}_\text{p}\hat{\gamma}_\text{s}-x)-x]S_L(S_Lx)^{K-2}dx\\
		\end{aligned}
	\end{equation}

After some algebra manipulations, we can obtain 
\begin{equation}\label{pout_cog4_bus_dayu11}
	\setlength{\abovedisplayskip}{3pt}
	\setlength{\belowdisplayskip}{3pt}
	\begin{aligned}
		\vec{\mathcal{P}}^{\text{CPA}}_\text{GUS}
		=&K(S_L\hat{\tau}_\text{s})^{K-1}-(K-1)(S_L\hat{\tau}_\text{s})^K+K(K-1)(S_L)^{K}\\
		&\times\left[(\hat{\tau}_\text{s}+\hat{\tau}_\text{p}+\hat{\tau}_\text{p}\hat{\gamma}_\text{s})\frac{m_{i2}^{K-1}-\hat{\tau}_{\text{s}}^{K-1}}{K-1}-\frac{1+\hat{\gamma}_{\text{s}}}{K\hat{\gamma}_{\text{s}}}((\hat{\tau}_{\text{s}}(1+\hat{\gamma}_{\text{p}}))^{K}-(\frac{\hat{\tau}_{\text{s}}}{1-\hat{\gamma}_{\text{s}}})^{K})\right.\\
		&+\left.\frac{\hat{\tau}_\text{s}+\hat{\tau}_\text{p}+\hat{\tau}_\text{p}\hat{\gamma}_\text{s}+\rho_m^{-1}}{K-1}[(\hat{\tau}_{\text{s}}(1+\hat{\gamma}_{\text{p}}))^{K-1}-(\frac{\hat{\tau}_{\text{s}}}{1-\hat{\gamma}_{\text{s}}})^{K-1}]-\frac{2}{K}(m_{i2}^{K}-\hat{\tau}_{\text{s}}^{K})\right]\\
	\end{aligned}
\end{equation}
for the case of $\hat{\gamma}_{\text{s}}< 1$ and $\hat{\gamma}_{\text{p}}>\frac{\hat{\gamma}_{\text{s}}}{1-\hat{\gamma}_{\text{s}}}$. By the same way, we can get 
\begin{equation}\label{pout_cog4_bus_dayu12}
	\setlength{\abovedisplayskip}{3pt}
	\setlength{\belowdisplayskip}{3pt}
	\begin{aligned}
		\vec{\mathcal{P}}^{\text{CPA}}_\text{GUS}
		=&K(S_L\hat{\tau}_\text{s})^{K-1}-(K-1)(S_L\hat{\tau}_\text{s})^K\\
		&+K(K-1)(S_L)^{K}\left[(\hat{\tau}_\text{s}+\hat{\tau}_\text{p}+\hat{\tau}_\text{p}\hat{\gamma}_\text{s})\frac{m_{i2}^{K-1}-\hat{\tau}_{\text{s}}^{K-1}}{K-1}-\frac{2}{K}(m_{i2}^{K}-\hat{\tau}_{\text{s}}^{K})\right]\\
	\end{aligned}
\end{equation}
for the case of  $\hat{\gamma}_{\text{s}}< 1$ and $\hat{\gamma}_{\text{p}}\leq\frac{\hat{\gamma}_{\text{s}}}{1-\hat{\gamma}_{\text{s}}}$; and get 
\begin{equation}\label{pout_cog4_approx}
	\setlength{\abovedisplayskip}{3pt}
	\setlength{\belowdisplayskip}{3pt}
	\begin{aligned}
		\vec{\mathcal{P}}^{\text{CPA}}_\text{GUS}
		=&K(S_L\hat{\tau}_\text{s})^{K-1}-(K-1)(S_L\hat{\tau}_\text{s})^K\\
		&+K(K-1)(S_L)^{K}\left[(\hat{\tau}_\text{s}+\hat{\tau}_\text{p}+\hat{\tau}_\text{p}\hat{\gamma}_\text{s})\frac{m_{i1}^{K-1}-\hat{\tau}_{\text{s}}^{K-1}}{K-1}-\frac{2}{K}(m_{i1}^{K}-\hat{\tau}_{\text{s}}^{K})\right]\\
	\end{aligned}
\end{equation}
for the case of $\hat{\gamma}_\text{s}\geq1$.

{\color{black}It can be observed that the first terms of (\ref{pout_cog4_bus_dayu11}), (\ref{pout_cog4_bus_dayu12}), and (\ref{pout_cog4_approx}) are proportional to $\rho_m^{-(K-1)}$, and other terms are proportional to $\rho_m^{-K}$. Hence, the high SNR approximations of all the three cases can be further approximated as $K(S_L\hat{\tau}_\text{s})^{K-1}$. By applying (\ref{define_DO}), we know that the secondary user U$_2$ can achieve a diversity order of $K-1$, and the proof is completed.}

\section{Proof of Theorem 2}\label{proof_theorem_cog_cdf}
By applying (\ref{pout_cog_cdf}), the outage probability of the second-largest CDF value user with the CPA strategy can be denoted as
\begin{equation}\label{pout_cog_cdf1}
	\setlength{\abovedisplayskip}{3pt}
	\setlength{\belowdisplayskip}{3pt}
	\begin{aligned}
		\mathcal{P}^{\text{CPA}}_\text{CUS}
		=&F_{Y}^{\text{CUS}}(\hat{\tau}_\text{s})+\int_{\hat{\tau}_{\text{s}}}^{\hat{\tau}_{\text{s}}(1+\hat{\gamma}_{\text{p}})}[F_{X,Y'}^{\text{CUS}}(\hat{\tau}_\text{s}+\hat{\tau}_\text{p}(1+\hat{\gamma}_\text{s})-y,y)-F_{X,Y'}^{\text{CUS}}(\frac{1}{\hat{\gamma}_\text{s}}y-\frac{1}{\rho_m},y)]dy\\
		=&F_{Y}^{\text{CUS}}(\hat{\tau}_\text{s})+\underbrace{\int_{\hat{\tau}_{\text{s}}}^{\hat{\tau}_{\text{s}}(1+\hat{\gamma}_{\text{p}})}F_{X,Y'}^{\text{CUS}}(\hat{\tau}_\text{s}+\hat{\tau}_\text{p}(1+\hat{\gamma}_\text{s})-y,y)dy}_{T_1}\\
		&-\underbrace{\int_{\hat{\tau}_{\text{s}}}^{\hat{\tau}_{\text{s}}(1+\hat{\gamma}_{\text{p}})}F_{X,Y'}^{\text{CUS}}(\frac{1}{\hat{\gamma}_\text{s}}y-\frac{1}{\rho_m},y)dy}_{T_2},\\
	\end{aligned}
\end{equation}
where  $F_{X,Y'}^{\text{CUS}}(x,y)=\frac{\partial F_{X,Y}^{\text{CUS}}(x,y)}{\partial y}$. $F_{Y}^{\text{CUS}}(\hat{\tau}_\text{s})$ can be calculated with (\ref{CDF_y}).

As shown in Remark 1, different values of $x$ and $y$ determine which CDF expression should be used. Assume $\hat{\tau}_\text{s}+\hat{\tau}_\text{p}(1+\hat{\gamma}_\text{s})-y=y(R^\alpha+1)$, we can obtain $y=\epsilon_1$. Let $\hat{\tau}_\text{s}+\hat{\tau}_\text{p}(1+\hat{\gamma}_\text{s})-y=y$, we have $y=\epsilon_2$. And assuming $(R^\alpha+1)(\hat{\tau}_\text{s}+\hat{\tau}_\text{p}(1+\hat{\gamma}_\text{s})-y)=y$, we can get $y=\epsilon_3$. Note that, $\epsilon_1$, $\epsilon_2$, and $\epsilon_3$ are the points deciding $\hat{\tau}_\text{s}+\hat{\tau}_\text{p}(1+\hat{\gamma}_\text{s})-y>y(R^\alpha+1)$, $y<\hat{\tau}_\text{s}+\hat{\tau}_\text{p}(1+\hat{\gamma}_\text{s})-y<y(R^\alpha+1)$, $(\hat{\tau}_\text{s}+\hat{\tau}_\text{p}(1+\hat{\gamma}_\text{s})-y)<y<(R^\alpha+1)(\hat{\tau}_\text{s}+\hat{\tau}_\text{p}(1+\hat{\gamma}_\text{s})-y)$, and $(R^\alpha+1)(\hat{\tau}_\text{s}+\hat{\tau}_\text{p}(1+\hat{\gamma}_\text{s})-y)<y$. We can also find that $ \epsilon_1< \epsilon_2< \epsilon_3$. Therefore, $T_1$ can be calculated with $F_{X,Y'}^{\text{CUS,I}}(x,y)$, if $y< \epsilon_1$, or with $F_{X,Y'}^{\text{CUS,II}}(x,y)$ if $ \epsilon_1<y< \epsilon_2$, or with $F_{X,Y'}^{\text{CUS,III}}(x,y)$ if $ \epsilon_2<y< \epsilon_3$, or 0 if $y>\epsilon_3$. We can see from the above analysis that, it is quite involved to obtain a closed-form expression for (\ref{pout_cog_cdf1}) with the joint CDFs in Lemma 1. To simplify the calculation process, we first calculate $T_1$ in (\ref{pout_cog_cdf1}). In the following, we will derive $T_1$ with three cases based on the values of $ \epsilon_2$, $\hat{\tau}_\text{s}$, and $\hat{\tau}_{\text{s}}(1+\hat{\gamma}_{\text{p}})$.

Firstly, we assume $ \epsilon_2>\hat{\tau}_{\text{s}}(1+\hat{\gamma}_{\text{p}})$. In this case, $T_1$ can be calculated with the following three sub-cases: 
\begin{enumerate}
	\item  if $ \epsilon_1>\hat{\tau}_{\text{s}}(1+\hat{\gamma}_{\text{p}})$: We can derive that $\hat{\tau}_\text{s}+\hat{\tau}_\text{p}(1+\hat{\gamma}_\text{s})-y>(R^\alpha+1)y$ for all $y\in(\hat{\tau}_{\text{s}},\hat{\tau}_{\text{s}}(1+\hat{\gamma}_{\text{p}}))$. Then, substituting $F_{X,Y'}^{\text{CUS,I}}(x,y)$ to $T_1$ and applying Gaussian-Chebyshev quadrature, we have  $T_1=\mathcal{G}[\hat{\tau}_{\text{s}},\hat{\tau}_{\text{s}}(1+\hat{\gamma}_{\text{p}});F_{X,Y'}^{\text{CUS,I}}(x,y)]$;
	\item if $\hat{\tau}_{\text{s}}< \epsilon_1<\hat{\tau}_{\text{s}}(1+\hat{\gamma}_{\text{p}})$: We have $\hat{\tau}_\text{s}+\hat{\tau}_\text{p}(1+\hat{\gamma}_\text{s})-y>(R^\alpha+1)y$ for $y\in(\hat{\tau}_{\text{s}}, \epsilon_1)$, and $y<\hat{\tau}_\text{s}+\hat{\tau}_\text{p}(1+\hat{\gamma}_\text{s})-y<(R^\alpha+1)y$ for $y\in( \epsilon_1,\hat{\tau}_{\text{s}}(1+\hat{\gamma}_{\text{p}}))$. Hence, substituting $F_{X,Y'}^{\text{CUS,I}}(x,y)$ ($F_{X,Y'}^{\text{CUS,II}}(x,y)$) to $T_1$ in integral region $(\hat{\tau}_{\text{s}}, \epsilon_1)$ ($\epsilon_1,\hat{\tau}_\text{s}(1+\hat{\gamma}_\text{p})$) and applying Gaussian-Chebyshev quadrature, we can get $T_1=\mathcal{G}[\hat{\tau}_\text{s}, \epsilon_1;F_{X,Y'}^{\text{CUS,I}}(x,y)]$ + $\mathcal{G} [\epsilon_1,\hat{\tau}_\text{s}(1+\hat{\gamma}_\text{p});F_{X,Y'}^{\text{CUS,II}}(x,y)]$;
	\item if $\epsilon_1<\hat{\tau}_{\text{s}}$: We know that $y<\hat{\tau}_\text{s}+\hat{\tau}_\text{p}(1+\hat{\gamma}_\text{s})-y<(R^\alpha+1)y$ for all $y\in(\hat{\tau}_{\text{s}},\hat{\tau}_{\text{s}}(1+\hat{\gamma}_{\text{p}}))$. Similarly, substituting $F_{X,Y'}^{\text{CUS,II}}(x,y)$ to $T_1$ and applying Gaussian-Chebyshev quadrature, we can obtain $T_1=\mathcal{G}[\hat{\tau}_\text{s},\hat{\tau}_\text{s}(1+\hat{\gamma}_\text{p});F_{X,Y'}^{\text{CUS,II}}(x,y)]$;
\end{enumerate}

Then, assuming $\hat{\tau}_{\text{s}}< \epsilon_2<\hat{\tau}_{\text{s}}(1+\hat{\gamma}_{\text{p}})$, $T_1$ can be calculated with the following four sub-cases:
\begin{enumerate}
	\item if $\epsilon_1<\hat{\tau}_{\text{s}}$ and $ \epsilon_3<\hat{\tau}_{\text{s}}(1+\hat{\gamma}_{\text{p}})$: We can derive that $y<\hat{\tau}_\text{s}+\hat{\tau}_\text{p}(1+\hat{\gamma}_\text{s})-y<(R^\alpha+1)y$ for $y\in(\hat{\tau}_{\text{s}}, \epsilon_2)$, $\hat{\tau}_\text{s}+\hat{\tau}_\text{p}(1+\hat{\gamma}_\text{s})-y<y<(R^\alpha+1)(\hat{\tau}_\text{s}+\hat{\tau}_\text{p}(1+\hat{\gamma}_\text{s})-y)$ for $y\in( \epsilon_2, \epsilon_3)$, and $y>(R^\alpha+1)(\hat{\tau}_\text{s}+\hat{\tau}_\text{p}(1+\hat{\gamma}_\text{s})-y)$ for $y\in( \epsilon_3,\hat{\tau}_{\text{s}}(1+\hat{\gamma}_{\text{p}}))$;
	\item if $\epsilon_1<\hat{\tau}_{\text{s}}$ and $ \epsilon_3>\hat{\tau}_{\text{s}}(1+\hat{\gamma}_{\text{p}})$: We have $y<\hat{\tau}_\text{s}+\hat{\tau}_\text{p}(1+\hat{\gamma}_\text{s})-y<(R^\alpha+1)y$ for $y\in(\hat{\tau}_{\text{s}}, \epsilon_2)$ and $\hat{\tau}_\text{s}+\hat{\tau}_\text{p}(1+\hat{\gamma}_\text{s})-y<y<(R^\alpha+1)(\hat{\tau}_\text{s}+\hat{\tau}_\text{p}(1+\hat{\gamma}_\text{s})-y)$ for $y\in( \epsilon_2,\hat{\tau}_{\text{s}}(1+\hat{\gamma}_{\text{p}}))$;
	\item if $\hat{\tau}_{\text{s}}< \epsilon_1< \epsilon_2$ and $ \epsilon_3<\hat{\tau}_{\text{s}}(1+\hat{\gamma}_{\text{p}})$: Thus $\hat{\tau}_\text{s}+\hat{\tau}_\text{p}(1+\hat{\gamma}_\text{s})-y>(R^\alpha+1)y$ for $y\in(\hat{\tau}_\text{s}, \epsilon_1)$, $y<\hat{\tau}_\text{s}+\hat{\tau}_\text{p}(1+\hat{\gamma}_\text{s})-y<(R^\alpha+1)y$ for $y\in( \epsilon_1, \epsilon_2)$, $\hat{\tau}_\text{s}+\hat{\tau}_\text{p}(1+\hat{\gamma}_\text{s})-y<y<(R^\alpha+1)(\hat{\tau}_\text{s}+\hat{\tau}_\text{p}(1+\hat{\gamma}_\text{s})-y)$ for $y\in( \epsilon_2, \epsilon_3)$, and $y>(R^\alpha+1)(\hat{\tau}_\text{s}+\hat{\tau}_\text{p}(1+\hat{\gamma}_\text{s})-y)$ for $y\in( \epsilon_3,\hat{\tau}_{\text{s}}(1+\hat{\gamma}_{\text{p}}))$;
	\item if $\hat{\tau}_{\text{s}}< \epsilon_1< \epsilon_2$ and $ \epsilon_3>\hat{\tau}_{\text{s}}(1+\hat{\gamma}_{\text{p}})$: Therefore, $\hat{\tau}_\text{s}+\hat{\tau}_\text{p}(1+\hat{\gamma}_\text{s})-y>(R^\alpha+1)y$ for $y\in(\hat{\tau}_\text{s}, \epsilon_1)$, $y<\hat{\tau}_\text{s}+\hat{\tau}_\text{p}(1+\hat{\gamma}_\text{s})-y<(R^\alpha+1)y$ for $y\in( \epsilon_1, \epsilon_2)$, $\hat{\tau}_\text{s}+\hat{\tau}_\text{p}(1+\hat{\gamma}_\text{s})-y<y<(R^\alpha+1)(\hat{\tau}_\text{s}+\hat{\tau}_\text{p}(1+\hat{\gamma}_\text{s})-y)$ for $y\in( \epsilon_2,\hat{\tau}_{\text{s}}(1+\hat{\gamma}_{\text{p}}))$.	
\end{enumerate}

At last, if $ \epsilon_2<\hat{\tau}_{\text{s}}$, $T_1$ can be calculated with the following three sub-cases:
\begin{enumerate}
	\item if $\epsilon_3<\hat{\tau}_{\text{s}}$: We have $y>(R^\alpha+1)(\hat{\tau}_\text{s}+\hat{\tau}_\text{p}(1+\hat{\gamma}_\text{s})-y)$ for $y\in(\hat{\tau}_{\text{s}},\hat{\tau}_{\text{s}}(1+\hat{\gamma}_{\text{p}}))$;
	\item if $\hat{\tau}_{\text{s}}< \epsilon_3<\hat{\tau}_{\text{s}}(1+\hat{\gamma}_{\text{p}})$: Then $\hat{\tau}_\text{s}+\hat{\tau}_\text{p}(1+\hat{\gamma}_\text{s})-y<y<(R^\alpha+1)(\hat{\tau}_\text{s}+\hat{\tau}_\text{p}(1+\hat{\gamma}_\text{s})-y)$ for $y\in(\hat{\tau}_{\text{s}}, \epsilon_3)$ and $y>(R^\alpha+1)(\hat{\tau}_\text{s}+\hat{\tau}_\text{p}(1+\hat{\gamma}_\text{s})-y)$ for $y\in( \epsilon_3,\hat{\tau}_{\text{s}}(1+\hat{\gamma}_{\text{p}}))$;
	\item if $ \epsilon_3>\hat{\tau}_{\text{s}}(1+\hat{\gamma}_{\text{p}})$: Hence, $\hat{\tau}_\text{s}+\hat{\tau}_\text{p}(1+\hat{\gamma}_\text{s})-y<y<(R^\alpha+1)\hat{\tau}_\text{s}+\hat{\tau}_\text{p}(1+\hat{\gamma}_\text{s})-y$ for $y\in(\hat{\tau}_{\text{s}},\hat{\tau}_{\text{s}}(1+\hat{\gamma}_{\text{p}}))$.
\end{enumerate}

Combining the above results and applying Gaussian-Chebyshev quadrature, we can obtain Table \ref{tableT1}.

Now we begin to calculate $T_2$. Let $\frac{1}{\hat{\gamma}_\text{s}}y-\frac{1}{\rho_m}=y(R^\alpha+1)$, we can get $y=\epsilon_4$. Assume $\frac{1}{\hat{\gamma}_\text{s}}y-\frac{1}{\rho_m}=y$, we have $y=\epsilon_5$. And let $(R^\alpha+1)(\frac{1}{\hat{\gamma}_\text{s}}y-\frac{1}{\rho_m})=y$, we can get $y=\epsilon_6$. Note that, $\epsilon_4$, $\epsilon_5$, and $\epsilon_6$ are the points deciding $\frac{1}{\hat{\gamma}_\text{s}}y-\frac{1}{\rho_m}>y(R^\alpha+1)$, $y<\frac{1}{\hat{\gamma}_\text{s}}y-\frac{1}{\rho_m}<y(R^\alpha+1)$, $(\frac{1}{\hat{\gamma}_\text{s}}y-\frac{1}{\rho_m})<y<(R^\alpha+1)(\frac{1}{\hat{\gamma}_\text{s}}y-\frac{1}{\rho_m})$, and $(R^\alpha+1)(\frac{1}{\hat{\gamma}_\text{s}}y-\frac{1}{\rho_m})<y$. Therefore, $T_2$ can be calculated with $F_{X,Y'}^{\text{CUS,I}}(x,y)$ if $y> \epsilon_4$, or with $F_{X,Y'}^{\text{CUS,II}}(x,y)$ if $ \epsilon_5<y< \epsilon_4$, or with $F_{X,Y'}^{\text{CUS,III}}(x,y)$ if $\epsilon_6<y< \epsilon_5$, or 0 if $y<\epsilon_6$. For ease of calculation, in the following, we will derive $T_2$ with two cases based on the values of $\hat{\gamma}_{\text{s}}$.

Firstly, we consider the case $\hat{\gamma}_{\text{s}}>1$, and in this case,  $\frac{1}{\hat{\gamma}_\text{s}}y-\frac{1}{\rho_m}<y$ always holds. Furthermore, we can derive that, if $R^\alpha+1<\hat{\gamma}_\text{s}$, $y>(R^\alpha+1)(\frac{1}{\hat{\gamma}_\text{s}}y-\frac{1}{\rho_m})$ always holds. In this case, $T_2=0$, since the partial derivative of (\ref{CDF_joint_yx0}) with respect to $y$ is $0$. On the other hand, if $R^\alpha+1>\hat{\gamma}_\text{s}$ and $y<\epsilon_6$, $y>(R^\alpha+1)(\frac{1}{\hat{\gamma}_\text{s}}y-\frac{1}{\rho_m})$ holds. Hence, we have $T_2=0$ if $R^\alpha+1>\hat{\gamma}_\text{s}$ and $\epsilon_6>\hat{\tau}_{\text{s}}(1+\hat{\gamma}_{\text{p}})$, and $T_2=\mathcal{G}[\epsilon_6,\hat{\tau}_{\text{s}}(1+\hat{\gamma}_{\text{p}});F_{X,Y'}^{\text{CUS,III}}(x,y)]$ if $R^\alpha+1>\hat{\gamma}_\text{s}$ and $\epsilon_6<\hat{\tau}_{\text{s}}(1+\hat{\gamma}_{\text{p}})$.

Now, we consider the case $\hat{\gamma}_\text{s}<1$. With some manipulations, we find that $\frac{1}{\hat{\gamma}_\text{s}}y-\frac{1}{\rho_m}>y$ holds when $y>\epsilon_5$, and vise versa. Since $\epsilon_5>\hat{\tau}_\text{s}$, we will calculate $T_2$ with two sub-cases according to the values of $\epsilon_5$ and $\hat{\tau}_{\text{s}}(1+\hat{\gamma}_{\text{p}})$.

The first sub-case is $\epsilon_5<\hat{\tau}_{\text{s}}(1+\hat{\gamma}_{\text{p}})$, namely, $\hat{\gamma}_\text{s}<\frac{\hat{\gamma}_\text{p}}{1+\hat{\gamma}_\text{p}}$. We first calculate the part of $T_2$ with integral region $y\in(\hat{\tau}_\text{s},\epsilon_5)$, represented as $T_2(\hat{\tau}_\text{s},\epsilon_5)$. Since $\hat{\tau}_\text{s}<\epsilon_6<\epsilon_5$, we have $T_2(\hat{\tau}_\text{s},\epsilon_5)=\mathcal{G}[\epsilon_6,\epsilon_5;F_{X,Y'}^{\text{CUS,III}}(x,y)]$. Now, we begin to calculate $T_2(\epsilon_5,\hat{\tau}_{\text{s}}(1+\hat{\gamma}_{\text{p}}))$. With some manipulations, we have the following observations.
\begin{enumerate}
	\item When $\hat{\gamma}_\text{s}>\frac{1}{R^\alpha+1}$, $\frac{1}{\hat{\gamma}_\text{s}}y-\frac{1}{\rho_m}<(R^\alpha+1)y$ always holds. Thus $y<\frac{1}{\hat{\gamma}_\text{s}}y-\frac{1}{\rho_m}<(R^\alpha+1)y$ for $y\in(\epsilon_5,\hat{\tau}_{\text{s}}(1+\hat{\gamma}_{\text{p}}))$, namely, $T_2(\epsilon_5,\hat{\tau}_{\text{s}}(1+\hat{\gamma}_{\text{p}}))=\mathcal{G}[\epsilon_5,\hat{\tau}_{\text{s}}(1+\hat{\gamma}_{\text{p}});F_{X,Y'}^{\text{CUS,II}}(x,y)]$.
	\item When $\frac{1}{(R^\alpha+1)(\hat{\gamma}_{\text{p}}+1)}<\hat{\gamma}_\text{s}<\frac{1}{R^\alpha+1}$, $y<\frac{1}{\hat{\gamma}_\text{s}}y-\frac{1}{\rho_m}<(R^\alpha+1)y$ holds for $y\in(\epsilon_5,\hat{\tau}_{\text{s}}(1+\hat{\gamma}_{\text{p}}))$ as well, namely, $T_2(\epsilon_5,\hat{\tau}_{\text{s}}(1+\hat{\gamma}_{\text{p}}))=\mathcal{G}[\epsilon_5,\hat{\tau}_{\text{s}}(1+\hat{\gamma}_{\text{p}});F_{X,Y'}^{\text{CUS,II}}(x,y)]$.
	\item When $\hat{\gamma}_\text{s}<\frac{1}{(R^\alpha+1)(\hat{\gamma}_{\text{p}}+1)}$,  $y<\frac{1}{\hat{\gamma}_\text{s}}y-\frac{1}{\rho_m}<(R^\alpha+1)y$ holds for $y\in(\epsilon_5,\epsilon_4)$ and $\frac{1}{\hat{\gamma}_\text{s}}y-\frac{1}{\rho_m}>(R^\alpha+1)y$ holds for $y\in(\epsilon_4,\hat{\tau}_{\text{s}}(1+\hat{\gamma}_{\text{p}}))$, namely, $T_2(\epsilon_5,\hat{\tau}_{\text{s}}(1+\hat{\gamma}_{\text{p}}))=\mathcal{G}[\epsilon_5,\epsilon_4;F_{X,Y'}^{\text{CUS,II}}(x,y)]+\mathcal{G}[\epsilon_4,\hat{\tau}_{\text{s}}(1+\hat{\gamma}_{\text{p}});F_{X,Y'}^{\text{CUS,I}}(x,y)]$.
\end{enumerate}
 
The second sub-case is $\epsilon_5>\hat{\tau}_{\text{s}}(1+\hat{\gamma}_{\text{p}})$, namely, $\hat{\gamma}_\text{s}>\frac{\hat{\gamma}_\text{p}}{1+\hat{\gamma}_\text{p}}$. In this case, if $\epsilon_6>\hat{\tau}_{\text{s}}(1+\hat{\gamma}_{\text{p}})$ (namely, $\hat{\gamma}_\text{s}>\frac{(R^\alpha+1)\hat{\gamma}_\text{p}}{1+\hat{\gamma}_\text{p}}$), we have $y>(\frac{1}{\hat{\gamma}_\text{s}}y-\frac{1}{\rho_m})(R^\alpha+1)$, thus $T_2=0$. On the other hand, if $\hat{\gamma}_\text{s}<\frac{(R^\alpha+1)\hat{\gamma}_\text{p}}{1+\hat{\gamma}_\text{p}}$, we have $y>(\frac{1}{\hat{\gamma}_\text{s}}y-\frac{1}{\rho_m})(R^\alpha+1)$ for $y\in(\hat{\tau}_{\text{s}},\epsilon_6)$
and $\frac{1}{\hat{\gamma}_\text{s}}y-\frac{1}{\rho_m}<y<(\frac{1}{\hat{\gamma}_\text{s}}y-\frac{1}{\rho_m})(R^\alpha+1)$ for $y\in(\epsilon_6,\hat{\tau}_{\text{s}}(1+\hat{\gamma}_{\text{p}}))$. Which means $T_2=\mathcal{G}[\epsilon_6,\hat{\tau}_{\text{s}}(1+\hat{\gamma}_{\text{p}});F_{X,Y'}^{\text{CUS,III}}(x,y)]$. Combine the above results, we can obtain Table \ref{tableT2}. The proof of Theorem 2 is completed.

\section{Proof of Corollary 2}\label{proof_corollary2}
	It can be observed from the proof of Theorem 2 that, it is quite involved to derive the achieved diversity order by first deriving the high SNR approximation of the outage probability expressions. As an alternative, in the following, the upper bound of the outage probability will be derived first and then be applied to derive the achieved diversity order. We can derive from (\ref{pout_cog_cdf}) that
	\begin{equation}\label{pout_cog_upbound}
		\setlength{\abovedisplayskip}{3pt}
		\setlength{\belowdisplayskip}{3pt}
		\begin{aligned}
			\mathcal{P}^{\text{CPA}}
			\leq&\mathbb{P}\left[|h_{\text{s}}|^2<\hat{\tau}_{\text{s}}\right]+\mathbb{P}\left[\hat{\tau}_{\text{s}}<|h_{\text{s}}|^2<\hat{\tau}_{\text{s}}(1+\hat{\gamma}_{\text{p}})\right]\\
			=&\mathbb{P}\left[|h_{\text{s}}|^2<\hat{\tau}_{\text{s}}(1+\hat{\gamma}_{\text{p}})\right].
		\end{aligned}
	\end{equation}
	Then, applying (\ref{CDF_y}), the upper bound of the secondary user in CDF-based scheduling can be expressed as
	\begin{equation}\label{CDF_cog}
		\setlength{\abovedisplayskip}{3pt}
		\setlength{\belowdisplayskip}{3pt}
		\begin{aligned}
			\mathcal{P}^{\text{CPA}}
			\leq&K\sum_{l_1=1}^{L}\Psi_{l_1}(1-e^{-\mu_{l_1}\hat{\tau}_{\text{s}}(1+\hat{\gamma}_{\text{p}})})^{K-1}-(K-1)\sum_{l_1=1}^{L}\Psi_{l_1}(1-e^{-\mu_{l_1}\hat{\tau}_{\text{s}}(1+\hat{\gamma}_{\text{p}})})^K.
		\end{aligned}
	\end{equation}
	When the transmit SNR $\rho_m\to \infty$, by using $1-e^{-x}\approx x$, we have
	\begin{equation}\label{CDF_cog2}
		\setlength{\abovedisplayskip}{3pt}
		\setlength{\belowdisplayskip}{3pt}
		\begin{aligned}
			\mathcal{P}^{\text{CPA}}
			\leq&K\sum_{l_1=1}^{L}\Psi_{l_1}(\mu_{l_1}\hat{\tau}_{\text{s}}(1+\hat{\gamma}_{\text{p}}))^{K-1}-(K-1)\sum_{l_1=1}^{L}\Psi_{l_1}(\mu_{l_1}\hat{\tau}_{\text{s}}(1+\hat{\gamma}_{\text{p}}))^K.
		\end{aligned}
	\end{equation}
	By applying the definition of diversity order, we know that the secondary user can achieve a diversity order of $K-1$, and the proof is completed.

\section{Proof of Theorem 3}\label{proof_theorem3}
	Substituting $|h_1|^2\geq|h_2|^2$ into (\ref{Pout_fair3}), the outage probability of U$_1$ can be expressed as
	\begin{equation}\label{Pout_BUS1}
		\setlength{\abovedisplayskip}{3pt}
		\setlength{\belowdisplayskip}{3pt}
		\begin{aligned}
			\mathcal{P}^{\text{FPA}}_{\text{GUS},1}
			=&\mathbb{P}\left[\hat{\tau}_1<|h_1|^2<\hat{\gamma}_1\hat{\tau}_1+\hat{\tau}_1,\frac{1}{\hat{\gamma}_1}|h_1|^2-\frac{1}{\rho_m}<|h_2|^2<\text{min}\{|h_1|^2,\hat{\tau}_1\hat{\gamma}_1+2\hat{\tau}_1-|h_1|^2\}\right]\\
			&+\mathbb{P}\left[|h_1|^2<\hat{\tau}_1\right].\\
		\end{aligned}
	\end{equation}
	
	To simplify (\ref{Pout_BUS1}), we should compare the values of $|h_1|^2$ and $\hat{\tau}_1\hat{\gamma}_1+2\hat{\tau}_1-|h_1|^2$ first. It can be derived that, $|h_1|^2>\hat{\tau}_1\hat{\gamma}_1+2\hat{\tau}_1-|h_1|^2$ holds if $|h_1|^2>\hat{\tau}_1+\frac{1}{2}\hat{\gamma}_1\hat{\tau}_1$, and vise versa.
	Note that, the condition for $\frac{1}{\hat{\gamma}_1}|h_1|^2-\frac{1}{\rho_m}<|h_1|^2$ in (\ref{Pout_BUS1}) is
	\begin{equation}\label{codition1}
		\setlength{\abovedisplayskip}{3pt}
		\setlength{\belowdisplayskip}{3pt}
		\begin{cases}
			|h_1|^2<\frac{\hat{\tau}_1}{1-\hat{\gamma}_1},& \text{ if}\ \hat{\gamma}_1< 1\\
			\text{No constraints},& \ \text{otherwise}
		\end{cases}.
	\end{equation}
	However, in the case  $\hat{\gamma}_1< 1$, we have $\hat{\gamma}_1\hat{\tau}_1+\hat{\tau}_1<\frac{\hat{\tau}_1}{1-\hat{\gamma}_1}$. Thus, based on ($\ref{codition1}$), we know that, no matter $\hat{\gamma}_1\geq 1$ or $\hat{\gamma}_1< 1$, $\frac{1}{\hat{\gamma}_1}|h_1|^2-\frac{1}{\rho_m}<|h_1|^2$ always holds for $\hat{\tau}_1<|h_1|^2<\hat{\gamma}_1\hat{\tau}_1+\hat{\tau}_1$.

	Hence, $\mathcal{P}^{\text{FPA}}_{\text{GUS},1}$ can be further denoted as
	\begin{equation}\label{Pout_BUS2}
		\setlength{\abovedisplayskip}{3pt}
		\setlength{\belowdisplayskip}{3pt}
		\begin{aligned}
			\mathcal{P}^{\text{FPA}}_{\text{GUS},1}
			=&\mathbb{P}\left[\hat{\tau}_1+\frac{1}{2}\hat{\gamma}_1\hat{\tau}_1<|h_1|^2<\hat{\tau}_1+\hat{\gamma}_1\hat{\tau}_1,\frac{1}{\hat{\gamma}_1}|h_1|^2-\frac{1}{\rho_m}<|h_2|^2<\hat{\tau}_1\hat{\gamma}_1+2\hat{\tau}_1-|h_1|^2\right]\\
			&+\mathbb{P}\left[\hat{\tau}_1<|h_1|^2<\hat{\tau}_1+\frac{1}{2}\hat{\gamma}_1\hat{\tau}_1,\frac{1}{\hat{\gamma}_1}|h_1|^2-\frac{1}{\rho_m}<|h_2|^2<|h_1|^2\right]\\
			&+\mathbb{P}\left[|h_1|^2<\hat{\tau}_1\right].\\
		\end{aligned}
	\end{equation}
	By applying (\ref{PDF_largest2}), the outage probability of $\text{U}_1$ can be expressed as
	\begin{equation}\label{Pout_fair4_greedy}
		\setlength{\abovedisplayskip}{3pt}
		\setlength{\belowdisplayskip}{3pt}
		\begin{aligned}
			\mathcal{P}^{\text{FPA}}_{\text{GUS},1}
			=&K\int_{\hat{\tau}_1+\frac{1}{2}\hat{\tau}_1\hat{\gamma}_1}^{\hat{\tau}_1+\hat{\tau}_1\hat{\gamma}_1}[F_{|h|^2}(\hat{\gamma}_1\hat{\tau}_1+2\hat{\tau}_1-y)^{K-1}-F_{|h|^2}(\frac{1}{\hat{\gamma}_1}y-\frac{1}{\rho_m})^{K-1}]f_{|h|^2}(y)dy\\
			&+K\int_{\hat{\tau}_1}^{\hat{\tau}_1+\frac{1}{2}\hat{\tau}_1\hat{\gamma}_1}[F_{|h|^2}(y)^{K-1}-F_{|h|^2}(\frac{1}{\hat{\gamma}_1}y-\frac{1}{\rho_m})^{K-1}]f_{|h|^2}(y)dy+F_{|h|^2}(\hat{\tau}_1)^K,\\
		\end{aligned}
	\end{equation}
	where the CDF of the largest order statistics $F_{|h_{1}|^2}(x)=F_{|h|^2}(x)^{K}$ \cite{2003_order_statistics} is applied. Substituting (\ref{F_F_M}) and (\ref{CDF_2largest}) into (\ref{Pout_fair4_greedy}), and after some manipulations, we can obtain  (\ref{thm_fair_GS1}).
	
	%

	The outage probability for U$_2$ can be derived by substituting $|h_1|^2\geq|h_2|^2$ into (\ref{Pout_fair23}), namely,
	\begin{equation}\label{Pout_fair24}
		\setlength{\abovedisplayskip}{3pt}
		\setlength{\belowdisplayskip}{3pt}
		\begin{aligned}
			\mathcal{P}^{\text{FPA}}_{\text{GUS},2}
			=&\mathbb{P}\left[\hat{\tau}_2<|h_2|^2<\hat{\gamma}_2\hat{\tau}_2+\hat{\tau}_2,\text{max}\{|h_2|^2,\frac{1}{\hat{\gamma}_2}|h_2|^2-\frac{1}{\rho_m}\}<|h_1|^2<\hat{\gamma}_2\hat{\tau}_2+2\hat{\tau}_2-|h_2|^2\right]\\
			&+\mathbb{P}\left[|h_2|^2<\hat{\tau}_2\right].\\
		\end{aligned}
	\end{equation}
	Note that, when $\hat{\tau}_2<|h_2|^2<\hat{\gamma}_2\hat{\tau}_2+\hat{\tau}_2$, $|h_2|^2$ is always lager than $\frac{1}{\hat{\gamma}_2}|h_2|^2-\frac{1}{\rho_m}$. Thus (\ref{Pout_fair24}) can be converted to
	\begin{equation}\label{Pout_fair25}
		\setlength{\abovedisplayskip}{3pt}
		\setlength{\belowdisplayskip}{3pt}
		\begin{aligned}
			\mathcal{P}^{\text{FPA}}_{\text{GUS},2}
			=&\mathbb{P}\left[\hat{\tau}_2<|h_2|^2<\hat{\gamma}_2\hat{\tau}_2+\hat{\tau}_2,|h_2|^2<|h_1|^2<\hat{\gamma}_2\hat{\tau}_2+2\hat{\tau}_2-|h_2|^2\right]+\mathbb{P}\left[|h_2|^2<\hat{\tau}_2\right].\\
		\end{aligned}
	\end{equation}
	
	The constraint of $|h_2|^2<\hat{\gamma}_2\hat{\tau}_2+2\hat{\tau}_2-|h_2|^2$ in (\ref{Pout_fair25}) requires $|h_2|^2<\hat{\tau}_2+\frac{1}{2}\hat{\gamma}_2\hat{\tau}_2$. Thus $\mathcal{P}^{\text{FPA}}_{\text{GUS},2}$ can be finally denoted as
	\begin{equation}\label{Pout_fair26}
		\setlength{\abovedisplayskip}{3pt}
		\setlength{\belowdisplayskip}{3pt}
		\begin{aligned}
			\mathcal{P}^{\text{FPA}}_{\text{GUS},2}
			=&\mathbb{P}\left[\hat{\tau}_2<|h_2|^2<\hat{\tau}_2+\frac{1}{2}\hat{\gamma}_2\hat{\tau}_2,|h_2|^2<|h_1|^2<\hat{\gamma}_2\hat{\tau}_2+2\hat{\tau}_2-|h_2|^2\right]+\mathbb{P}\left[|h_2|^2<\hat{\tau}_2\right].\\
		\end{aligned}
	\end{equation}
	By applying (\ref{PDF_largest2}) and (\ref{CDF_2largest}), the outage probability of U$_2$ can be expressed as
	\begin{equation}\label{Pout_fair7_greedy}
		\setlength{\abovedisplayskip}{3pt}
		\setlength{\belowdisplayskip}{3pt}
		\begin{aligned}
			\mathcal{P}^{\text{FPA}}_{\text{GUS},2}
			=&K(K-1)\int_{\hat{\tau}_2}^{\hat{\tau}_2+\frac{1}{2}\hat{\tau}_2\hat{\gamma}_2}[F_{|h|^2}(\hat{\gamma}_2\hat{\tau}_2+2\hat{\tau}_2-x)-F_{|h|^2}(x)]f_{|h|^2}(x)F_{|h|^2}(x)^{K-2}dx\\
			&+KF_{|h|^2}(\hat{\tau}_2)^{K-1}-(K-1)F_{|h|^2}(\hat{\tau}_2)^K.\\
		\end{aligned}
	\end{equation}
	Following the same steps of deriving (\ref{pout_cog4_bus5}),  $\mathcal{P}^{\text{FPA}}_{\text{GUS},2}$ can be finally expressed as (\ref{thm_fair_GS2}). The proof is complete.

\section{Proof of Theorem 4}\label{proof_theorem_fair_CDF}
By applying (\ref{Pout_fair3}), the outage probability of the user with the largest CDF value in fairness-oriented power allocation scheme can be denoted as
\begin{equation}\label{pout_fair_cdf1}
	\setlength{\abovedisplayskip}{3pt}
	\setlength{\belowdisplayskip}{3pt}
	\begin{aligned}
		\mathcal{P}^{\text{FPA}}_\text{CUS}
		=&F_{X}^{\text{CUS}}(\hat{\tau}_i)+\int_{\hat{\tau}_i}^{\hat{\tau}_i(1+\hat{\gamma}_i)}[F_{X',Y}^{\text{CUS}}(x,2\hat{\tau}_i+\hat{\tau}_i\hat{\gamma}_i-x)-F_{X',Y}^{\text{CUS}}(x,\frac{1}{\hat{\gamma}_i}x-\frac{1}{\rho_m})]dx\\
		=&F_{X}^{\text{CUS}}(\hat{\tau}_i)+\underbrace{\int_{\hat{\tau}_i}^{\hat{\tau}_i(1+\hat{\gamma}_i)}F_{X',Y}^{\text{CUS}}(x,2\hat{\tau}_i+\hat{\tau}_i\hat{\gamma}_i-x)dx}_{T_3}\\
		&-\underbrace{\int_{\hat{\tau}_i}^{\hat{\tau}_i(1+\hat{\gamma}_i)}F_{X',Y}^{\text{CUS}}(x,\frac{1}{\hat{\gamma}_i}x-\frac{1}{\rho_m})dx}_{T_4},\\
	\end{aligned}
\end{equation}
where  $F_{X',Y}^{\text{CUS}}(x,y)=\frac{\partial F_{X,Y}^{\text{CUS}}(x,y)}{\partial x}$. Note that, $F_{X}^{\text{CUS}}(\hat{\tau}_i)$ in (\ref{pout_fair_cdf1}) can be calculated with (\ref{CDF_joint_yx0}).

In the following, we will derive $T_3$ first, where $y=2\hat{\tau}_i+\hat{\tau}_i\hat{\gamma}_i-x$. When $x\left(R^{\alpha}+1\right)=2\hat{\tau}_i+\hat{\tau}_i\hat{\gamma}_i-x$, we have $x=\epsilon_7$; let $x=2\hat{\tau}_i+\hat{\tau}_i\hat{\gamma}_i-x$, we have $x=\epsilon_8$; and when $x=\left(R^{\alpha}+1\right)(2\hat{\tau}_i+\hat{\tau}_i\hat{\gamma}_i-x)$, we can get $x=\epsilon_9$. Note that, $\epsilon_7<\epsilon_8<\epsilon_9$. Since $\hat{\tau}_i<\epsilon_8<\hat{\tau}_i+\hat{\tau}_i\hat{\gamma}_i$, the calculation of $T_3$ can be divided into two parts, namely, $T_3\left(\hat{\tau}_i,\epsilon_8\right)$ and $T_3\left(\epsilon_8,\hat{\tau}_i+\hat{\tau}_i\hat{\gamma}_i\right)$.

Let us derive $T_3\left(\hat{\tau}_i,\epsilon_8\right)$ first, which can be calculated with the following two cases:
\begin{enumerate}
	\item if $\epsilon_7<\hat{\tau}_i$ (namely, $R^{\alpha}>\hat{\gamma}_i$): We have $x<2\hat{\tau}_i+\hat{\tau}_i\hat{\gamma}_i-x<x\left(R^{\alpha}+1\right)$ for $x\in\left(\hat{\tau}_i,\epsilon_8\right)$.
	\item if $\epsilon_7>\hat{\tau}_i$ (namely, $R^{\alpha}<\hat{\gamma}_i$): We have $2\hat{\tau}_i+\hat{\tau}_i\hat{\gamma}_i-x>x\left(R^{\alpha}+1\right)$ for $x\in\left(\hat{\tau}_i,\epsilon_7\right)$, and  $x<2\hat{\tau}_i+\hat{\tau}_i\hat{\gamma}_i-x<x\left(R^{\alpha}+1\right)$ holds for $x\in\left(\epsilon_7,\epsilon_8\right)$.
\end{enumerate}

Similarly, the calculation of $T_3\left(\epsilon_8,\hat{\tau}_i+\hat{\tau}_i\hat{\gamma}_i\right)$ can also be divided into two cases:
\begin{enumerate}
	\item if $\epsilon_9>\hat{\tau}_i\left(1+\hat{\gamma}_i\right)$ (namely, $R^{\alpha}>\hat{\gamma}_i$): We have $\left(2\hat{\tau}_i+\hat{\tau}_i\hat{\gamma}_i-x\right)\left(R^{\alpha}+1\right)>x>2\hat{\tau}_i+\hat{\tau}_i\hat{\gamma}_i-x$ for $x\in\left(\epsilon_8,\hat{\tau}_i+\hat{\tau}_i\hat{\gamma}_i\right)$.
	\item if $\epsilon_9<\hat{\tau}_i\left(1+\hat{\gamma}_i\right)$ (namely, $R^{\alpha}<\hat{\gamma}_i$): We have $2\hat{\tau}_i+\hat{\tau}_i\hat{\gamma}_i-x<x<\left(2\hat{\tau}_i+\hat{\tau}_i\hat{\gamma}_i-x\right)\left(R^{\alpha}+1\right)$ for $x\in\left(\epsilon_8,\epsilon_9\right)$, and  $x>\left(2\hat{\tau}_i+\hat{\tau}_i\hat{\gamma}_i-x\right)\left(R^{\alpha}+1\right)$ holds for $x\in\left(\epsilon_9,\hat{\tau}_i+\hat{\tau}_i\hat{\gamma}_i\right)$.
\end{enumerate}

In the following, we will derive $T_4$, where $y=\frac{1}{\hat{\gamma}_i}x-\frac{1}{\rho_m}$. After some manipulations, we can find that $x=\frac{1}{\hat{\gamma}_i}x-\frac{1}{\rho_m}$ holds, if $x=\epsilon_{10}$. Since $\epsilon_{10}>\hat{\tau}_i+\hat{\tau}_i\hat{\gamma}_i$, we know that $x>\frac{1}{\hat{\gamma}_i}x-\frac{1}{\rho_m}$ always holds for all $x\in\left(\hat{\tau}_i,\hat{\tau}_i+\hat{\tau}_i\hat{\gamma}_i\right)$. Moreover, we know that, $x=\left(R^{\alpha}+1\right)\left(\frac{1}{\hat{\gamma}_i}x-\frac{1}{\rho_m}\right)$ holds when $x=\epsilon_{11}$, and $\epsilon_{11}>\hat{\tau}_i$. Thus, $T_4$ can be calculated in the following two cases:
\begin{enumerate}
	\item if $\epsilon_{11}<\hat{\tau}_i\left(1+\hat{\gamma}_i\right)$ (namely, $R^{\alpha}>\hat{\gamma}_i$): We have $x>\left(\frac{1}{\hat{\gamma}_i}x-\frac{1}{\rho_m}\right)\left(R^{\alpha}+1\right)$ for $x\in\left(\hat{\tau}_i,\epsilon_{11}\right)$, and  $\frac{1}{\hat{\gamma}_i}x-\frac{1}{\rho_m}<x<\left(\frac{1}{\hat{\gamma}_i}x-\frac{1}{\rho_m}\right)\left(R^{\alpha}+1\right)$ holds for $x\in\left(\epsilon_{11},\hat{\tau}_i+\hat{\tau}_i\hat{\gamma}_i\right)$.
	\item if $\epsilon_{11}>\hat{\tau}_i\left(1+\hat{\gamma}_i\right)$ (namely, $R^{\alpha}<\hat{\gamma}_i$): We have $x>\left(\frac{1}{\hat{\gamma}_i}x-\frac{1}{\rho_m}\right)\left(R^{\alpha}+1\right)$ for all $x\in\left(\hat{\tau}_i,\hat{\tau}_i+\hat{\tau}_i\hat{\gamma}_i\right)$.
\end{enumerate}
Finally, combining all the above results and applying Gaussian-Chebyshev quadrature, we can get (\ref{Pout_fair_CDF}) and (\ref{Pout_fair_CDF2}). It can be observed that (\ref{Pout_fair23}) is a special case of (\ref{pout_cog_cdf}), hence the outage probability of the user with the second largest CDF value in FPA scheme can be calculated with the results of Theorem 2. The proof of Theorem 4 is completed.



\begin{thebibliography}{99}
\setlength{\baselineskip}{1.3em} 



%

%



\bibitem{Clerckx_2016_RS_COMM}
B. Clerckx, H. Joudeh, C. Hao, M. Dai, and B. Rassouli, ``Rate splitting for MIMO wireless networks: A promising PHY-layer strategy for LTE evolution,'' \textit{IEEE Commun. Mag.}, vol. 54, no. 5, pp. 98\textendash105, May 2016.


\bibitem{2021_Clerckx_OJCS_Critical}
B. Clerckx \textit{et al.}, ``Is NOMA efficient in multi-antenna networks? A critical look at next generation multiple
access techniques,'' \textit{IEEE Open J. Commun. Soc.}, vol. 2, pp. 1310\textendash1343, Jun. 2021.


\bibitem{2017_JSAC_Ding_survey_NOMA}
Z. Ding \textit{et al.}, ``A survey on non-orthogonal multiple access for 5G networks: Research challenges and future trends,'' \textit{IEEE J. Sel. Areas Commun.}, vol. 35, no. 10, pp. 2181\textendash2195, Oct. 2017.

\bibitem{2019_TMC_Cai}
H. Cao, J. Cai, S. Huang, and Y. Lu, ``Online adaptive transmission strategy for buffer-aided cooperative NOMA systems,'' \textit{IEEE Trans.
Mobile Comput.}, vol. 18, no. 5, pp. 1133\textendash1144, May 2019.

\bibitem{2019_TVT_Zaidi_UAV_Ground}
{\color{black}S. K. Zaidi, S. F. Hasan, and X. Gui, ``Outage analysis of ground-aerial NOMA with distinct instantaneous channel gain ranking," \textit{IEEE Trans. Veh. Technol.}, vol. 68, no. 11, pp. 10775\textendash10790, Nov. 2019.}

\bibitem{2019_Mao_RS_multiantenna}
Y. Mao, B. Clerckx, and V. O. K. Li, ``Rate-splitting for multiantenna non-orthogonal unicast and multicast transmission: Spectral
and energy efficiency analysis,'' \textit{IEEE Trans. Commun.}, vol. 67, no. 12, pp. 8754\textendash8770, Dec. 2019.



\bibitem{1996_Remoldi_TIT}
B. Rimoldi and R. Urbanke, ``A rate-splitting approach to the Gaussian multiple-access channel,'' \textit{IEEE Trans. Inf. Theory}, vol. 42, no. 2, pp. 364\textendash375, Mar. 1996.


{\color{black}\bibitem{2022_Mao_RSMA_SURVEY}
Y. Mao, O. Dizdar, B. Clerckx, R. Schober, P. Popovski, and H. V. Poor, ``Rate-splitting multiple access: Fundamentals, survey, and future research trends,'' \textit{IEEE Commun. Surveys Tuts.}, vol. 24, no. 4, pp. 2073\textendash2126, 4th Quart., 2022.}


\bibitem{2005_Tse}
D. Tse and P. Viswanath, \textit{Fundamentals of Wireless Communication}. Cambridge, U.K.: Cambridge Univ. Press, 2005.


\bibitem{2017_Zhu_CD_RS}
Y. Zhu, Z. Zhang, X. Wang, and X. Liang, ``A low-complexity nonorthogonal multiple access system based on rate splitting,'' in \textit{Proc. Int. Conf. Wireless Commun. Signal Process. (WCSP)}, Oct. 2017, pp. 1\textendash6.



 
\bibitem{2016_Joudeh_TCOM_Partial_CSIT}
H. Joudeh and B. Clerckx, ``Sum-rate maximization for linearly precoded
downlink multiuser MISO systems with partial CSIT: A rate-splitting approach,'' \textit{IEEE Trans. Commun.}, vol. 64, no. 11, pp. 4847\textendash4861,
Nov. 2016.


\bibitem{2020_Clarckx_WCL_unifying}
B. Clerckx \textit{et al.}, ``Rate-splitting unifying SDMA, OMA, NOMA, and multicasting in MISO broadcast channel: A simple two-user rate analysis,'' \textit{IEEE Wireless Commun. Lett.}, vol. 9, no. 3, pp. 349\textendash353, Mar. 2020.


\bibitem{2021_Yang_TCOM_Optimization}
Z. Yang, M. Chen, W. Saad, and M. Shikh-Bahaei, ``Optimization of rate allocation and power control for rate splitting multiple access (RSMA),'' \textit{IEEE Trans. Commun.}, vol. 69, no. 9, pp. 5988\textendash6002, Sep. 2021.



\bibitem{2019_Zeng_up_RS}
J. Zeng, T. Lv, W. Ni, R. P. Liu, N. C. Beaulieu, and Y. J. Guo, ``Ensuring max-min fairness of UL SIMO-NOMA: A rate splitting approach,'' \textit{IEEE Trans. Veh. Technol.}, vol. 68, no. 11, pp. 11080\textendash11093, Nov. 2019.

\bibitem{2020_Yang_TMC_Sum-rate_optimization_UL}
Z. Yang, M. Chen, W. Saad, W. Xu, and M. Shikh-Bahaei, ``Sum-rate maximization of uplink rate splitting multiple access (RSMA) communication,'' \textit{IEEE Trans. Mobile Comput.}, vol. 21, no. 7, pp. 2596\textendash2609, Jul. 2022.


\bibitem{2017_Zhu_RS}
Y. Zhu, X. Wang, Z. Zhang, X. Chen, and Y. Chen, ``A rate-splitting nonorthogonal multiple access scheme for uplink transmission,'' in \textit{Proc. Int. Conf. Wireless Commun. Signal Process. (WCSP)}, Oct. 2017, pp. 1\textendash6.

\bibitem{2020_Liu_RS}
H. Liu, T. A. Tsiftsis, K. J. Kim, K. S. Kwak, and H. V. Poor, ``Rate splitting for uplink NOMA with enhanced fairness and outage performance,'' \textit{IEEE Trans. Wireless Commun.}, vol. 19, no. 7, pp. 4657\textendash4670, Jul. 2020.

\bibitem{2021_Singh_WCL_UAV-RSMA}
S. K. Singh, K. Agrawal, K. Singh, and C. -P. Li, ``Outage probability and throughput analysis of UAV-assisted rate-splitting multiple access," \textit{IEEE Wireless Commun. Lett.}, vol. 10, no. 11, pp. 2528\textendash2532, Nov. 2021.

\bibitem{2022_Kong_WCL_throughput_analysis}
H. Kong, M. Lin, Z. Wang, J. -Y. Wang, W. -P. Zhu, and J. Wang, ``Performance analysis for rate splitting uplink NOMA transmission in high throughput satellite systems,'' \textit{IEEE Wireless Commun. Lett.}, vol. 11, no. 4, pp. 816\textendash820, Apr. 2022.

{\color{black}\bibitem{2022_Abbasi_arxiv_C-RSMA}
O. Abbasi and H. Yanikomeroglu, ``Transmission scheme, detection and power allocation for uplink user cooperation with NOMA and RSMA,'' \textit{IEEE Trans. Wireless Commun.}, vol. 22, no. 1, pp. 471\textendash485, Jan. 2023.}

\bibitem{2022_Tegos_CL_Performance_of_UL_RSMA}
S. A. Tegos, P. D. Diamantoulakis, and G. K. Karagiannidis, ``On the performance of uplink rate-splitting multiple access,'' \textit{IEEE Commun. Lett.}, vol. 26, no. 3, pp. 523\textendash527, Mar. 2022.

\bibitem{2022_Liu_TVT_RS-CR-NOMA}
H. Liu, Z. Bai, H. Lei, G. Pan, K. J. Kim, and T. Tsiftsis, ``A new rate splitting strategy for uplink CR-NOMA systems,'' \textit{IEEE Trans. Veh. Technol.}, vol. 71, no. 7, pp. 7947\textendash7951, Jul. 2022.


\bibitem{2020_Demarchou_ICC_Spatial_randomness}
E. Demarchou, C. Psomas, and I. Krikidis, ``Channel statistics-based rate splitting with spatial randomness,'' in \textit{Proc. Int. Conf. Commun. Workshops (ICC Workshops)}, Jun. 2020, pp. 1\textendash6.

\bibitem{2021_Demarchou_ISWCS_MISO}
E. Demarchou, C. Psomas, and I. Krikidis, ``On the sum rate of MISO rate splitting with spatial randomness,'' in \textit{Proc. Int. Symp. Wireless Commun. Syst. (ISWCS)}, Sep. 2021, pp. 1\textendash5.

\bibitem{2022_Demarchou_TCOM_RS_Caching}
E. Demarchou, C. Psomas, and I. Krikidis, ``Rate splitting with wireless edge caching: A system-level-based co-design,'' \textit{IEEE Trans. Commun.}, vol. 70, no. 1, pp. 664\textendash679, Jan. 2022.

\bibitem{2023_Chen_JSAC_RSMA-MEC}
{\color{black}P. Chen, H. Liu, Y. Ye, L. Yang, K. J. Kim, and T. A. Tsiftsis, ``Rate-splitting multiple access aided mobile edge computing with randomly deployed users," \textit{IEEE J. Sel. Areas Commun.}, 2023. DOI: 10.1109/JSAC.2023.3240786}


\bibitem{2020_Lu_TVT}
H. Lu, X. Xie, Z. Shi, and J. Cai, ``Outage performance of CDF-based scheduling in downlink and uplink NOMA systems,'' \textit{IEEE Trans. Veh. Technol.}, vol. 69, no. 12, pp. 14945\textendash14959, Dec. 2020.

\bibitem{2016_Liu_cooperative_SWIPT_NOMA}
Y. Liu, Z. Ding, M. Elkashlan, and H. V. Poor, ``Cooperative non-orthogonal multiple access with simultaneous wireless information and power transfer,'' \textit{IEEE J. Sel. Areas Commun.}, vol. 34, no. 4, pp. 938\textendash953, Apr. 2016.

\bibitem{2018Yue_TCOM_unified}
X. Yue, Z. Qin, Y. Liu, S. Kang, and Y. Chen, ``A unified framework for non-orthogonal multiple access,'' \textit{IEEE Trans. Commun.}, vol. 66, no. 11, pp. 5346\textendash5359, Nov. 2018.
		
\bibitem{2017Afshang_TWC_BPP}
M. Afshang and H. S. Dhillon, ``Fundamentals of modeling finite wireless networks using binomial point process,'' \textit{IEEE Trans. Wireless Commun.}, vol. 16, no. 5, pp. 3355\textendash3370, May 2017.

\bibitem{2012_Haenggin_stochastic_geometry}
M. Haenggi, \textit{Stochastic Geometry for Wireless Networks}. Cambridge, U.K.: Cambridge Univ. Press, 2012.

\bibitem{2003_order_statistics}
H. A. David and H. N. Nagaraja, \textit{Order Statistics}. John Wiley, New York, 3rd ed., 2003.

\bibitem{2014_Ding_random_deployed}
Z. Ding, Z. Yang, P. Fan, and H. V. Poor, ``On the performance of non-orthogonal multiple access in 5G systems with randomly deployed users,'' \textit{IEEE Signal Process. Lett.}, vol. 21, no. 12, pp. 1501\textendash1505, Dec. 2014.

\bibitem{Gaussian_Chebyshev}
E. Hildebrand, \textit{Introduction to Numerical Analysis}. New York, NY, USA:	Dover, 1987.

\bibitem{2020_Ali_Partial}
K. S. Ali, E. Hossain, and M. J. Hossain, ``Partial non-orthogonal multiple access (NOMA) in downlink poisson networks,'' \textit{IEEE Trans. Wireless Commun.}, vol. 19, no. 11, pp. 7637\textendash7652, Nov. 2020.



\bibitem{2015_Tabassum_TWC_user_scheduling}
{\color{black}H. Tabassum, E. Hossain, M. J. Hossain, and D. I. Kim, ``On the spectral efficiency of multiuser scheduling in RF-powered uplink cellular networks," \textit{IEEE Trans. Wireless Commun.}, vol. 14, no. 7, pp. 3586\textendash3600, Jul. 2015.}

\bibitem{2013_Sediq_TWC}
A. B. Sediq, R. H. Gohary, R. Schoenen, and H. Yanikomeroglu, “Optimal tradeoff between sum-rate efficiency and jain’s fairness index in resource allocation,” \textit{IEEE Trans. Wireless Commun.}, vol. 12, no. 7, pp. 3496\textendash3509, Jul. 2013.

\bibitem{2014_Shi_CST}
H. Shi, R. V. Prasad, E. Onur, and I. G. M. M. Niemegeers, ``Fairness in wireless networks: Issues, measures and challenges,'' \textit{IEEE Commun. Surveys Tuts.}, vol. 16, no. 1, pp. 5\textendash24, 1st Quart., 2014.

\bibitem{2005_Park_packet_CDF}
D. Park, H. Seo, H. Kwon, and B. G. Lee, ``Wireless packet admitting based on the cumulative distribution function of user transmission rate,'' \textit{IEEE Trans. Commun.}, vol. 53, no. 11, pp. 1919\textendash1929, Nov. 2005.
		
\bibitem{2015-Jin-Fundamental_limits_CDF}
H. Jin \textit{et al.}, ``Fundamental limits of CDF-based scheduling: Throughput, fairness, and feedback overhead,'' \textit{IEEE/ACM Trans. Netw.}, vol. 23, no. 3, pp. 894\textendash907, Jun. 2015.

\bibitem{2006-Bletsas-a-simple}
A. Bletsas, A. Khisti, D. P. Reed, and A. Lippman, ``A simple cooperative diversity method based on network path selection,'' \textit{IEEE J. Sel. Areas Commun.}, vol. 24, no. 3, pp. 659\textendash672, Mar. 2006.

\bibitem{2005-Zhao-Opprotunistic}
Q. Zhao and L. Tong, ``Opportunistic carrier sensing for energy-efficient information retrieval in sensor networks," \textit{EURASIP J. Wireless Commun. Netw.}, vol. 2, pp. 231\textendash241, Apr. 2005.

\bibitem{2017_Choi_ALOHA}
J. Choi, ``NOMA-based random access with multichannel ALOHA,'' \textit{IEEE J. Sel. Areas Commun.}, vol. 35, no. 12, pp. 2736\textendash2743, Dec. 2017.

\bibitem{2014_Nguyen_TSP_Leaning_methods}
A. H. Nguyen, Y. Huang, and B. D. Rao, ``Learning methods for CDF scheduling in multiuser heterogeneous systems,'' \textit{IEEE Trans. Signal Process.}, vol. 62, no. 15, pp. 3727\textendash3740, Aug. 2014.

\bibitem{2006_Bletsas_network_path_selection}
A. Bletsas, A. Khisti, D. P. Reed, and A. Lippman, ``A simple cooperative diversity method based on network path selection,'' \textit{IEEE J. Sel. Areas Commun.}, vol. 24, no. 3, pp. 659\textendash672, Mar. 2006.


\bibitem{2016_Ding_impact_of_pairing}
Z. Ding, P. Fan, and H. V. Poor, ``Impact of user pairing on 5G nonorthogonal multiple-access downlink transmissions,'' \textit{IEEE Trans. Veh. Technol.}, vol. 65, no. 8, pp. 6010\textendash6023, Aug. 2016.


\bibitem{diversity_order}
L. Zheng and D. N. C. Tse, ``Diversity and multiplexing: a fundamental tradeoff in multiple antenna channels,'' \textit{IEEE Trans. Inf. Theory}, vol. 49, pp. 1073\textendash1096, May 2003.


\bibitem{2020_Lim_OP_CS_NOMA}
B. Lim, S. S. Nam, Y.-C. Ko, and M.-S. Alouini, ``Outage analysis for downlink non-orthogonal multiple access (NOMA) with CDF-based scheduling,'' \textit{IEEE Wireless Commun. Lett.}, vol. 9, no. 6, pp. 822\textendash825, Jun. 2020.



\bibitem{2021_Yang_opportunistic_TWC}
L. Yang \textit{et al.}, ``Opportunistic adaptive non-orthogonal multiple access in multiuser wireless systems: Probabilistic user scheduling and performance analysis,'' \textit{IEEE Trans. Wireless Commun.}, vol. 19, no. 9, pp. 6065\textendash6082, Sep. 2020.


\bibitem{2018_Janghel_adaptive_COML}
K. Janghel and S. Prakriya, ``Performance of adaptive OMA/cooperative NOMA scheme with user selection,'' \textit{IEEE Commun. Lett.}, vol. 22, no. 10, pp. 2092\textendash2095, Oct. 2018.

\bibitem{2020_Ye_TCOM_MEC_NNNF}
{\color{black}Y. Ye, R. Q. Hu, G. Lu, and L. Shi, ``Enhance latency-constrained computation in MEC networks using uplink NOMA," \textit{IEEE Trans. Commun.}, vol. 68, no. 4, pp. 2409\textendash2425, Apr. 2020.}

\bibitem{2020_Wei_Gain_NOMA_OMA}
Z. Wei, L. Yang, D. W. K. Ng, J. Yuan, and L. Hanzo, ``On the performance gain of NOMA Over OMA in uplink communication systems,'' \textit{IEEE Trans. Commun.}, vol. 68, no. 1, pp. 536\textendash568, Jan. 2020.

\bibitem{2021_Sun_HSIC}
Y. Sun, Z. Ding, and X. Dai, ``A new design of hybrid SIC for improving transmission robustness in uplink NOMA,'' \textit{IEEE Trans. Veh. Technol.}, vol. 70, no. 5, pp. 5083\textendash5087, May 2021.

\bibitem{2013_Tabassum_ICI}
{\color{black}H. Tabassum, F. Yilmaz, Z. Dawy, and M.-S. Alouini, ``A statistical model of uplink inter-cell interference with slow and fast power control mechanisms," \textit{IEEE Trans. Commun.}, vol. 61, no. 9, pp. 3953\textendash3966, Sep. 2013.}


\bibitem{2017_Wei_fairness_comparision}
Z. Wei, J. Guo, D. W. K. Ng, and J. Yuan, ``Fairness comparison of uplink NOMA and OMA,'' in \textit{Proc. IEEE Veh. Technol. Conf. (VTC Spring)}, Jun. 2017, pp. 1\textendash6.



\bibitem{1984_jain}
R. Jain, D. Chiu, and W. Hawe, ``A quantitative measure of fairness and discrimination for resource allocation in shared systems'' Digit. Equip. Corp., Maynard, MA, USA, Tech. Rep. DEC-TR-301, 1984.

\bibitem{1993_Kingman_PP}
J. F. C. Kingman, \textit{Poisson Processes}. Oxford, U.K.: Oxford Univ. Press, 1993.




%
%
%
%
%
%
%
%

%
%
%
%
%
%
%
%
%
%
%
%
%



%
%
%
%



%
%
%
%
%
%
%
%
%
%
%
%
%
%
%
%
%
%
%
%
%
%
%
%
%
%
%
%
%
%
%
%
%
%
%
%
%
%

%
%
%
%
%
%
%
%
%
%
%
%
%
%
%
%
%
%
%
%
%
%
%
%
%
%
%
%
%
%
%

%
%
%
%
%
%
%

%
%
%
%
%
%



	\end{thebibliography}
	
\end{document}